\documentclass[12pt, letterpaper]{article} 
\usepackage[english]{babel}
\usepackage[latin1]{inputenc}
\usepackage{graphicx}       
\usepackage{listings}
\usepackage{url}
\usepackage{hyperref}
\usepackage{inputenc}
\usepackage{listings}
\usepackage{epsfig}
\usepackage{amsmath,amssymb,amsfonts} 
\usepackage{relsize}
\usepackage{pstricks}
\usepackage{multirow}
\usepackage[all]{xypic}
\usepackage{colortbl}
\usepackage{rotating}
\usepackage{subfigure}
\usepackage{float}
\usepackage{fancyhdr}
\usepackage{ae}
\usepackage{natbib}
\usepackage{multicol}
\usepackage{enumerate}
\usepackage{wrapfig}
\usepackage[hang,flushmargin]{footmisc}
\usepackage{bold-extra}

\newtheorem{definition}{Definition}
\newtheorem{theorem}{Theorem}

\usepackage[explicit]{titlesec}
\usepackage[normalem]{ulem}

\usepackage{ccaption}
\captionnamefont{\bfseries}

\bibpunct{(}{)}{;}{a}{}{,}

\usepackage[margin=1.0in]{geometry}

\newcommand{\betaABBB}{\beta^{\scriptsize\arraycolsep=0.0pt\def\arraystretch{0.0}\begin{array}{@{}l@{}l@{}} & \\\framebox(5.0,5.0){} & \end{array}}}
\newcommand{\betaAABB}{\beta^{\scriptsize\arraycolsep=0.0pt\def\arraystretch{0.0}\begin{array}{@{}l@{}l@{}}\framebox(5.0,5.0){} & \framebox(5.0,5.0){} \\ &\end{array}}}
\newcommand{\betaABAB}{\beta^{\scriptsize\arraycolsep=0.0pt\def\arraystretch{0.0}\begin{array}{@{}l@{}l@{}} \framebox(5.0,5.0){}& \\ \framebox(5.0,5.0){}  & \end{array}}}
\newcommand{\betaABBA}{ \beta^{\scriptsize\arraycolsep=0.0pt\def\arraystretch{0.0}\begin{array}{@{}l@{}l@{}}\framebox(5.0,5.0){} & \\ &\framebox(5.0,5.0){}\end{array}}}
\newcommand{\betaBAAB}{ \beta^{\scriptsize\arraycolsep=0.0pt\def\arraystretch{0.0}\begin{array}{@{}l@{}l@{}} &\framebox(5.0,5.0){} \\ \framebox(5.0,5.0){} & \end{array}}}

\newcommand{\betaBAAA}{ \beta^{\scriptsize\arraycolsep=0.0pt\def\arraystretch{0.0}\begin{array}{@{}l@{}l@{}} &\framebox(5.0,5.0){} \\ \framebox(5.0,5.0){} & \framebox(5.0,5.0){} \end{array}}}
\newcommand{\betaABAA}{ \beta^{\scriptsize\arraycolsep=0.0pt\def\arraystretch{0.0}\begin{array}{@{}l@{}l@{}} \framebox(5.0,5.0){}& \\ \framebox(5.0,5.0){} & \framebox(5.0,5.0){} \end{array}}}
\newcommand{\betaAABA}{ \beta^{\scriptsize\arraycolsep=0.0pt\def\arraystretch{0.0}\begin{array}{@{}l@{}l@{}} \framebox(5.0,5.0){} &\framebox(5.0,5.0){} \\ & \framebox(5.0,5.0){} \end{array}}}
\newcommand{\betaAAAB}{ \beta^{\scriptsize\arraycolsep=0.0pt\def\arraystretch{0.0}\begin{array}{@{}l@{}l@{}} \framebox(5.0,5.0){} &\framebox(5.0,5.0){} \\ \framebox(5.0,5.0){} & \end{array}}}

\newcommand{\betaAAAA}{ \beta^{\scriptsize\arraycolsep=0.0pt\def\arraystretch{0.0}\begin{array}{@{}l@{}l@{}} \framebox(5.0,5.0){} &\framebox(5.0,5.0){} \\ \framebox(5.0,5.0){} & \framebox(5.0,5.0){} \end{array}}}

\newcommand{\confBBBB}{\scriptsize \def\arraystretch{0.3}\begin{array}{@{}c@{}c@{}} 0 & 0 \\ 0 & 0 \end{array}}
\newcommand{\confABBB}{\scriptsize \def\arraystretch{0.3}\begin{array}{@{}c@{}c@{}} 1 & 0 \\ 0 & 0 \end{array}}
\newcommand{\confBABB}{\scriptsize \def\arraystretch{0.3}\begin{array}{@{}c@{}c@{}} 0 & 1 \\ 0 & 0 \end{array}}
\newcommand{\confBBAB}{\scriptsize \def\arraystretch{0.3}\begin{array}{@{}c@{}c@{}} 0 & 0 \\ 1 & 0 \end{array}}
\newcommand{\confBBBA}{\scriptsize \def\arraystretch{0.3}\begin{array}{@{}c@{}c@{}} 0 & 0 \\ 0 & 1 \end{array}}
\newcommand{\confAABB}{\scriptsize \def\arraystretch{0.3}\begin{array}{@{}c@{}c@{}} 1 & 1 \\ 0 & 0 \end{array}}
\newcommand{\confBBAA}{\scriptsize \def\arraystretch{0.3}\begin{array}{@{}c@{}c@{}} 0 & 0 \\ 1 & 1 \end{array}}
\newcommand{\confABAB}{\scriptsize \def\arraystretch{0.3}\begin{array}{@{}c@{}c@{}} 1 & 0 \\ 1 & 0 \end{array}}
\newcommand{\confBABA}{\scriptsize \def\arraystretch{0.3}\begin{array}{@{}c@{}c@{}} 0 & 1 \\ 0 & 1 \end{array}}
\newcommand{\confABBA}{\scriptsize \def\arraystretch{0.3}\begin{array}{@{}c@{}c@{}} 1 & 0 \\ 0 & 1 \end{array}}
\newcommand{\confBAAB}{\scriptsize \def\arraystretch{0.3}\begin{array}{@{}c@{}c@{}} 0 & 1 \\ 1 & 0 \end{array}}
\newcommand{\confBAAA}{\scriptsize \def\arraystretch{0.3}\begin{array}{@{}c@{}c@{}} 0 & 1 \\ 1 & 1 \end{array}}
\newcommand{\confABAA}{\scriptsize \def\arraystretch{0.3}\begin{array}{@{}c@{}c@{}} 1 & 0 \\ 1 & 1 \end{array}}
\newcommand{\confAABA}{\scriptsize \def\arraystretch{0.3}\begin{array}{@{}c@{}c@{}} 1 & 1 \\ 0 & 1 \end{array}}
\newcommand{\confAAAB}{\scriptsize \def\arraystretch{0.3}\begin{array}{@{}c@{}c@{}} 1 & 1 \\ 1 & 0 \end{array}}
\newcommand{\confAAAA}{\scriptsize \def\arraystretch{0.3}\begin{array}{@{}c@{}c@{}} 1 & 1 \\ 1 & 1 \end{array}}

\newcommand{\confcABAB}{\scriptsize \def\arraystretch{0.3}\begin{array}{@{}c@{}} 1 \\ 1 \end{array}}
\newcommand{\confcABBA}{\scriptsize \def\arraystretch{0.3}\begin{array}{@{}c@{}c@{}} 1 &  \\  & 1 \end{array}}
\newcommand{\confcBAAB}{\scriptsize \def\arraystretch{0.3}\begin{array}{@{}c@{}c@{}}  & 1 \\ 1 &  \end{array}}
\newcommand{\confcBAAA}{\scriptsize \def\arraystretch{0.3}\begin{array}{@{}c@{}c@{}}  & 1 \\ 1 & 1 \end{array}}
\newcommand{\confcABAA}{\scriptsize \def\arraystretch{0.3}\begin{array}{@{}c@{}c@{}} 1 & \\ 1 & 1 \end{array}}
\newcommand{\confcAABA}{\scriptsize \def\arraystretch{0.3}\begin{array}{@{}c@{}c@{}} 1 & 1 \\  & 1 \end{array}}
\newcommand{\confcAAAB}{\scriptsize \def\arraystretch{0.3}\begin{array}{@{}c@{}c@{}} 1 & 1 \\ 1 &  \end{array}}
\newcommand{\confcAAAA}{\scriptsize \def\arraystretch{0.3}\begin{array}{@{}c@{}c@{}} 1 & 1 \\ 1 & 1 \end{array}}

\newcommand{\confAABABBBBB}{\scriptsize \def\arraystretch{0.2}\begin{array}{@{}c@{}c@{}c@{}} 1 & 1 & 0 \\ 1 & 0 & 0 \\ 0 & 0 & 0 \end{array}}
\newcommand{\confBAABABBBB}{\scriptsize \def\arraystretch{0.2}\begin{array}{@{}c@{}c@{}c@{}} 0 & 1 & 1 \\ 0 & 1 & 0 \\ 0 & 0 & 0 \end{array}}
\newcommand{\confBBBAABABB}{\scriptsize \def\arraystretch{0.2}\begin{array}{@{}c@{}c@{}c@{}} 0 & 0 & 0 \\ 1 & 1 & 0 \\ 1 & 0 & 0 \end{array}}
\newcommand{\confBBBBAABAB}{\scriptsize \def\arraystretch{0.2}\begin{array}{@{}c@{}c@{}c@{}} 0 & 0 & 0 \\ 0 & 1 & 1 \\ 0 & 1 & 0 \end{array}}

\newcommand\vv{\cellcolor{gray!25}}

\usepackage{xr}
\begin{document}
	
	

{\noindent{\LARGE \textbf{Fully Bayesian binary Markov random field \\
      models: 
Prior specification and posterior\\ simulation} \vspace{1cm} \\} 
{\Large \textsc{Petter Arnesen}}\\{\it Department of
    Mathematical Sciences, Norwegian University of Science and
    Technology}\vspace{1cm}\\{\Large \textsc{H\aa kon Tjelmeland}}\\{\it Department of
    Mathematical Sciences, Norwegian University of Science and
    Technology}\vspace{1cm}
}	
                
{   \bf \noindent ABSTRACT:
We propose a flexible prior model for the
parameters of binary Markov random fields (MRF) defined on rectangular lattices and with maximal cliques defined from a template
maximal clique.
The prior model allows higher-order interactions to be
included.
We also define a reversible jump Markov chain Monte Carlo (RJMCMC)
algorithm to sample from the associated posterior distribution. 
The number of possible parameters for an MRF with for instance $k\times l$ maximal cliques becomes high even for small values of $k$ and $l$. To get 
a flexible model 
which may adapt to the structure of a particular observed image 
we do not put any absolute restrictions on the parametrisation. 
Instead we define a parametric form for the MRF where the 
parameters have interpretation as potentials for the 
various clique configurations, and limit the effective number 
of parameters by assigning apriori discrete probabilities 
for events where groups of parameter values are equal. 

To run our RJMCMC algorithm we have to cope with the 
computationally intractable normalising constant of MRFs. 
For this we adopt a previously defined approximation for
binary MRFs, but we also briefly discuss other alternatives. 
We demonstrate the flexibility of our prior formulation with 
simulated and real data examples.}
 
\vspace{0.5cm}
\vspace{-0.1cm}

\renewcommand{\baselinestretch}{1.5}   \small\normalsize

\section{Introduction}
Markov random fields (MRF) are frequently used as prior distributions
in spatial statistics. A common situation is that we have an observed 
or latent field $x$ which we model as an MRF, $p(x|\phi)$, conditioned 
on a vector of model parameters $\phi$. The most common situation in
the literature is to consider $\phi$ as fixed, see for instance
examples in \citet{Besag1986} and \citet{moller2003}, but several
articles have also considered a fully Bayesian approach by assigning a 
prior on $\phi$. A fully Bayesian model is computationally 
simplest when $x|\phi$ is a Gaussian Markov random field (GMRF)
and this case is therefore especially well developed. A flexible
implementation of the GMRF case is given in the integrated
nested Laplace approximation (INLA) software, see
\citet{art129} and \citet{art143}. The case when the components of
$x$ are discrete variables is computationally much harder and therefore 
less developed in the literature. However, some articles have considered
the fully Bayesian approach also in this case, see in particular the 
early \citet{art109} and \citet{art144} and the more recent 
\citet{art120}, \citet{art117}, \citet{phd6}, \citet{art145} and \citet{tjelmeland2012}.

MRFs is a very flexible class of models. Formally, any distribution is an MRF with 
respect to a neighbourhood system where all nodes are neighbours of 
each other. For the MRF formulation to be of any help, however, reasonably small 
neighbourhoods must be adopted. The typical choice in the literature is 
to assume each node to have the nearest four, eight or 24 other nodes as 
neighbours. Moreover, in the model specification it is common to restrict 
oneself to models that include interactions between pairs of nodes only.
Such pairwise interaction priors are just token priors, unable to 
specify more spatial structure than that nodes close to each other
should tend to have the same value. In the literature it is often argued
that such token priors are sufficient in many applications, as the information that neighbour nodes should tend to have the 
same value is the information lacking in the observed data. In 
particular one typically gets much better results based on such a token prior
than by not including any spatial prior information in the analysis at all.
The main reason for resorting to pairwise interaction priors, in addition
to the argument that these are good enough, is that the class of 
MRFs with higher-order interactions is so large that it becomes 
difficult both to select a reasonable parametric form for the prior
and to specify associated parameter values, not to mention the 
specification of a hyper-prior if a fully Bayesian approach is 
adopted. However, \citet{DescombesEtAl} and \citet{TjelmelandBesag}
demonstrate that it is possible to 
specify MRFs with higher-order interactions that are able to model 
more spatial structure than a pairwise interaction MRF, 
and where the model parameters have a reasonable interpretation.

In this article we consider the fully Bayesian approach 
and for simplicity we limit the attention to the case 
where the components of $x$ are binary. Our focus is 
on the specification of a prior distribution and on 
simulation from the associated posterior distribution.
We define priors both on the parametric form of the MRF 
and on the parameter values. To the best of our knowledge
this is the first attempt on putting a prior on the 
parametric form of a discrete MRF. Other articles
considering a fully Bayesian approach in such a setting,
are using a fixed parametric model and put 
a prior on the parameter values only. One should note that
by assigning a prior to the parametric structure of the model, 
including the number of parameters, we get an automatic model 
choice when simulating from the posterior distribution. 

To be able to define a reasonable prior it is essential
to adopt a model where the parameters have a natural 
interpretation. In this article we consider two ways to
parametrise the MRF. The first approach is inspired by 
the so-called $u$-parameters commonly used in 
the log-linear and graphical model literature 
for contingency tables
\citep{Bishop1975,Forster1999,massam2009,overstall2012}. Here the 
parameters are interactions of different orders. To 
limit the complexity of the model is easy by 
restricting some of the parameters to be zero,
but we argue that the interpretation of the 
parameters is difficult. The second parametrisation 
we consider is inspired by the MRF formulation in
\citet{TjelmelandBesag}. The parameters then represent 
potentials for configurations in maximal cliques, 
and we limit the model complexity by restricting
different configurations to have the same potential.
In \citet{TjelmelandBesag} this grouping of 
configurations is done manually, whereas we assign a 
prior to the grouping so that it is done 
automatically in the posterior simulation. Thereby we 
do not need, for example, to specify apriori whether or not the field 
is isotropic. We argue that the interpretation of the
configuration potentials is much easier than
for the interactions, and unless any particular
prior information is available and suggest the opposite,
it is natural to assume the configuration potentials
to be on the same scale.

To explore the resulting posterior distribution we construct
a reversible jump MCMC (RJMCMC) algorithm \citep{green1995}.
To run this algorithm we have to cope with the computationally 
intractable normalising constant of the MRF. In the literature
several strategies for handling this have been proposed.
We adopt an approximation strategy for binary MRFs 
introduced in \citet{phd6}, where a partially ordered
Markov model (POMM), see \citet{art119}, approximation to the MRF is defined.
We simply replace the MRF with the corresponding POMM approximation.

The article has the following organisation. In Section \ref{sec:MRF}
we discuss the two parametrisations of binary MRFs, and in 
particular we identify the maximal number of free parameters for a
model with specified maximal cliques. In Section \ref{sec:prior}
we define our prior for $\phi$, and in Section \ref{sec:sampling}
we discuss how to handle the computationally intractable 
normalising constant and describe our RJMCMC algorithm for 
simulating from the posterior distribution.
In Section \ref{sec:results} we present results for one simulated data example 
and for one real data example. One additional simulated example is
given in the supplementary materials. Finally, 
some closing remarks are provided in Section \ref{sec:discussion}.

\section{MRF}
\label{sec:MRF}
In this section we give a brief introduction to MRFs, see
\cite{cressie1993} and \cite{moller2003} for more details, and
in particular we focus on binary MRFs and the parametrisation in this
case. We close with one example of a binary MRF, the Ising model. This section provides
the theoretical background needed in order to understand the
construction of our prior distribution in Section \ref{sec:prior}, and
the RJMCMC algorithm given in Section \ref{sec:sampling}.    

\subsection{Binary MRF}
\label{sec:binaryMRF}

Consider a rectangular lattice of dimension $n\times m$, and let the
nodes be identified by $(i,j)$ where $i=0,...,n-1$ and $j=0,...,m-1$. To each node $(i,j)\in
S=\{(i,j);i=0,...,n-1,j=0,...,m-1\}$ we associate a binary variable
$x_{i,j}\in \{0,1\}$, and let $x=(x_{i,j};(i,j)\in S)$ be the collection of these
binary variables. We let $x_A=(x_{i,j};(i,j)\in A)$ denote the collection of 
the binary variables with indices belonging to an index set
$A\subseteq S$, and let
$x_{-(i,j)}=x_{S\setminus \{(i,j)\}}$. Associating zero with black and one with white we may 
thereby say the $x$ specifies a colouring of the nodes. We let 
$\mathcal{N}=\{\mathcal{N}_{0,0},...,\mathcal{N}_{n-1,m-1}\}$ be
a neighbourhood system on $S$, where $\mathcal{N}_{i,j}\subseteq S \setminus \{(i,j)\}$ is 
the set of neighbour
nodes of node $(i,j)$. We assume symmetry in the neighbour sets, so if $(i,j)\in \mathcal{N}_{t,u}$,
then also $(t,u)\in \mathcal{N}_{i,j}$. Now, $x$ is a binary MRF if $p(x)>0$ for all
$x$, and $p(x_{i,j}|x_{-(i,j)})$ fulfils the Markov property 
\begin{equation}
p(x_{i,j}|x_{-(i,j)})=p(x_{i,j}|x_{\mathcal{N}_{i,j}}) \text{ for all}
\ (i,j)\in S.
\end{equation}
A clique is defined to be a set $\lambda\subseteq
S$, where $(i,j)\in \mathcal{N}_{t,u}$ for all distinct pair of
nodes $(i,j),(t,u)\in \lambda$,
and we denote the set of all cliques by $\mathcal{L}$. Note that by this
definition sets containing only one node and the empty set are cliques. A
maximal clique is defined to be a clique that is not a subset of
another clique, and we denote the set of all maximal cliques by
$\mathcal{L}_m$. Moreover, for $\lambda\in \mathcal{L}$ we let 
$\mathcal{L}_m^\lambda$ denote the set of all maximal 
cliques that contains $\lambda$, i.e. $\mathcal{L}_m^\lambda=\{\Lambda \in \mathcal{L}_m;\lambda \subseteq
\Lambda\}$.
In the following we use $\Lambda$ and $\Lambda^\star$ to denote
maximal cliques, i.e. $\Lambda,\Lambda^* \in \mathcal{L}_m$, whereas we use $\lambda$ and
$\lambda^\star$ to denote cliques that do not need to be maximal,
i.e. $\lambda,\lambda^\star\in \mathcal{L}$. To denote an $x$ where $x_{i,j}=1$
for all $(i,j)\in A$ for some $A\subseteq S$ and $x_{i,j}=0$ otherwise,
we use $\bold 1^A = \{ x_{i,j}=I( (i,j)\in A); (i,j)\in S\}$. Thereby a colouring
of the nodes in a maximal clique $\Lambda\in\mathcal{L}_m$ may be specified
by $\bold 1^\lambda_\Lambda$, where $\lambda\subseteq \Lambda$
specifies the set of nodes in $\Lambda$ that has the value one.

According to the Hammersley-Clifford theorem
\citep{Clifford}, the most general form the distribution $p(x)$ of an
MRF can take is 
\begin{equation}
p(x)=Z\exp\{ U(x)\} \mbox{~~with~~} \ U(x)=\sum_{\Lambda \in \mathcal{L}_m}
    V_{\Lambda}(x_{\Lambda}),
\label{eq:cliqueRep}
\end{equation}
where $Z$ is the
computationally demanding normalising constant,
$U(x)$ is frequently called the energy function, 
and $V_{\Lambda}(x_{\Lambda})$ is a potential function for
$\Lambda$. A naive parametrisation of $V_{\Lambda}(x_{\Lambda})$ is to
introduce one parameter for each
possible $\Lambda\in \mathcal{L}_m$ and $x_{\Lambda}\in
\{0,1\}^{|\Lambda|}$ by setting
\begin{equation}\label{eq:phi}
V_{\Lambda}(\bold 1^\lambda_{\Lambda})=\phi^\lambda_{\Lambda}.
\end{equation} 
It is a well known fact the $\phi^\lambda_\Lambda$ parameters do not
constitute a unique representation of $U(x)$. Thereby, in the resulting parametric model $p(x)$ the $\phi^\lambda_\Lambda$ parameters
are not identifiable, meaning
that different choices for the $\phi^\lambda_\Lambda$ parameters may give the same
model $p(x)$. For example, adding the same value to all $\phi^\lambda_\Lambda$ 
parameters will not change the model, 
as this will be compensated for by a corresponding change in the normalising 
constant $Z$. If the set of maximal cliques $\mathcal{L}_m$
consists of, for example, all $2\times 2$ blocks of nodes a perhaps less 
obvious way to change the parameter values without changing
the model and neither the normalising constant, is to add an 
arbitrary value to $\phi^{\{ (i,j)\}}_\Lambda$ for some
$(i,j)\in\Lambda\in\mathcal{L}_m$, and to subtract the same 
value from $\phi^{\{(i,j)\}}_{\Lambda^\star}$ for some 
$\Lambda^\star\in\mathcal{L}_m$, $\Lambda^\star\neq \Lambda$
for which $(i,j)\in\Lambda^\star$.

An alternative way to represent an MRF is through a parametrisation of
the cliques. The energy function $U(x)$ is a pseudo-Boolean function and 
when it is given as in (\ref{eq:cliqueRep}) \cite{tjelmeland2012} show that 
it can be represented as
\begin{equation}
U(x)=\sum_{\lambda \in
    \mathcal{L}}\beta^{\lambda}\prod_{(i,j)\in\lambda}x_{i,j},
\label{eq:DAGrep}
\end{equation} 
where $\beta^{\lambda}$ is referred to as the interaction parameter for
clique $\lambda$, which is said to be of $|\lambda|$'th order. More
details on pseudo-Boolean functions and their properties can be found
in \cite{Grabisch2000} and \cite{Hammer1992}. Since
this representation consists of linearly independent functions of $x$,
it is clear that the set of interaction parameters is a unique representation of $U(x)$. 
Furthermore, in the corresponding parametric model $p(x)$ the $\beta^\lambda$ 
parameters become
identifiable if fixing $\beta^\emptyset$ to zero (say). 
We 
note in passing that \citet{Besag1974} uses the 
representation in \eqref{eq:DAGrep} in a proof for 
the Hammersley--Clifford theorem.

In the following we define a set of constraints on the $\phi^\lambda_\Lambda$
parameters in \eqref{eq:cliqueRep} and show that subject to these 
constraints there is a one-to-one relation between the 
$\phi^\lambda_\Lambda$ parameters and the
interaction parameters $\beta^\lambda$.
The constrained $\phi^\lambda_\Lambda$ parameters thereby constitute an 
alternative unique representation of $U(x)$.

\begin{definition}\label{def:constraints}
The constrained set of $\phi^\lambda_\Lambda$ parameters are defined by 
requiring that $\phi^\lambda_\Lambda=\phi^\lambda_{\Lambda^\star}$ for all
$\Lambda,\Lambda^\star\in\mathcal{L}_m$, $\lambda\subseteq \Lambda \cap \Lambda^\star$.
To simplify the notation we then write $\phi^\lambda_\Lambda=\phi^\lambda$.
\end{definition}
To understand the implication of the constraint one may again consider the 
situation where the set of maximal cliques $\mathcal{L}_m$ consists of all 
$2\times 2$ blocks of nodes, and focus on the two overlapping maximal cliques
$\Lambda = \{ (i,j-1),(i+1,j-1),(i,j),(i+1,j)\}$ and 
$\Lambda^\star = \{(i,j),(i+1,j),(i,j+1),(i+1,j+1)\}$. For $\lambda=
\{ (i,j),(i+1,j)\}$ the constraint is that the potential $V_\Lambda(x_\Lambda)$ for the colouring $\confBABA$
in $\Lambda$ is the same as the potential $V_{\Lambda^\star}(x_{\Lambda^\star})$ for the
colouring
$\confABAB$
in $\Lambda^\star$. One should also note that the constraint implies that 
$\phi^\emptyset_\Lambda$ is the same for all $\Lambda\in\mathcal{L}_m$, so in 
the $2\times 2$ maximal cliques case the potential for the colouring 
$\confBBBB$
must be the same for all maximal cliques.

\begin{theorem}\label{th:onetoone}
Consider an MRF and constrain the $\phi$ parametrisation of the potential functions as described in
Definition \ref{def:constraints}. Then there is a one-to-one relation between 
$\{\beta^\lambda;\lambda\in\mathcal{L}\}$
and $\{\phi^\lambda;\lambda\in\mathcal{L}\}$. 
\end{theorem}
The proof is given in the supplemental material, and the result is 
shown by establishing recursive equations
showing how to compute the $\beta^\lambda$'s from 
the $\phi^\lambda$'s and vice versa.

To simplify the definition of a prior for the parameter
vector of an MRF in the next section, we first limit
the attention to {\em stationary} MRFs defined on 
a rectangular  $n\times m$ lattice, and to obtain 
stationarity we assume {\em torus boundary conditions}.
In the following we
define the concepts of stationarity and 
torus boundary conditions and states two 
theorems which identify what properties the
$\{\beta^\lambda;\lambda\in\mathcal{L}\}$ parameters
and the $\{ \phi^\lambda;\lambda\in\mathcal{L}\}$ 
parameters must have
for the MRF to be stationary.
\begin{definition}\label{def:torus}
If, for a rectangular lattice
$S = \{ (i,j);i=0,\ldots,n-1,
j=0,\ldots,m-1\}$, the translation of a node
$(i,j)\in S$ with an amount $(t,u)\in S$
is defined to be
\begin{equation*}
(i,j) \oplus (t,u) = (i+t \mod n, j+u \mod m),
\end{equation*}
we say that the lattice has torus boundary conditions.
\end{definition}
To denote translation of a set of nodes $A\subseteq S$ by an
amount $(t,u)\in S$ we write
$A \oplus (t,u) = \{ (i,j)\oplus (t,u) ; (i,j)\in A\}$.
With this notation stationarity of an MRF defined
on a rectangular lattice with torus boundary conditions 
can be defined as follows.
\begin{definition}\label{def:stationarity}
An MRF $x$ defined on a rectangular lattice
$S$ with torus boundary
conditions is said to be stationary if and only if
$p(\bold 1^A) = p(\bold 1^{A \oplus (t,u)})$ for all
$A\subseteq S$ and $(t,u)\in S$.
\end{definition}
To explore what restrictions the stationarity assumption
puts on the $\beta^\lambda$ and $\phi^\lambda$ parameters
we assume the set of maximal cliques to consist of all 
possible translations of a given nonempty template set $\Lambda_0\subset S$, i.e.
\begin{equation}\label{eq:Lm}
\mathcal{L}_m = \{ \Lambda_0 \oplus (t,u); (t,u)\in S\}.
\end{equation}
For example,
with $\Lambda_0 = 
\{ (0,0),(0,1),(1,0),(1,1)\}$ the set of maximal cliques
will consist of all $2\times 2$ blocks of nodes. One 
should note that with the
torus boundary assumption there is always $|\mathcal{L}_m|=nm$ 
maximal cliques.
\begin{theorem}\label{th:betaStationarity}
An MRF $x$ defined on a rectangular lattice
$S=\{ (i,j);i=0,\ldots,n-1,j=0,\ldots,m-1\}$ with 
torus boundary conditions and $\mathcal{L}_m$ given
in \eqref{eq:Lm} is stationary if and only 
if $\beta^{\lambda}=\beta^{\lambda\oplus (t,u)}$ for all 
$\lambda \in \mathcal{L}$, $(t,u)\in S$. We then say
that $\beta^\lambda$ is translational invariant.
\end{theorem}
The proof is again given in the supplemental material. We proof the 
if part of the theorem by induction on $|\lambda|$, and the only if part
by direct manipulation with the expression for the energy function.

To better understand the effect of the theorem we can again consider the 
$2\times 2$ maximal clique case, i.e. $\mathcal{L}_m$ is 
given by \eqref{eq:Lm} with $\Lambda_0=\{ (0,0),(0,1),(1,0),(1,1)\}$. 
The translational invariance means that all first-order interactions 
$\{\beta^{\{ (i,j)\}},(i,j)\in S\}$ must be equal and in the following we 
denote their common value by $\betaABBB$, 
where the idea is that the superscript represents any node 
$(i,j)\in S$. Correspondingly we use $\betaAABB$, where the superscript
represent any two horizontally first-order neighbours, to denote 
the common value for all 
$\{\beta^{\{ (0,0),(0,1)\}\oplus (t,u)},(t,u)\in S\}$.
Continuing in this way we get, in addition to $\betaABBB$, $\betaAABB$ 
and the constant term $\beta^\emptyset$,
the parameters
$\betaABAB$, $\betaABBA$, $\betaBAAB$, $\betaAAAB$, $\betaAABA$, 
$\betaABAA$, $\betaBAAA$ and $\betaAAAA$.
We collect the eleven parameter values necessary to represent $U(x)$
in this stationary MRF case 
into a vector which we denote by $\beta$, i.e.
\begin{equation}
\beta = \left(
\beta^\emptyset,\betaABBB,\betaAABB,\betaABAB,\betaABBA,\betaBAAB,
\betaAAAB,\betaAABA,\betaABAA,\betaBAAA,\betaAAAA\right).
\label{eq:beta2x2}
\end{equation}

The next theorem gives a similar result for the $\phi^\lambda$ 
parameters as Theorem \ref{th:betaStationarity} did for the 
interaction parameters $\beta^\lambda$.
\begin{theorem}\label{th:phiStationarity}
An MRF $x$ defined on a rectangular lattice 
$S=\{ (i,j);i=0,\ldots,n-1,j=0,\ldots,m-1\}$ with torus boundary
conditions and $\mathcal{L}_m$ given in 
\eqref{eq:Lm} is stationary if and only if
$\phi^{\lambda}=\phi^{\lambda\oplus (t,u)}$ for all 
$\lambda \in \mathcal{L}$ and $(t,u)\in S$. We then say
that $\phi^\lambda$ is translational invariant. 
\end{theorem}
The proof is again given in the supplemental material. Given the result in 
Theorem \ref{th:betaStationarity} it is 
sufficient to show that $\phi^\lambda$ is translational invariant
if and only if $\beta^\lambda$ is translational invariant, and we show
this by induction on $|\lambda|$.

It should be noted that the 
interpretation of the $\phi^\lambda$ parameters is very different from the 
interpretation of the $\beta^\lambda$ parameters. Whereas the $\beta^\lambda$ 
parameters relates to cliques $\lambda$ of different sizes, all the 
$\phi^\lambda$'s represent the potential of a maximal clique $\Lambda\in\mathcal{L}_m$,
which are all of the same size. The effect of the above theorem is that we
get groups of configurations in maximal cliques that must be assigned the same potential, hereafter referred to as configuration sets. 
We let $\cal C$ denote the set of these configuration sets. In the $2\times 2$ maximal clique
case for example, we get
\begin{eqnarray} \nonumber
{\cal C} &=& \left\{ \left\{ \left[\confBBBB\right]\right\},
\left\{ \left[\confABBB\right],\left[ \confBABB\right],\left[\confBBAB\right],\left[\confBBBA\right]\right\},
\left\{ \left[\confAABB\right],\left[\confBBAA\right]\right\},
\left\{ \left[\confABAB\right],\left[\confBABA\right]\right\},
\right.\\
&& \left. 
\left\{\left[ \confABBA\right]\right\},
\left\{\left[ \confBAAB\right]\right\},
\left\{\left[ \confAAAB\right]\right\},
\left\{\left[ \confAABA\right]\right\},
\left\{\left[ \confABAA\right]\right\},
\left\{\left[ \confBAAA\right]\right\},
\left\{\left[ \confAAAA\right]\right\}
\right\}
\label{eq:confsubset}
\end{eqnarray}
We denote these sets of configurations by $c^0$, $c^1$, $c^{11}$, $c^{\confcABAB}$,
$c^{\confcABBA}$, $c^{\confcBAAB}$, $c^{\confcAAAB}$, $c^{\confcAABA}$, $c^{\confcABAA}$, $c^{\confcBAAA}$ and $c^{\confAAAA}$
when listed in the same order as in \eqref{eq:confsubset}, where the idea of the notation is that the 
$1$'s in the superscript can be placed anywhere inside a maximal clique and the remaining nodes takes the value of 
zero. One should note that a similar notation can be used in other sets of maximal cliques. In the $3\times 3$
maximal clique case we have for example
\[
c^{\confcAAAB} = \left\{\left[\confAABABBBBB\right],\left[\confBAABABBBB\right],\left[\confBBBAABABB\right],
\left[\confBBBBAABAB\right]\right\}.
\]
Associated to each member $c\in {\cal C}$ we thus have a corresponding parameter value $\phi(c)$ 
which is the potential assigned
to any maximal clique configuration in the set $c$. We use corresponding superscripts for the $\phi$ parameters as we did
for the sets $c\in{\cal C}$. In the $2\times 2$ maximal clique case we thereby get the parameter vector
\[
\phi = \left (\phi^0, \phi^1, \phi^{11}, \phi^{\confcABAB},\phi^{\confcABBA},\phi^{\confcBAAB},\phi^{\confcAAAB},\phi^{\confcAABA},\phi^{\confcABAA}, 
\phi^{\confcBAAA},\phi^{\confAAAA} \right),
\]
where for example $\phi^1$ is the potential for the four maximal clique configurations in $c^1$.

We end this section with a discussion on how the above stationary MRF
defined with torus boundary condition 
can be modified in the free boundary case. Using the same template maximal clique $\Lambda_0$
as before, the set of maximal cliques $\mathcal{L}_m$ now has to be redefined relative to the torus boundary condition case.
In the free boundary case we let $\mathcal{L}_m$ contain all translations of $\Lambda_0$ that are 
completely inside our $n\times m$ lattice, i.e.
\[
\mathcal{L}_m = \{ \Lambda_0 + (t,u); t=-n,\ldots,n,u=-m,\ldots,m, \Lambda_0 + (t,u)\subseteq S\},
\]
where $\Lambda_0 + (t,u) = \{ (i+t,j+u); (i,j)\in \Lambda_0\}$.
One should note that for a free boundary MRF the translational invariance property 
of the $\phi^\lambda$ parameters identified in Theorem \ref{th:phiStationarity} no 
longer apply, and neither will such a model be stationary. However,
the extra free $\phi$ parameters that may be introduced in the 
free boundary case will only model properties sufficiently close to
a boundary of the lattice. Our strategy in the free boundary case 
is to keep the same $\phi$ parameter vector as in the torus case,
to adopt translational invariant potential functions $V_\Lambda(\bold 1_\Lambda^\lambda)=\phi^\lambda$
for all maximal cliques $\Lambda\in\mathcal{L}_m$ just as in the torus case, but to add non-zero 
potential functions for some (non-maximal) cliques at the 
boundaries of the lattice. Our motivation for this is to reduce the
boundary effect and, hopefully, to get a model which is less non-stationary. To define our 
non-zero potential functions at the boundaries, imagine that our
$n\times m$ lattice is included in a much larger lattice and that 
this extended lattice also has maximal cliques that are translations of $\Lambda_0$. 
We then
include a non-zero potential function for every maximal clique
in the extended lattice which is partly inside and partly outside 
our original $n\times m$ lattice. In such a maximal clique in the extended lattice, let $\lambda$ denote 
the set of nodes that are inside our $n\times m$ lattice, and let $\lambda^\star$
denote the set of nodes outside. As we have assumed
that the maximal clique is partly inside and partly outside 
our original $n\times m$ lattice, $\lambda$ and $\lambda^\star$ are both non-empty and 
$\lambda\cup\lambda^\star$ is clearly a maximal clique in the extended 
lattice. For the (non-maximal) clique $\lambda$ we define 
the potential function
\begin{equation}\label{Uborder}
V_\lambda(x_\lambda) = \frac{1}{2^{|\lambda^\star|}}\sum_{x_{\lambda^\star}}
V_{\lambda\cup\lambda^\star}(x_{\lambda\cup\lambda^\star}),
\end{equation}
where $V_{\lambda\cup\lambda^\star}(x_{\lambda\cup\lambda^\star})$
is the same (translational invariant) potential function we are 
using for maximal cliques inside our $n\times m$ lattice. One can note 
that (\ref{Uborder}) corresponds to averaging over the 
values in the nodes outside our lattice, assuming them to 
be independent, and to take the values $0$ or $1$ with 
probability a half for each.

\subsection{Example: The Ising model}
\label{sec:Ising}
The Ising model \citep{Besag1986} is given by 
\begin{equation}
p(x)=Z\exp \left \{- \omega \sum_{(i,j)\sim (t,u)}I(x_{i,j}\neq x_{t,u})\right\},
\label{eq:Ising}
\end{equation}
where the sum is over all horizontally and vertically adjacent sites, 
and $\omega$ is a parameter controlling the probability of
adjacent sites having the same value. We use the Ising model as an
example also later in the paper, and in particular we fit an MRF with
$2\times 2$ maximal cliques to data simulated from this model. 
Assuming torus boundary conditions and using that for binary variables
we have $I(x_{i,j}\neq x_{t,u})=x_{i,j}+x_{t,u}-2x_{i,j}x_{t,u}$, we
can rewrite \eqref{eq:Ising} as 
\begin{equation*}
p(x)=Z\exp \left \{-4\omega\sum_{(i,j)\in S}x_{i,j} +2\omega\sum_{(i,j)\sim(t,u)}x_{i,j}x_{t,u} \right\}.
\end{equation*}
Thus, the $\beta^\emptyset$ can be given any value as this will be
compensated for by the normalising constant, whereas
$\beta^{\{(i,j)\}}=-4\omega$, $\beta^{\{(i,j),(i,j)\oplus(1,0)\}}=2\omega$ and
$\beta^{\{(i,j),(i,j)\oplus(0,1)\}}=2\omega$, and $\beta^{\lambda}=0$ for
all other cliques $\lambda$. The corresponding
$\phi^{\lambda}$ parameters can then be found using the recursive
equation (S2) identified in the proof of Theorem
\ref{th:onetoone}. Using the notation introduced above for the $2\times 2$
maximal clique case this gives $\phi^0=\phi^{\confcAAAA}=\eta$, $\phi^1=\phi^{11}=\phi^{\confcABAB}=\phi^{\confcAAAB}=\phi^{\confcAABA}=
\phi^{\confcABAA}=\phi^{\confcBAAA}=-\omega+\eta$ and $\phi^{\confcABBA}=\phi^{\confcBAAB}=-2\omega+\eta$,
where $\eta$ is an arbitrary value originating from the
arbitrary value that can be assign to $\beta^\emptyset$.

\section{Prior specification}
\label{sec:prior}    

In this section we define a generic prior for the parameters 
of an MRF with maximal cliques specified as in \eqref{eq:Lm}. 
The first step in the specification is to choose what parametrisation of the MRF to consider. 
In the previous section we discussed two
parametrisations for the MRF, with parameter vectors 
$\beta$ and $\phi$, respectively. When choosing between the two
parametrisations and defining the prior we primarily have the torus
version of the MRF in mind. However, as the free boundary version of 
the model is using the same parameter vectors, the prior we end up with 
can also be used in that case. It should be remembered that 
the parametrisations using $\beta$ and $\phi$ are non-identifiable,
but that it is sufficient to add one restriction to make them 
identifiable. The perhaps easiest way to do this is to 
restrict one of the parameters to equal zero, but other alternatives also
exist. We return to this issue below. 
The dimension of the $\beta$ and $\phi$ parameter vectors grows rapidly with 
the number of elements in the set $\Lambda_0$ defining the set of maximal 
cliques. Table \ref{tab:ktimesl} 
\begin{figure*}[h]\center
Table \ref{tab:ktimesl} approximately here.
\end{figure*}
gives the number of parameters in the
identifiable models, which we in the following 
denote by $N_{\Lambda_0}$, when $\Lambda_0$ is a $k\times l$ block of nodes.
We see that the number of parameters grows rapidly with the size of $\Lambda_0$.
It is therefore natural to 
look for prior formulations which include the possibility for a reduced
number of free parameters. For the $\beta$ parametrisation the 
perhaps most natural strategy to do this is to assign positive prior 
probability to the event that one or several of the interaction 
parameters are exactly zero.
The interpretation of the $\phi$ parameters
is different from the interpretation of the $\beta$ parameters,
and it is not natural to assign positive probability for elements
of the $\phi$ vector to be exactly zero. A more reasonable scheme here
is instead to set a positive prior probability for the event that groups of 
$\phi$ parameters have exactly the same value. 

In the Bayesian contingency tables literature
the $\beta$ parametrisation is popular, see for example
\citet{Forster1999}, \citet{massam2009} and \citet{overstall2012} and
references therein, where the second article develops a conjugate prior 
for this parametrisation. 
However, these results do not
directly apply for an MRF where one restricts the potential functions
to be translational invariant. More importantly, however, the various
$\beta$ parameters relates to cliques of different sizes and this makes 
the interpretation of the parameters difficult. In \citet{Forster1999} and
in \citet{overstall2012} effort is made in order to create a
reasonable multinormal prior for the $\beta$ parameters. In contrast, the $\phi$ 
parameters all represent the potential of a configuration of a maximal clique, 
which is all of the same size. Unless particular prior information is available and suggests the opposite, 
it is therefore natural to assume that all $\phi$ parameters are on the
same scale. A tempting option is therefore first to assign identical and independent normal
distributions to these parameters, and obtain identifiability by
constraining the sum of the parameters to be zero. Thereby the elements of $\phi$ are exchangeable \citep{diaconis1980}. Note that the $\beta$ parameters
become multinormal also in our case, see for instance (S3) in the
supplementary materials.  In the following we therefore focus on specifying a prior for $\phi$. 
We first introduce notation necessary to define the groups of 
configuration parameters $\phi$ that should have the same value and 
thereafter discuss possibilities for how to define the prior.

To define groups of configuration set parameters that should have the 
same value, let $C_1,\ldots,C_r$ be a partition of the configuration sets in $\mathcal{C}$ with $C_i\neq \emptyset$ for $i=1,\ldots,r$. 
Thus, $C_i\cap C_j = \emptyset$ for $i\neq j$ and 
$C_1\cup\ldots\cup C_r=\mathcal{C}$. For each $i=1,\ldots,r$ we thereby assume 
$\phi(c)$ to be equal for all $c\in C_i$, and we denote this common value by $\varphi_i$.
Setting $z=\{ (C_i,\varphi_i),i=1,\ldots,r\}$ we thus can write the resulting potential
functions as
\begin{equation}\label{eq:VFinal}
V_\Lambda(x_\Lambda|z) = \sum_{(C,\varphi)\in z} \varphi I\left(x_\Lambda \in \bigcup_{c\in C}  c\right).
\end{equation}
We define 
a prior on the $\phi$ parameters by specifying a prior for $z$. An
alternative to this construction would be to build up $\{C_1,...,C_r\}$ in a non-random fashion, 
constraining the $\phi$ parameters
according to properties like symmetry and rotational invariance. However, our goal is that such properties can be
inferred from observed data. 

Given all configuration sets, we want to
assign positive probability to the event that groups of configuration
sets have exactly the same parameter value. For instance, the
three groups in Section \ref{sec:Ising} is an example of such a
grouping for a $2\times 2$ maximal clique. Since we do not allow empty groups $C_i$, the maximum number of
groups one can get is $N_{\Lambda_0}+1$.
Our prior distribution for $z$ is on
the form
\begin{equation*}
p(z)=p(\{C_1,....,C_r\})p(\{\varphi_1,...,\varphi_r\}|r)
\end{equation*}
where $p(\{C_1,...,C_r\})$ is
a prior for the grouping of the configuration sets, while $p(\{\varphi_1,...,\varphi_r\}|r)$
is a prior for the group parameters given the number of groups $r$. Two possibilities
for $\{C_1,...,C_r\}$ immediately comes to
mind. The first is to assume a uniform distribution on the groupings, i.e.
\begin{equation*}
p_1(\{C_1,...,C_r\})\propto const,
\end{equation*}
meaning that each grouping is apriori equally likely. However for $p(r)$, the marginal probability of the number of
groups, this
means that most of the probability is put on groupings with approximately
$(N_{\Lambda_0}+1)/2$ groups. In fact the probability $p(r)$ becomes equal to
\begin{equation*}
p(r)=\frac{g(N_{\Lambda_0}+1,r)}{\sum_{i=1}^{N_{\Lambda_0}+1}g(N_{\Lambda_0}+1,i)},
\end{equation*}
where $g(N_{\Lambda_0}+1,r)$ is the number of ways $N_{\Lambda_0}+1$ configuration sets can be
organised into $r$ unordered groups, remembering that no empty groups
are allowed. The function $g(N_{\Lambda_0}+1,r)$ can be written as
\begin{equation*}
g(N+1,r)=\frac{1}{r!}\sum_{i=0}^r\binom{r}{i}(-1)^{r-i}i^{N+1},
\end{equation*}
and is known as the Stirling number of the second kind \citep{Graham1988}. 
For the $2\times 2$ maximal clique this means for instance
that $p(r=1)=p(r=11)\approx 10^{-6}$ while $p(r=5)=0.36$. An
alternative for $p(\{C_1,...,C_r\})$ is to make the
distribution for the number of groups uniform. This can be done by defining the probability distribution
\begin{equation*}
p_2(\{C_1,...,C_r\})=\frac{1}{(N_{\Lambda_0}+1)g(N_{\Lambda_0}+1,r)}.
\end{equation*}
With this prior a particular grouping with many or few groups will have a larger
probability than a particular grouping with approximately
$(N_{\Lambda_0}+1)/2$ groups. In the $2\times 2$ case for example, the
probability of the grouping where all
configuration sets are assigned to the same group or the grouping
with 11 groups is $p(\{C_1\})=p(\{C_1,...,C_{11}\})=0.09$, while
the probability of a particular grouping with 5 groups is
$p(\{C_1,...,C_5\})\approx 10^{-7}$. Observe however, that with both
priors we have that the groups are uniformly distributed when the number of groups is given.
As a compromise between the two prior
distributions we propose  
\begin{equation*}
p(\{C_1,...,C_r\})\propto p_1(\{C_1,...,C_r\})^{1-\gamma}p_2(\{C_1,...,C_r\})^{\gamma},
\end{equation*}
where $0 \leq\gamma\leq 1$. 

As also discussed above, to get an identifiable model we need to put one additional restriction
on the elements of $\phi$, or alternatively on the $\varphi_i$ parameters. As we want the 
distribution $p(\varphi_1,\ldots,\varphi_r|r)$ to be exchangeable we want the restriction also
to be exchangeable in the $\varphi_i$ parameters, and set
\begin{equation}
\sum_{(C,\varphi)\in z}\varphi=0.
\label{eq:identifiability}
\end{equation}
Under this sum-to-zero restriction we assume the $\varphi_i$ apriori 
to be independent normal with zero mean and with a common variance
$\sigma_{\varphi}^2$. 
This fully defines
the prior for $z$, except that we have not specified values for the
two hyper-parameters $\gamma$ and $\sigma_{\varphi}^2$.
   
\section{Posterior sampling}\label{sec:sampling}
In this section we first discuss different strategies proposed
in the literature for how to handle the computationally intractable 
normalising constant in discrete MRFs, and in particular
discuss their applicability in our situation. Thereafter
we describe the RJMCMC algorithm we adopt for simulating
from our posterior distribution.

\subsection{Handling of the normalising constant}
\label{sec:normconstant}
Discrete MRFs contain a computationally intractable 
normalising constant and this makes the fully Bayesian approach
problematic. Three strategies have been proposed to 
circumvent or solve this problem. The first alternative is to replace 
the MRF likelihood with a computationally tractable 
approximation. The early \citet{art109} use the pseudo-likelihood for 
this, \citet{art117} and \citet{art145} adopt a reduced dependency 
approximation (RDA), and \citet{phd6} and \citet{tjelmeland2012}
construct a POMM approximation by making use of theory for 
pseudo-Boolean functions. The second strategy, used in 
\citet{art144}, is to adopt an estimate of the normalisation constant obtained
by some Markov chain Monte Carlo (MCMC) procedure prior to 
simulating from the posterior, and the third alternative is to include an auxiliary variable sampled 
from the MRF $p(x|\phi)$ in the posterior simulation algorithm. 
\citet{art120} is the first article using the third approach, and 
the exchange algorithm of \citet{pro20} falls within the same
class. \cite{CaimoFriel2011} and \citet{Everitt2012} adopt an 
approximate version of this third approach, by replacing
perfect sampling from $p(x|\phi)$ with approximate sampling
via an MCMC algorithm.

The three approaches all have their advantages and disadvantages.
First of all, only the third approach is without approximations
in the sense that it defines an MCMC
algorithm with limiting distribution exactly equal to the 
posterior distribution of interest. However, for this approach 
to be feasible perfect sampling from $p(x|\phi)$ must 
be possible, and computationally reasonably efficient,
for all values of $\phi$. The strategy used in the second 
class requires in practice that the parameter vector
$\phi$ is low dimensional. The approximation 
strategy does not have restrictions on the dimension of $\phi$
and perfect sampling from $p(x|\phi)$ is not needed. In that 
sense this approach is more flexible, but of course the 
approximation quality may depend on the the parametric form 
of the MRF and the value of $\phi$.

In principle 
any of the approaches discussed above may be used in our
situation, but the
complexity of the parameter space makes the prior estimation of the 
normalisation constant approach impractical. Moreover, the 
accuracy of the pseudo-likelihood approximation is known to be quite poor, and
in simulation exercises we found that perfect sampling from $p(x|\phi)$
was in practice infeasible for many of the
higher-order interaction models visited by our RJMCMC algorithm.
The approximate version in \citet{CaimoFriel2011} is, however,
a viable alternative.
We are thereby left with the RDA approach, the POMM approximation,
and the strategy proposed in \citet{CaimoFriel2011}.
In our simulation examples we adopt the second of these, but the other two
could equally well have been used. In fact, in one of our simulation 
examples we use also the strategy from \citet{CaimoFriel2011} to check
the approximation quality obtained when replacing the MRF with the 
POMM approximation.

\subsection{MCMC algorithm}
Assume that an observed binary $n\times m$ image is 
available. We consider this image as a realisation from our MRF 
with the free boundary conditions defined in Section \ref{sec:MRF}. As
a prior for the MRF parameters we adopt the prior specified in 
Section \ref{sec:prior}.
The focus in this section is then on how to sample from the 
resulting posterior distribution. One should note that in 
this section we formulate the algorithm as if one can evaluate
the MRF likelihood, including the normalising constant. This is 
of course not feasible in practice, so when running the algorithm 
we replace the MRF likelihood with the corresponding POMM 
approximation discussed above.

Letting $x$ denote the observed
image, the posterior distribution we want to sample from is 
given by
\begin{equation*}
p(z|x) \propto p(x|z) p(z),
\end{equation*}
where $p(x|z)$ and $p(z)$ are the MRF defined by \eqref{eq:VFinal}
and the prior defined in Section
\ref{sec:prior}, respectively. 
To simulate from this posterior we adopt a reversible 
jump Markov chain Monte Carlo (RJMCMC) algorithm \citep{green1995} with 
three types of updates. The detailed proposal mechanisms are specified
in the supplementary materials, here we just give a brief description
of our proposal strategies.

The first proposal in our algorithm is simply first to propose a
change in an existing $\varphi$ parameter by a random walk proposal
with variance $\sigma^2$,
and thereafter to subtract the same value from all $\varphi$
parameters to commit with the sum-to-zero constraint. In the second proposal we draw a pair of groups and propose to move
one configuration set from the first group to the second group,
ensuring that the two groups are still non-empty. In the last proposal
type, we propose a new state by either increasing or
decreasing the number of groups with one. When increasing the number of
groups by one we randomly choose a configuration set from a
randomly chosen group and propose this configuration set to be a new
group. When proposing to reduce the number of parameters with
one, we randomly choose a group with only one configuration
set and propose to merge this group into another group. In the
trans-dimensional proposals we ensure that the proposed parameters
commit with the sum-to-zero constrain by subtracting the same value
from all $\varphi$ parameters.

\section{Simulation examples}
\label{sec:results}

In this section we first present an example based on a simulated
data set from the Ising model, and thereafter present results for a data set of census 
counts of red deer in the Grampians Region of north-east Scotland. In
addition, another example based on simulated data is included in the
supplementary materials. In all the simulation experiments we use the prior distribution
as defined in Section \ref{sec:prior}. In this prior the values of the two 
hyper-parameters $\sigma_\varphi$ and $\gamma$ must be specified.
We have fixed $\sigma_\varphi=10$ and tried $\gamma=0$, $0.5$ and
$1$. When discussing simulation results we first present results for 
$\gamma=0.5$. As the likelihood function we use the MRF discussed in 
Section \ref{sec:MRF} and we use $2\times 2$ maximal cliques except in the last part of
the real data example where we also discuss results for $3\times 3$ 
maximal cliques. To cope with the computationally intractable normalising 
constant of the MRF likelihoods, we adopt the approximation strategy of
\cite{tjelmeland2012}. The MRF is then approximated with a 
partially ordered Markov model (POMM), see \citet{art119}, 
where the conditional distribution of one variable given all 
previous variables is allowed to depend on maximally $\nu$ 
previous variables. We have tried different values for $\nu$
and found that in all our examples $\nu=7$ is sufficient to obtain
very good approximations, so all the results presented here
are based on this value of $\nu$. To simulate from
posterior distributions we use the reversible jump 
MCMC algorithm defined in Section \ref{sec:sampling}. In our sampling algorithm we have
an algorithmic tuning parameter $\sigma^2$ as the variance in Gaussian
proposals. Based on the results of
some preliminary runs we set $\sigma=0.3$. One
iteration of our sampling algorithm is defined to be one proposal of each type. Lastly we note that parallel
computing was used in order to reduce computational time, and the
technique that is used is explained in the supplementary materials.


\subsection{The Ising model}
\label{sec:IsingResults}

We generated a realisation from the Ising model given in
Section \ref{sec:Ising} with $\omega=0.4$ on a $100\times 100$ lattice, 
consider this as our observed data $x$ and simulate by the RJMCMC algorithm
from the resulting posterior distribution. The $x$ was obtained using 
the perfect sampler presented in \cite{propp1996}. From
the calculations in Section \ref{sec:Ising} we ideally want the correct
groups,
$\{c^0,c^{\confcAAAA}\}$ $\{c^1,c^{11},c^{\confcABAB},c^{\confcAAAB},
c^{\confcAABA},c^{\confcABAA},c^{\confcBAAA}\}$, and $\{c^{\confcABBA},
c^{\confcBAAB}\}$,
to be visited frequently by our sampler. Note that due to our
identifiability restriction in
\eqref{eq:identifiability} the configuration set parameters
should be close to the values given in Section \ref{sec:Ising} with
$\eta=\omega$. We run our sampler for 20000
iterations and study the simulation
results after convergence. A small convergence study is included in
the supplementary materials for the other simulated data set. The acceptance rate for the parameter value proposals is 19\%, whereas the acceptance
rates for the other two types of proposals are both around 1\%. The
estimated distribution for the number of groups is 94\%, 5\% and 1\%,
for 3, 4 and 5 groups respectively.  

In Figure \ref{fig:groupingMatrixIsing} we have plotted the
matrix representing the estimated posterior probability of two
configuration sets being
assigned to the same group. 
\begin{figure*}[h]\center
Figure \ref{fig:groupingMatrixIsing} approximately here.
\end{figure*}
As we can see in this
figure, the configuration sets are separated into 3 groups, and these
groups correspond to the correct grouping shown in grey. About $94\%$ of the realisations is assigned to
this particular grouping, and almost all other groupings that are simulated correspond to
groupings where the middle group is split in various ways, while some
very few are splits of the groups $\{c^0,c^{\confcAAAA}\}$ and $\{c^{\confcABBA},c^{\confcBAAB}\}$. Every one
of these alternative groupings
have an estimated posterior probability of less than 0.5\%. 

\begin{figure}
  \centering
\subfigure[][$\betaABBB\  (-1.62,-1.38)$]{\includegraphics[scale=0.5]{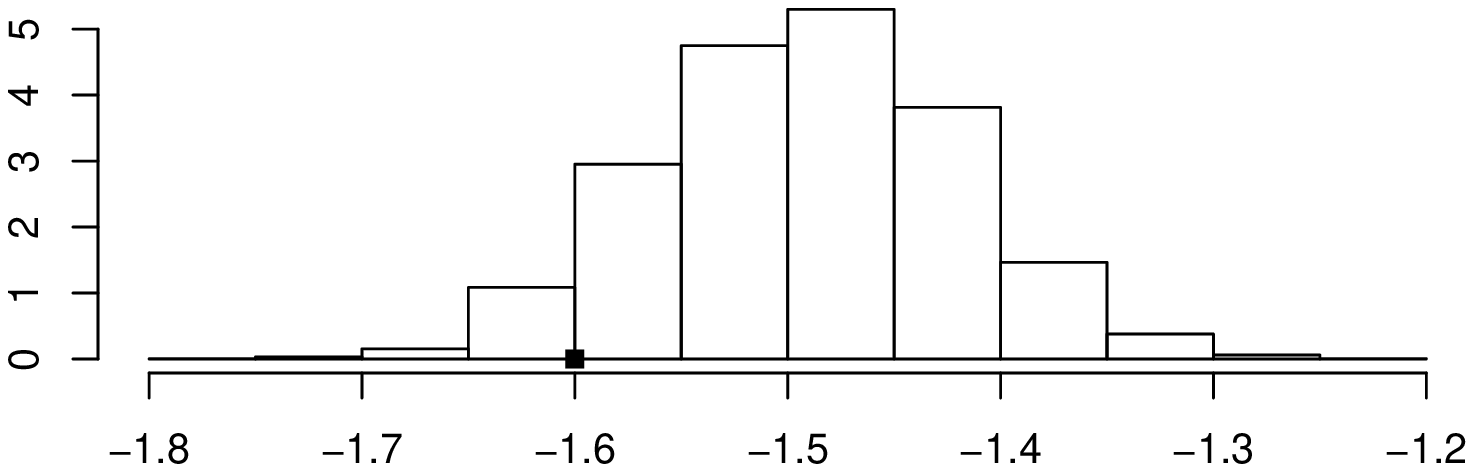}}
\subfigure[][$\betaAABB\ (0.69,0.81)$]{\label{fig:Isingi2}\includegraphics[scale=0.5]{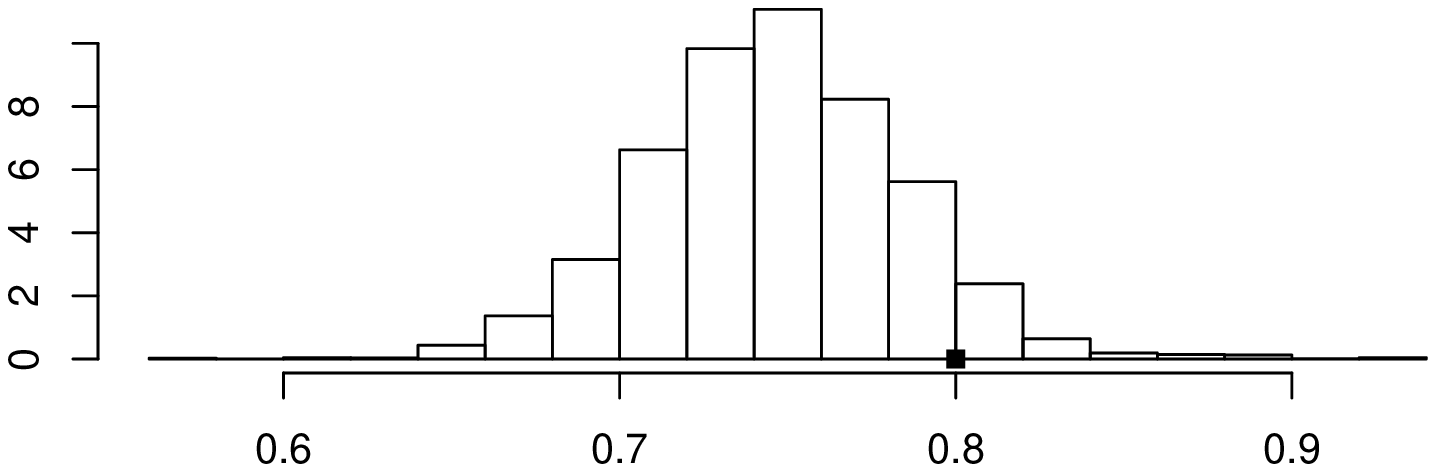}}\\
\subfigure[][$\betaABAB\ (0.69,0.81)$]{\label{fig:Isingi3}\includegraphics[scale=0.5]{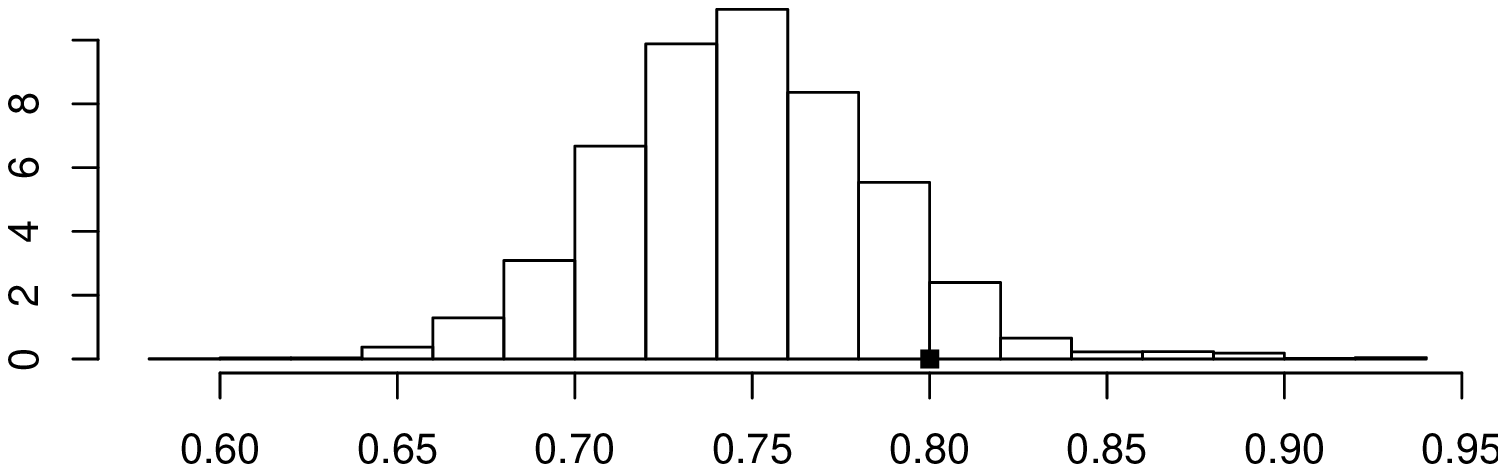}}
\subfigure[][$\betaABBA\ (-0.09,0.09)$]{\label{fig:Isingi4}\includegraphics[scale=0.5]{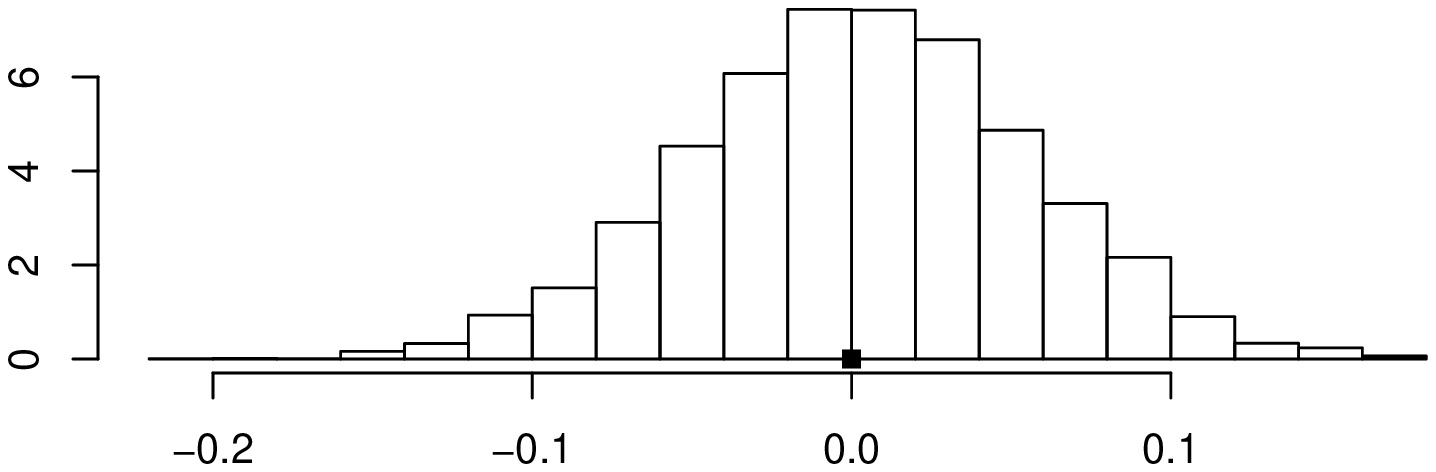}}\\
\subfigure[][$\betaBAAB\ (-0.09,0.09)$]{\label{fig:Isingi5}\includegraphics[scale=0.5]{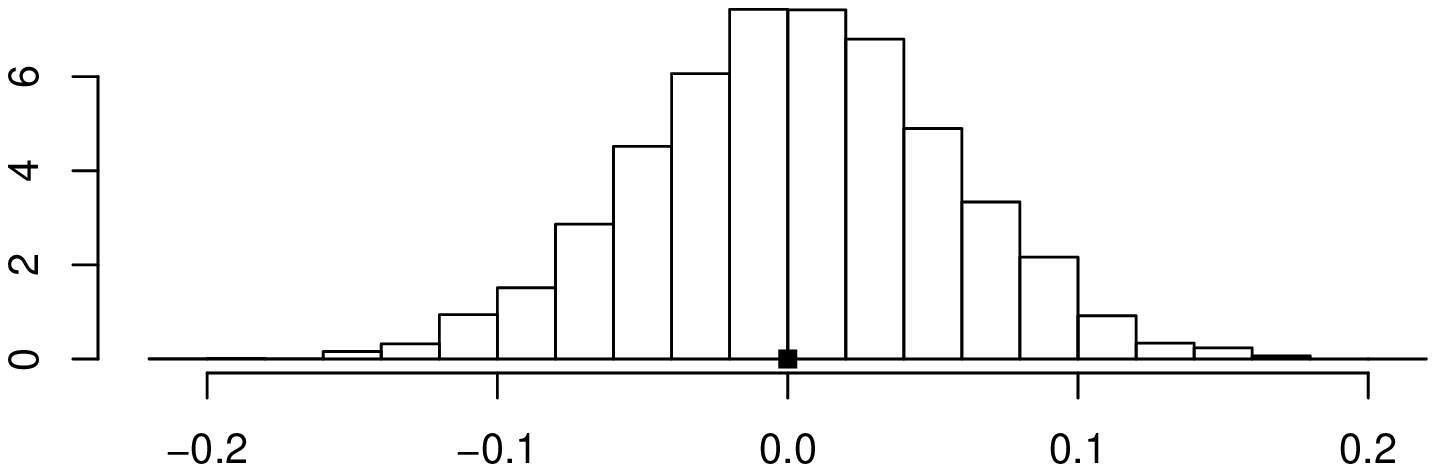}}
\subfigure[][$\betaAAAB\ (-0.10,0.08)$]{\label{fig:Isingi9}\includegraphics[scale=0.5]{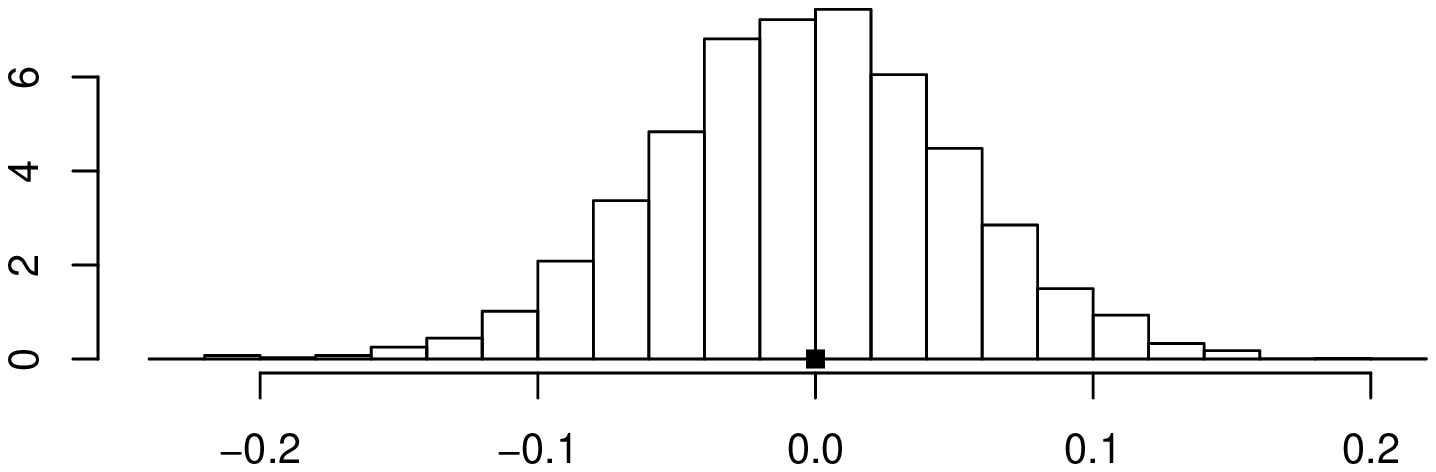}}\\
\subfigure[][$\betaAABA\ (-0.09,0.09)$]{\label{fig:Isingi8}\includegraphics[scale=0.5]{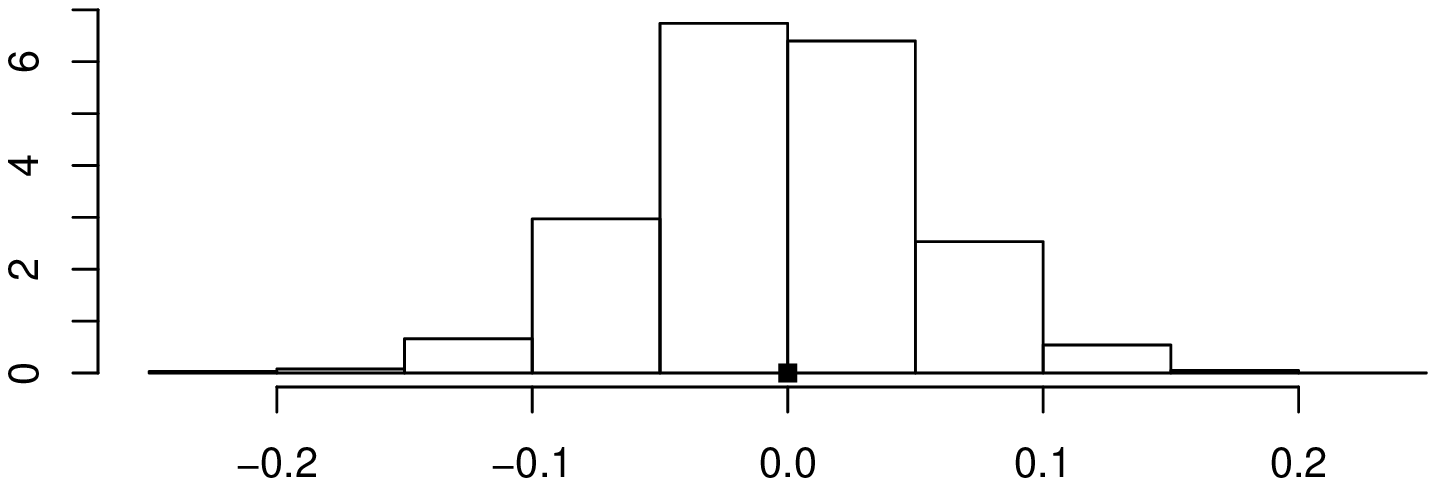}}
\subfigure[][$\betaABAA\  (-0.10,0.09)$]{\label{fig:Isingi7}\includegraphics[scale=0.5]{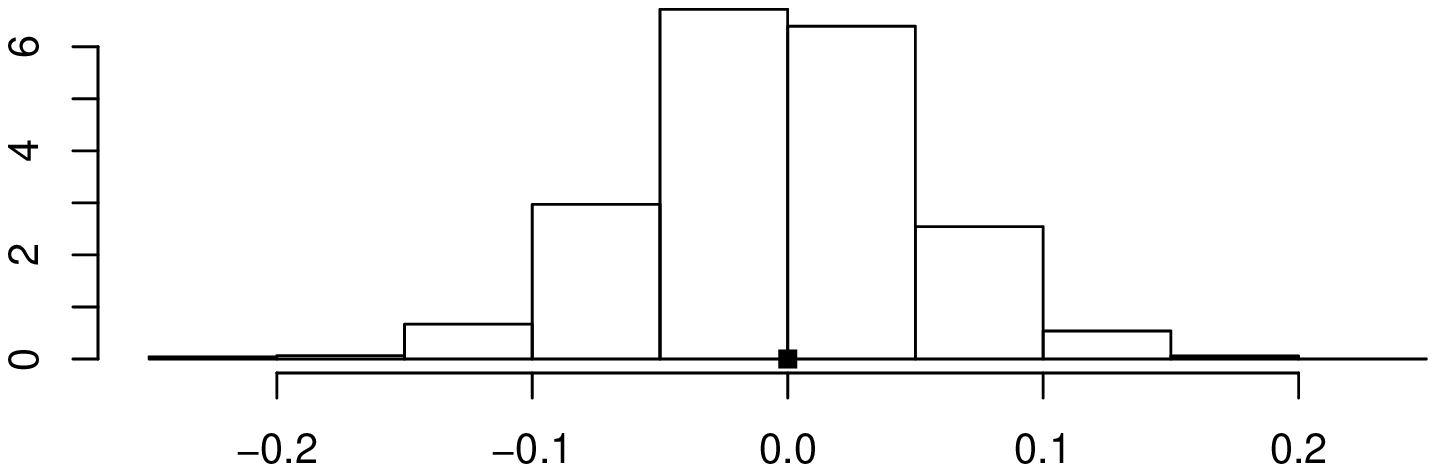}}\\
\subfigure[][$\betaBAAA\ (-0.09,0.09)$]{\label{fig:Isingi6}\includegraphics[scale=0.5]{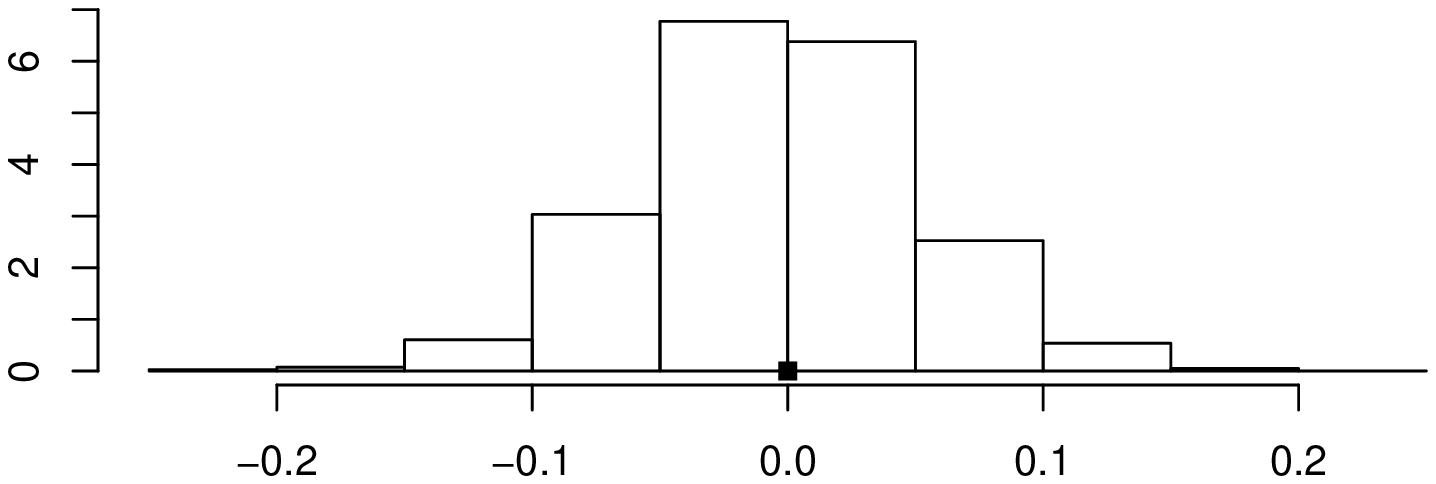}}
\subfigure[][$ \betaAAAA\  (-0.17,0.19)$]{\includegraphics[scale=0.5]{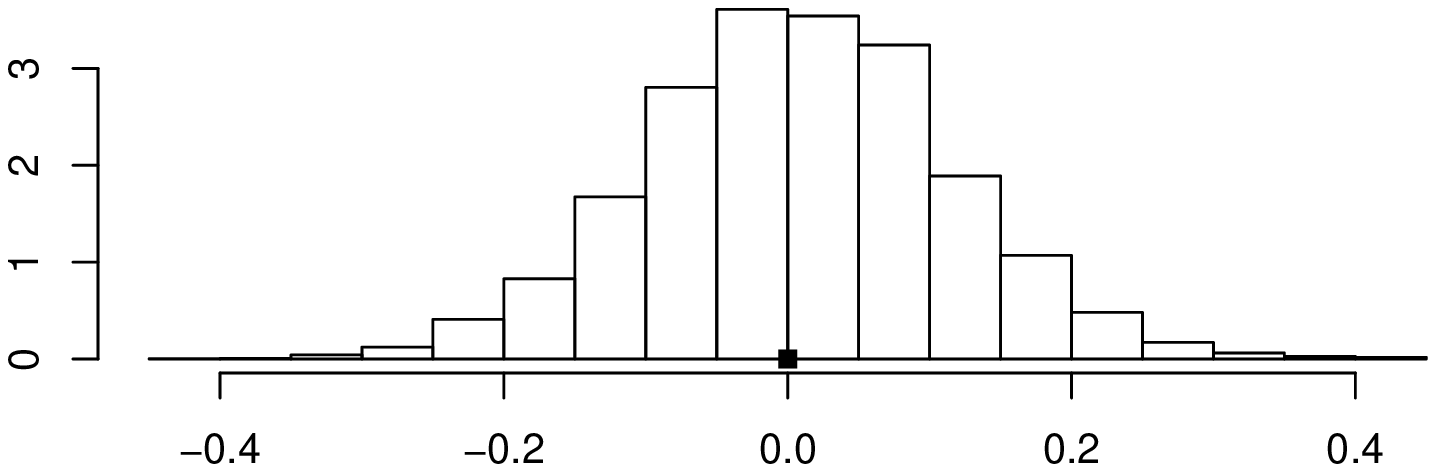}}
\caption[Caption For LOF]%
{\label{fig:interactionIsing}Ising model example: Estimated marginal posterior distribution for the
    interaction parameters. True values are shown with a
    black dot and estimated 95\% credibility
    intervals are given for each parameter.}
\end{figure}

One informative way to look at the result of the simulation is to estimate the posterior
distribution for the interaction parameters $\beta$. Histograms and
estimated 95\% credibility intervals for each of the parameters are
given in Figure \ref{fig:interactionIsing}.
\begin{figure*}[h]\center
Figure \ref{fig:interactionIsing} approximately here.
\end{figure*}
As we can see, all the
true values of the interaction parameters are within the estimated
credibility intervals, however the modes of the distributions for the
pairwise horizontal and vertical second order interactions, see Figure
\ref{fig:Isingi2} and \ref{fig:Isingi3}, seem to be somewhat lower than the
correct value. 

\begin{figure}
  \centering
\subfigure[][$g(x)=\sum_ix_i$]{\label{fig:is1}\includegraphics[scale=0.5]{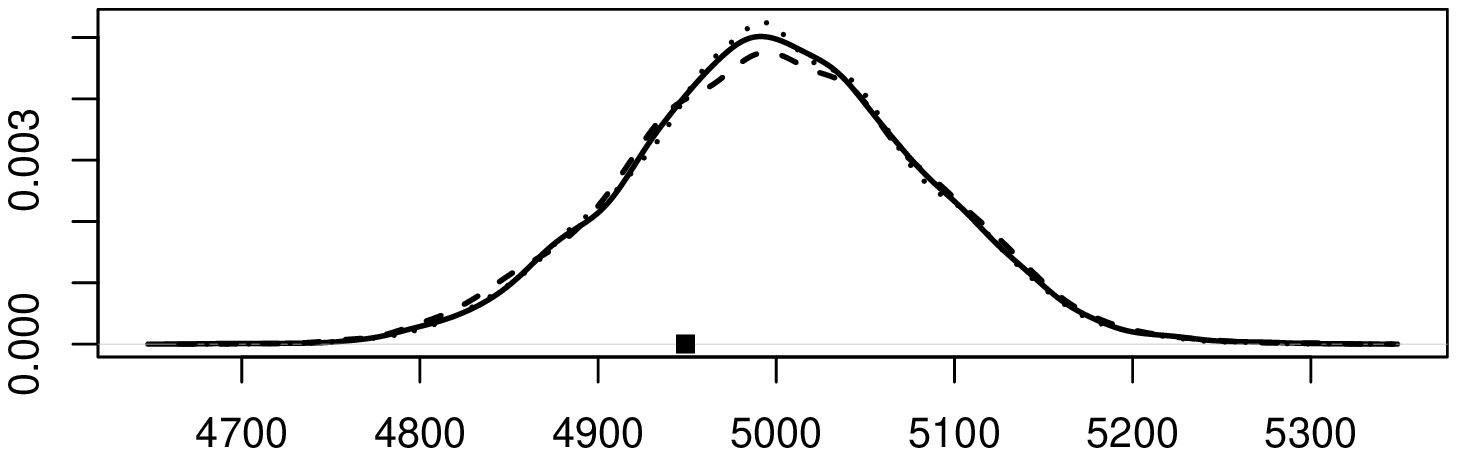}}
\subfigure[][$g(x)=\sum_{i,j:\text{vertical adjacent sites}}I(x_i=x_j)$]{\label{fig:is2}\includegraphics[scale=0.5]{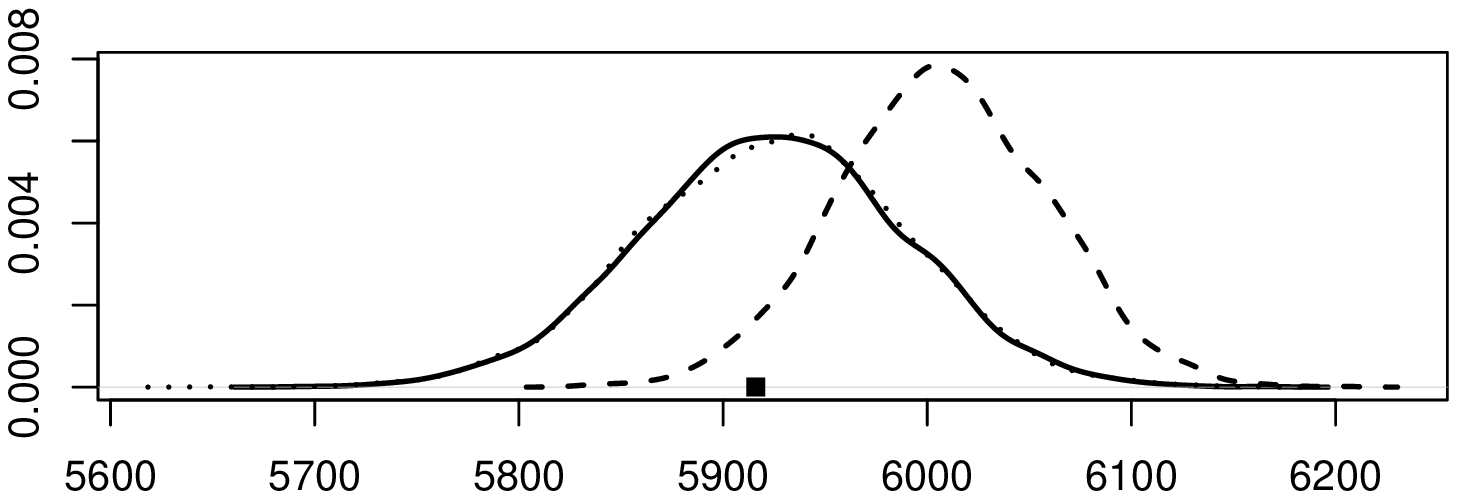}}\\
\subfigure[][$g(x)=\sum_{i,j:\text{horizontal adjacent sites}}I(x_i=x_j)$]{\label{fig:is3}\includegraphics[scale=0.5]{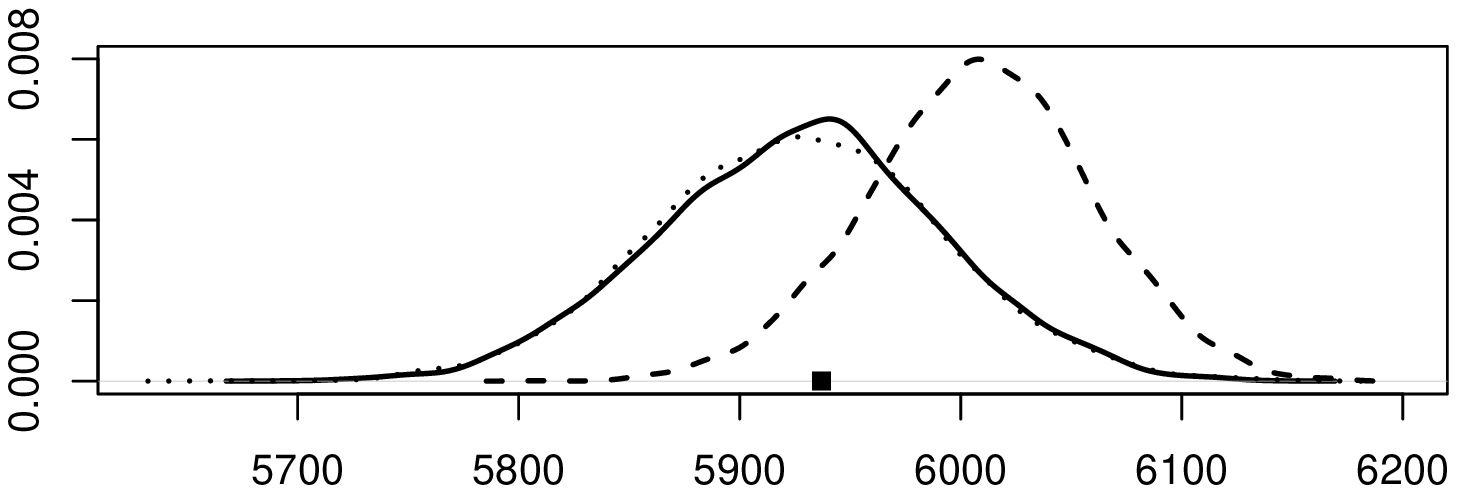}}
\subfigure[][$g(x)=\sum_{\Lambda\in\mathcal{L}_m}I\left ( x_{\Lambda}=
  { \left[\confBBBB \right]}\right)$]{\label{fig:is4}\includegraphics[scale=0.5]{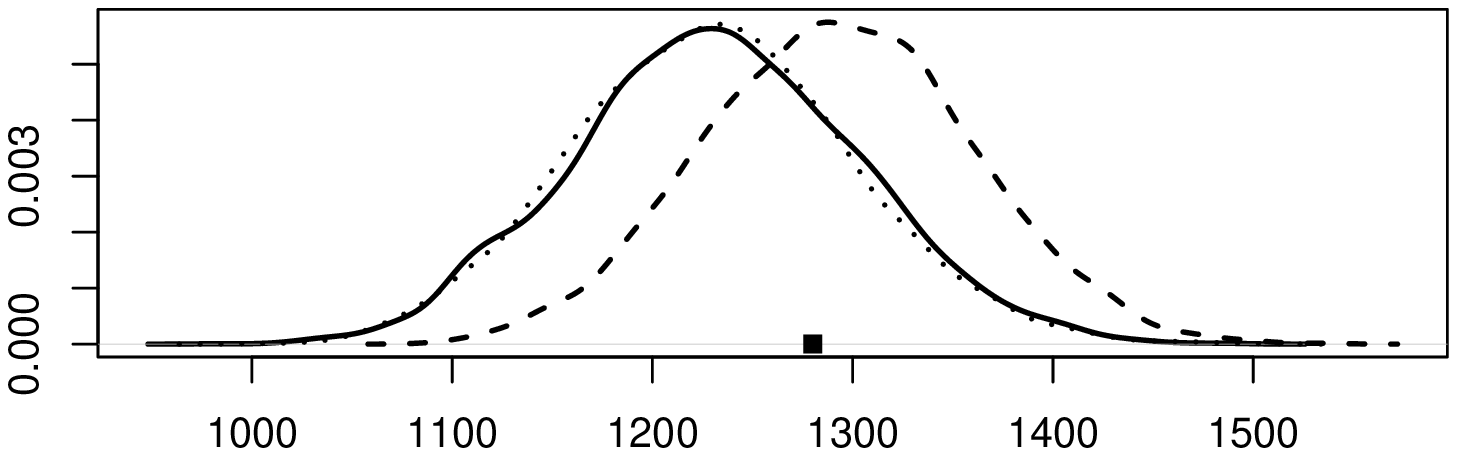}}\\
\subfigure[][$g(x)=\sum_{\Lambda\in\mathcal{L}_m}I\left ( x_{\Lambda}=
  {\left[\confBAAB\right]}\right)$]{\label{fig:is5}\includegraphics[scale=0.5]{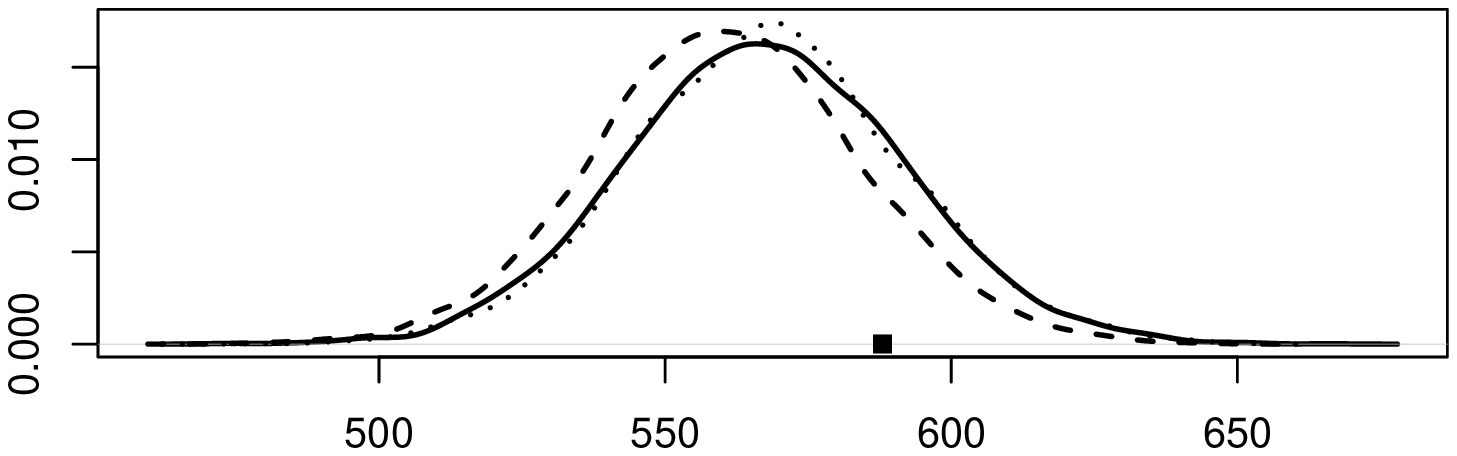}}
\subfigure[][$g(x)=\sum_{\Lambda\in\mathcal{L}_m}I\left ( x_{\Lambda}=
  {\left[\confAAAB\right]}\right)$]{\label{fig:is6}\includegraphics[scale=0.5]{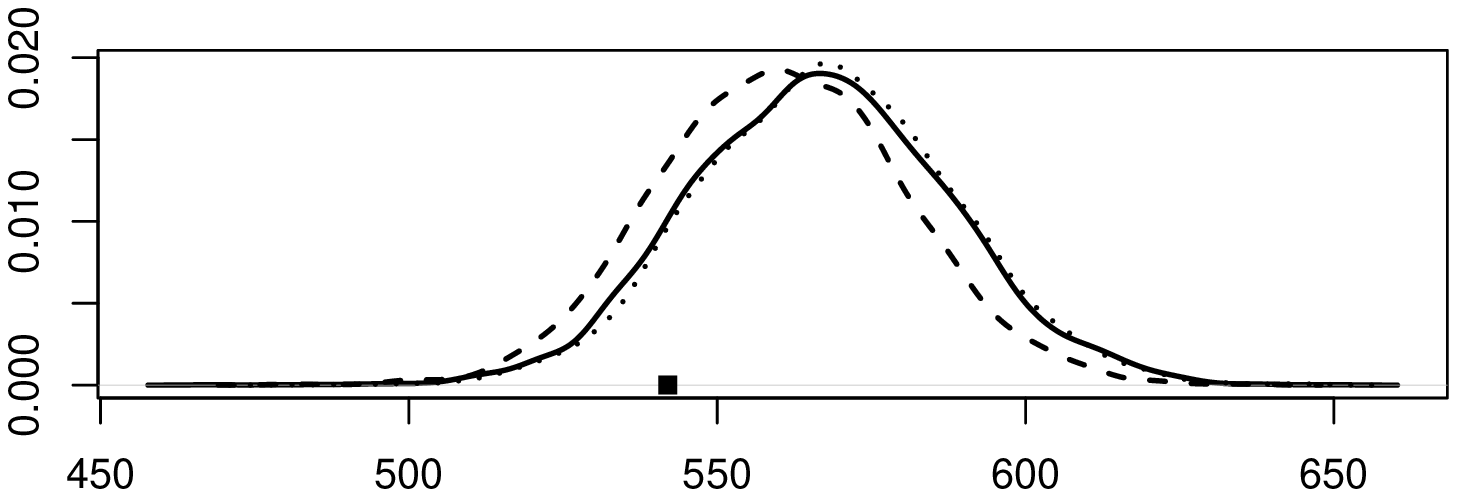}}\\
\subfigure[][$g(x)=\sum_ix_i$]{\label{fig:is_MCMC1}\includegraphics[scale=0.5]{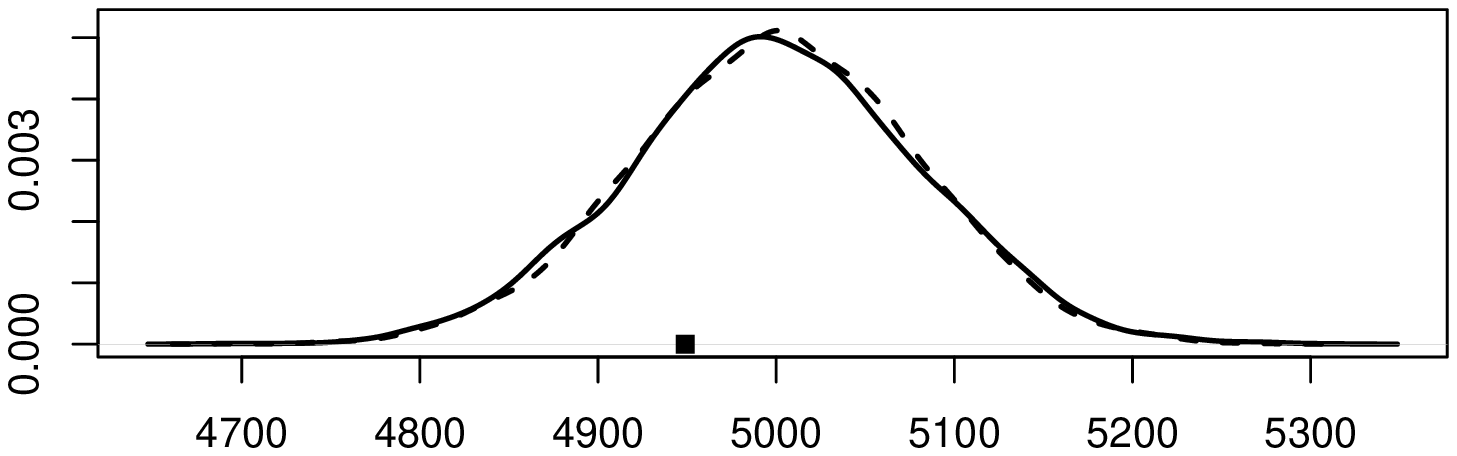}}
\subfigure[][$g(x)=\sum_{\Lambda\in\mathcal{L}_m}I\left ( x_{\Lambda}=
  {\left[\confAAAB\right]}\right)$]{\label{fig:is_MCMC15}\includegraphics[scale=0.5]{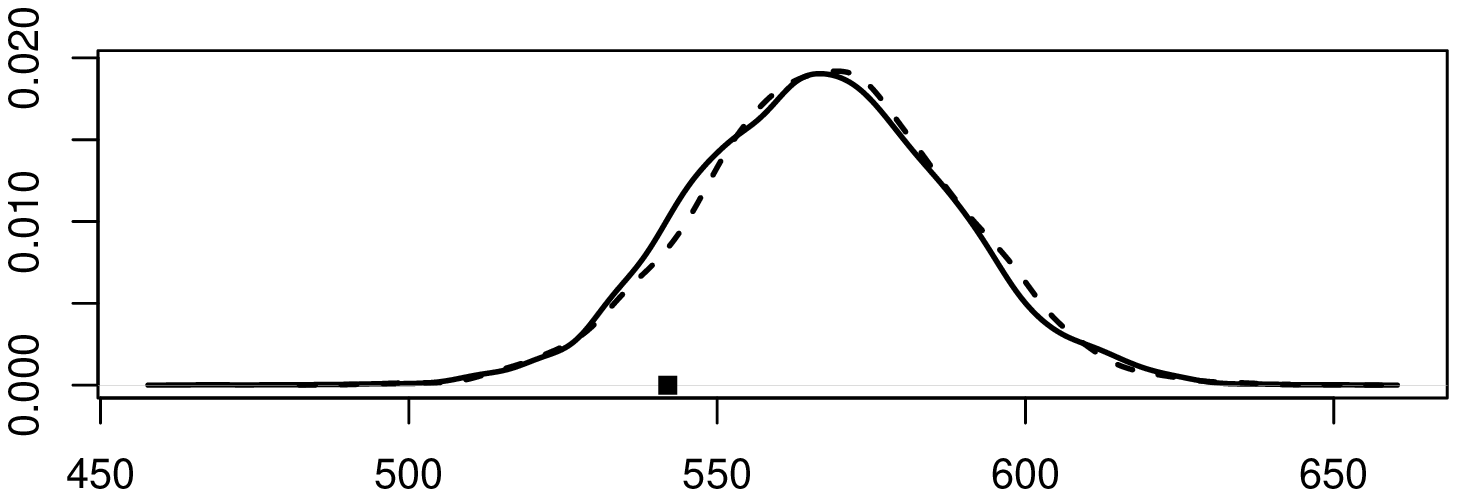}}
\caption{\label{fig:compareSimIsing}Ising model example: (a)-(f) Distribution
  of six statistics of
  realisations from our $2\times 2$ model with posterior samples of
  $z$ (solid), the Ising model with correct parameter value (dashed), and the Ising model with posterior samples of the
parameter value (dotted). In (g) and (h) we compare two of these
statistics with results obtained using the exchange algorithm with
MCMC samples as auxiliary variables (dashed) instead of the approximation. The data evaluated with
each statistic is shown with a black dot.}
\end{figure}

To study the properties of the MRF $p(\cdot |z)$ when 
$z$ is a sample from the posterior $p(z|x)$ we take $5000$ samples
from the MCMC run for $p(z|x)$ and generate for each of these
a corresponding realisation from the MRF $p(\cdot |z)$. 
To analyse these $5000$ images we use six statistics describing
local properties of the images. The statistics used and resulting
density estimates (solid) of the distribution of these statistics
are shown in Figures \ref{fig:compareSimIsing} (a)-(f).
\begin{figure*}[h]\center
Figure \ref{fig:compareSimIsing} approximately here.
\end{figure*}
In the same figures we also show 
density estimates of the same statistics when images are generated 
from the Ising model with the true parameter value (dashed),
and when images are generated from the Ising model with 
parameter value $\omega$ generated from the posterior distribution
given our observed image $x$ (dotted). In this last case, a zero mean
Gaussian prior with standard deviation equal to ten is
used for $\omega$. In these figures we also see that the data we use for posterior sampling
(black dots) of
$z$ is a realisation from the Ising model with low values for 
the number of equal horizontal and vertical
adjacent sites, see Figure \ref{fig:is2} and \ref{fig:is3}, which
causes, as already observed above, our simulations of the second order
interactions between horizontal and vertical adjacent sites to be 
somewhat lower than the true values.
In fact we can see that the simulations from the
Ising model using posterior samples for the parameter value closely
follows that of our $2 \times 2$ model. This means that the results
from our model is
as accurate as the result one gets when knowing that the true model is the
Ising model without knowing the model parameter. 

To evaluate the quality of the POMM approximation
in this example, we also simulate from the posterior
distribution with the same RJMCMC algorithm using the 
approximate exchange algorithm of \cite{pro20}, as discussed 
in Section \ref{sec:normconstant}.
We compare in Figures \ref{fig:compareSimIsing} (g) and (h) the 
results using the POMM approximation (solid) to the results from 
the approximate 
exchange algorithm (dashed) using two of our six statistics. We 
observe that the differences are minimal for these two, and indeed we 
get as accurate results for the four other statistics as well. 
That these two very different approximation strategies produces
essentially the same results strongly indicate that both 
procedures are very accurate.     

All the above results are for $\gamma=0.5$, but as mentioned in the introduction of this section we also investigate
the results for $\gamma=0$ and $1$. For $\gamma=0$ the configuration sets
are organised into 3 (66\%), 4 (31\%) or 5 (3\%) groups, and for $\gamma=1$ we get 3 (96\%) or 4
(4\%) groups. From these numbers we see the effect of varying $\gamma$. In particular when increasing $\gamma$ from
$0.5$ to $1.0$ the tendency to group
more configuration sets together becomes stronger for this data set.

\subsection{Red deer census count data}
\label{sec:redDeer}
\begin{figure}
  \centering
  \includegraphics[scale=0.5]{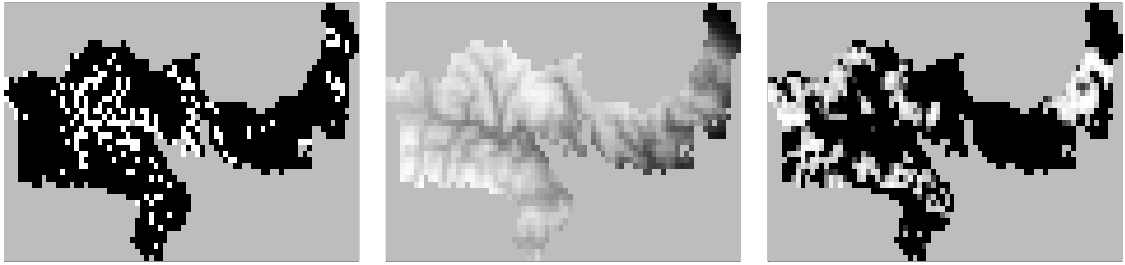} 
  \caption{\label{fig:redDeerData}Red deer example: The presence/absence of red deer (left),
    altitude (middle), and mires (right) in the Grampians Region of north-east Scotland.}
\end{figure}

In this section we analyse a data set of census counts of
red deer in the Grampians Region of north-east Scotland. A full
description of the data set is found in \cite{augustin1996} and
\cite{buckland1993}. The data is obtained by dividing the region of
interest into $n=1277$ grid cells on a lattice and observing the presence or
absence of red deer in each cell. In our notation this is our observed
image $x$, but in this example we also have the four covariates
altitude, mires, north coordinate and east coordinate available in each
grid cell. The
binary data $x$ and the two first covariates are shown in Figure
\ref{fig:redDeerData}. 
\begin{figure*}[h]\center
Figure \ref{fig:redDeerData} approximately here.
\end{figure*}
We denote the covariate $k$ at a location $(i,j)$ by
$y_{i,j,k}, \ j=1,2,3,4$, and model them into the likelihood function in
the following way 
\begin{equation}\label{eq:likelihoodDeer}
p(x|z,\theta^C,y)=Z\exp{\left(
    \sum_{\Lambda\in \mathcal{L}_m}V_\Lambda(x_{\Lambda}|z)+\sum_{(i,j)\in S}x_{i,j}\sum_{k=1}^4\theta^C_ky_{i,j,k}\right )},
\end{equation}  
where $\theta^C=(\theta^C_1,...,\theta^C_4)$ are the parameters for
the covariates.

We put independent zero mean Gaussian prior distributions with standard
deviation equal to 10 on $\theta^C_j$, $j=1,...,4$. In the sampling algorithm
these covariates are updated using random walk, i.e. we uniformly choose
one of the four covariates to update and propose a new value using a
Gaussian distribution with the old parameter value as
the mean and a standard deviation of $0.1$.

\begin{figure}
  \centering
\subfigure[][$\betaABBB\ (-3.52,-2.85)$]{\label{fig:Deerb1}\includegraphics[scale=0.5]{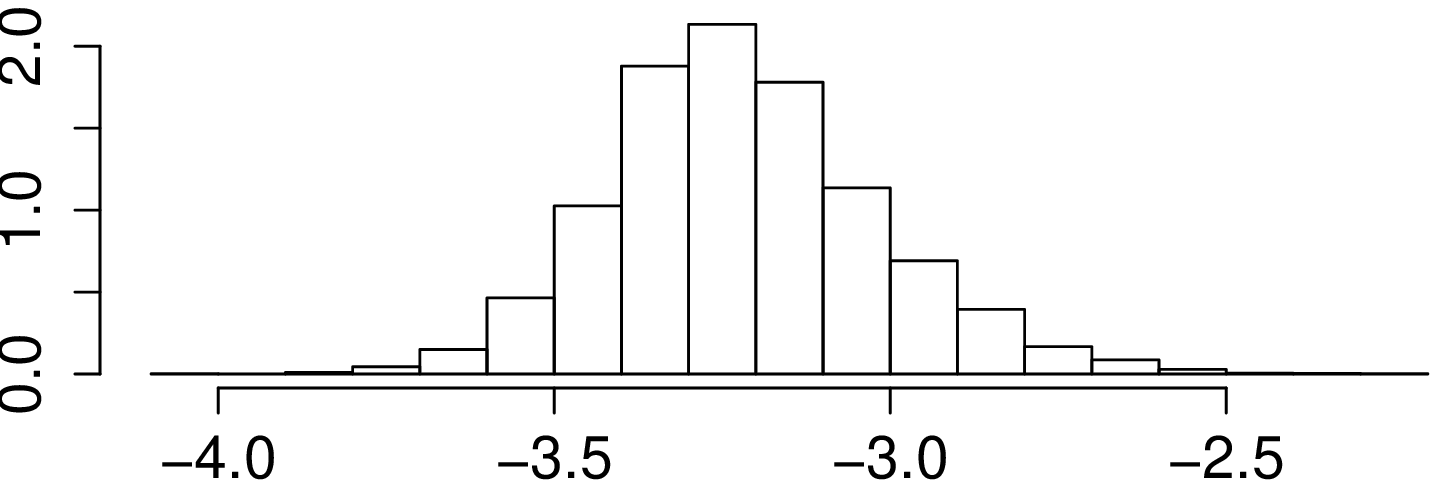}}
\subfigure[][$\betaAABB\ (0.92,1.77)$]{\label{fig:Deerb2}\includegraphics[scale=0.5]{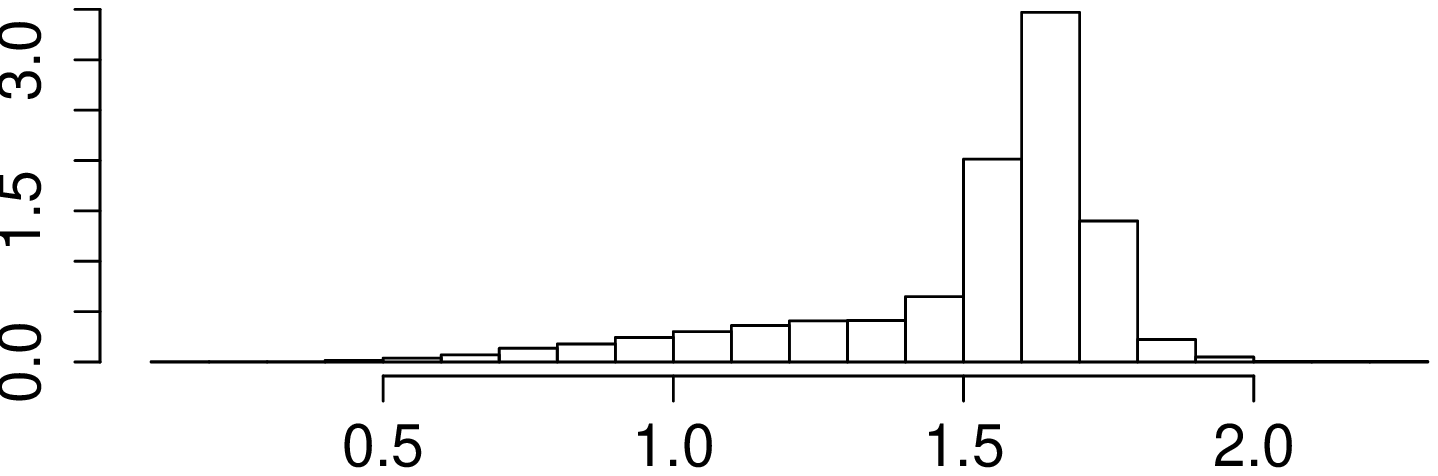}}\\
\subfigure[][$\betaABAB\ (0.91,1.77)$]{\label{fig:Deerb3}\includegraphics[scale=0.5]{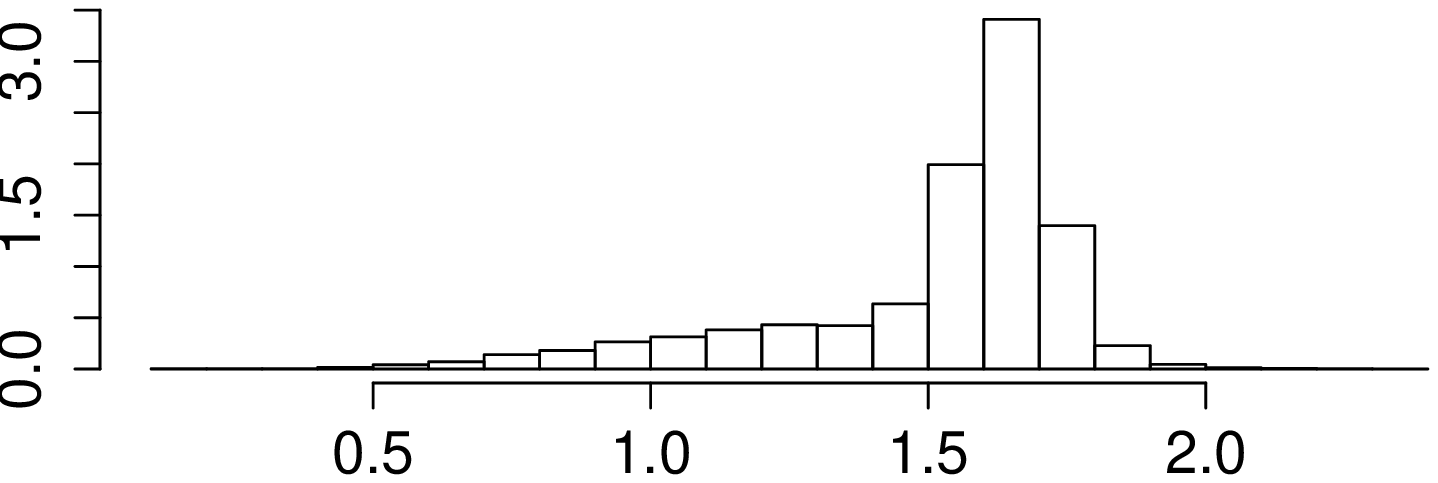}}
\subfigure[][$\betaABBA\ (0.62,1.54)$.]{\label{fig:Deerb4}\includegraphics[scale=0.5]{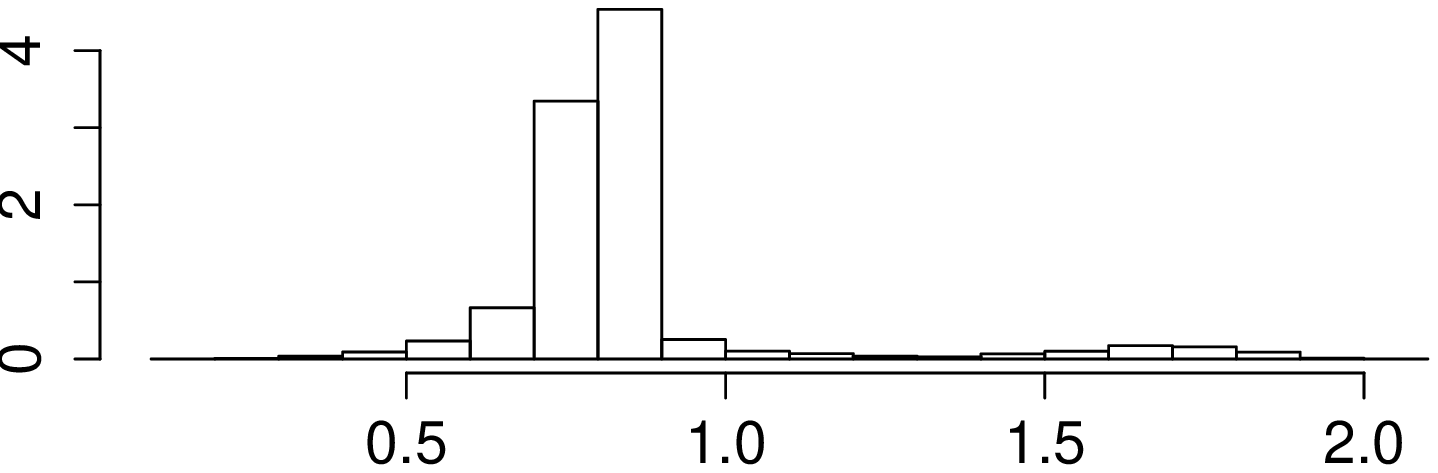}}\\
\subfigure[][$\betaBAAB\ (0.14,0.88)$]{\label{fig:Deerb5}\includegraphics[scale=0.5]{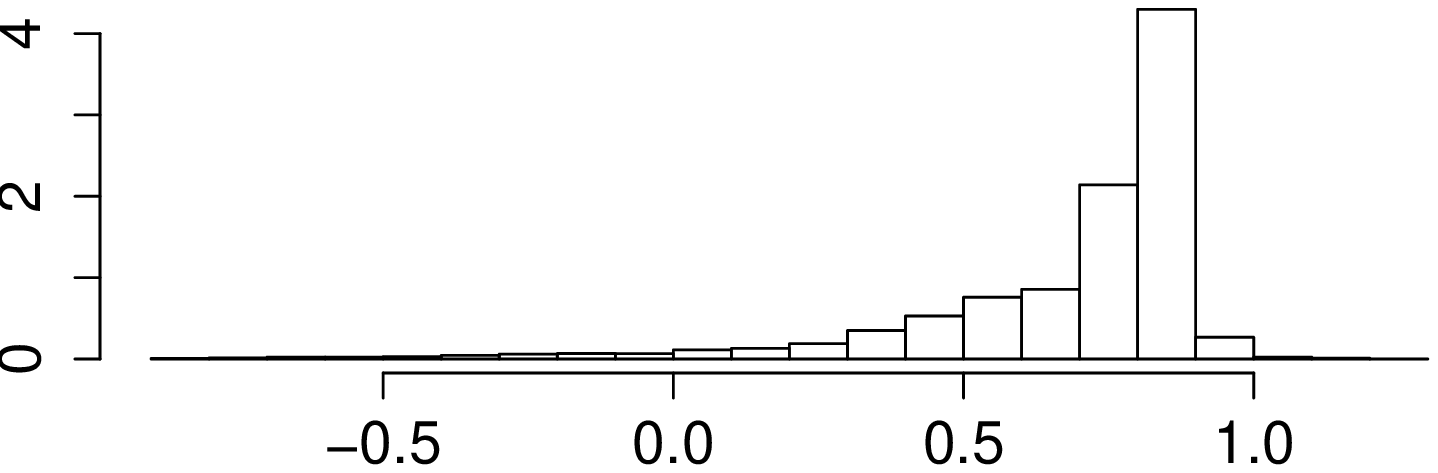}}
\subfigure[][$\betaAAAB\  (-1.08,-0.09)$]{\label{fig:Deerb9}\includegraphics[scale=0.5]{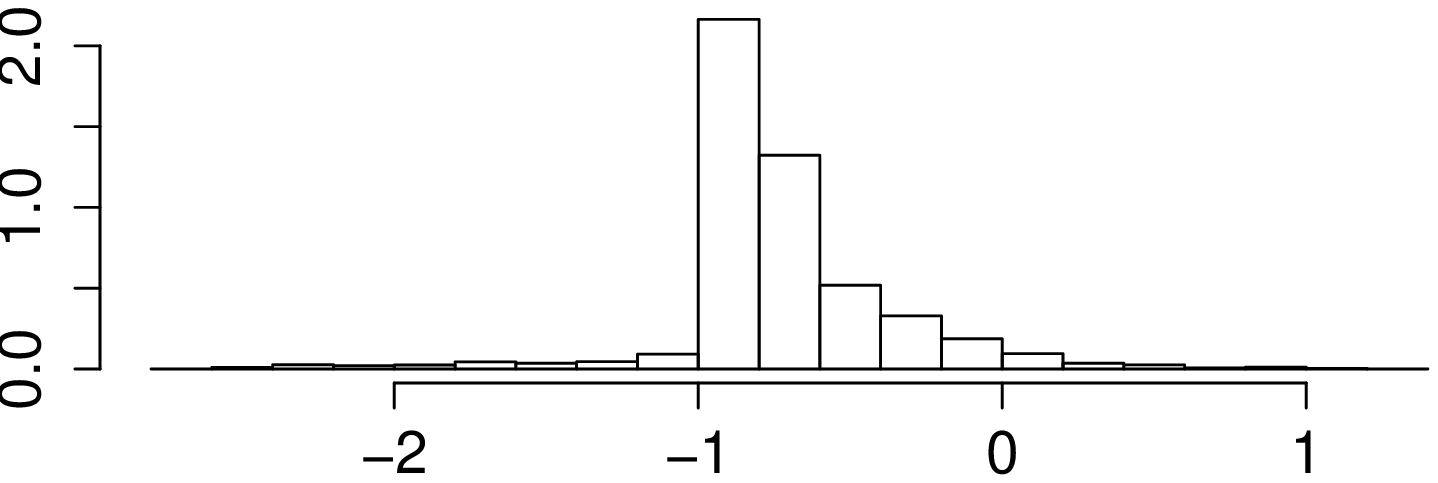}}\\
\subfigure[][$\betaAABA\ (-1.54,-0.25)$]{\label{fig:Deerb8}\includegraphics[scale=0.5]{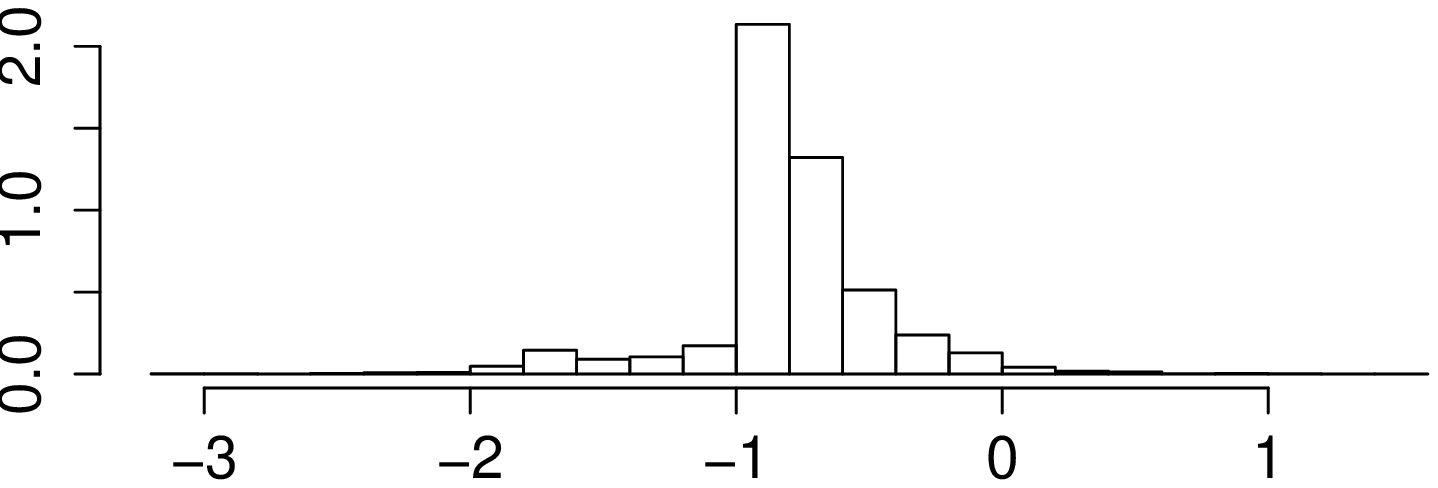}}
\subfigure[][$\betaABAA\ (-1.55,-0.28)$]{\label{fig:Deerb7}\includegraphics[scale=0.5]{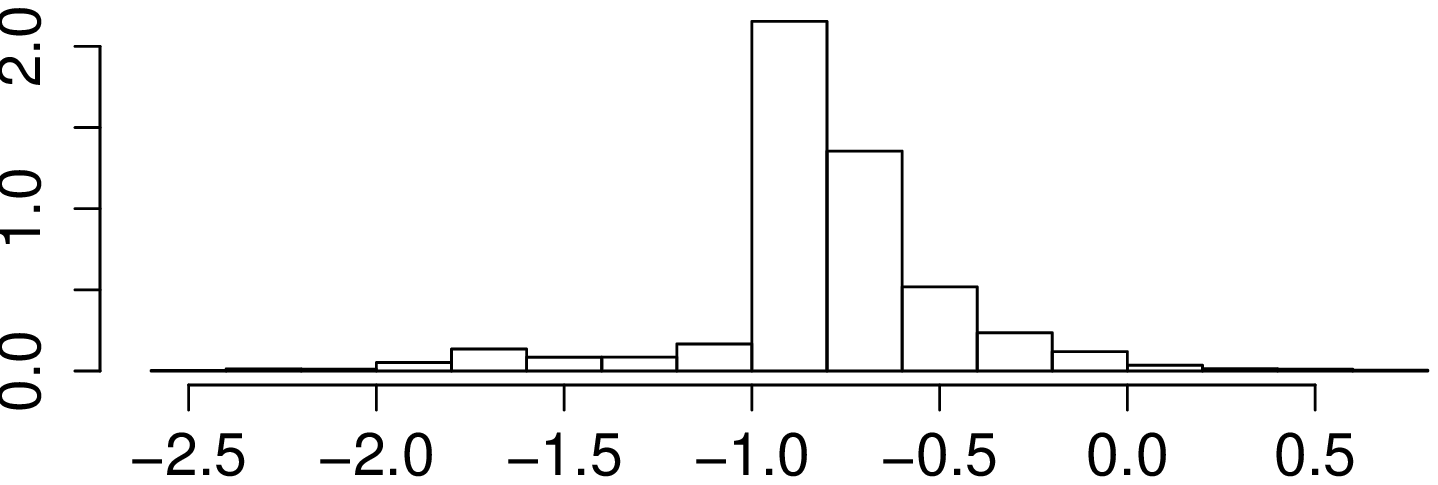}}\\
\subfigure[][$\betaBAAA\  (-0.88,0.47)$]{\label{fig:Deerb6}\includegraphics[scale=0.5]{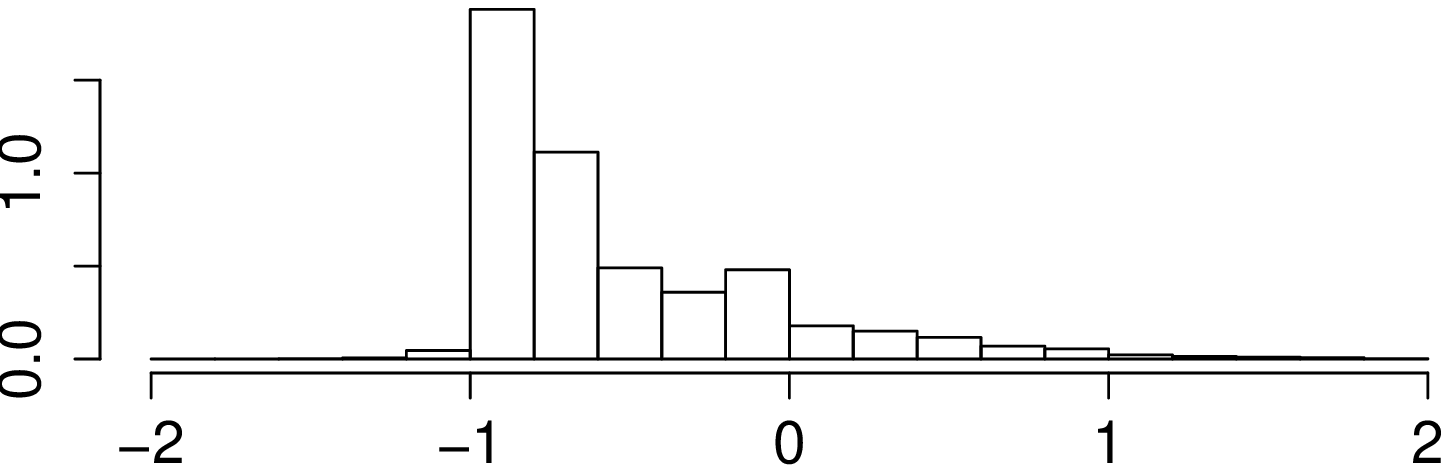}}
\subfigure[][$ \betaAAAA\  (-3.70,0.29)$]{\label{fig:Deerb10}\includegraphics[scale=0.5]{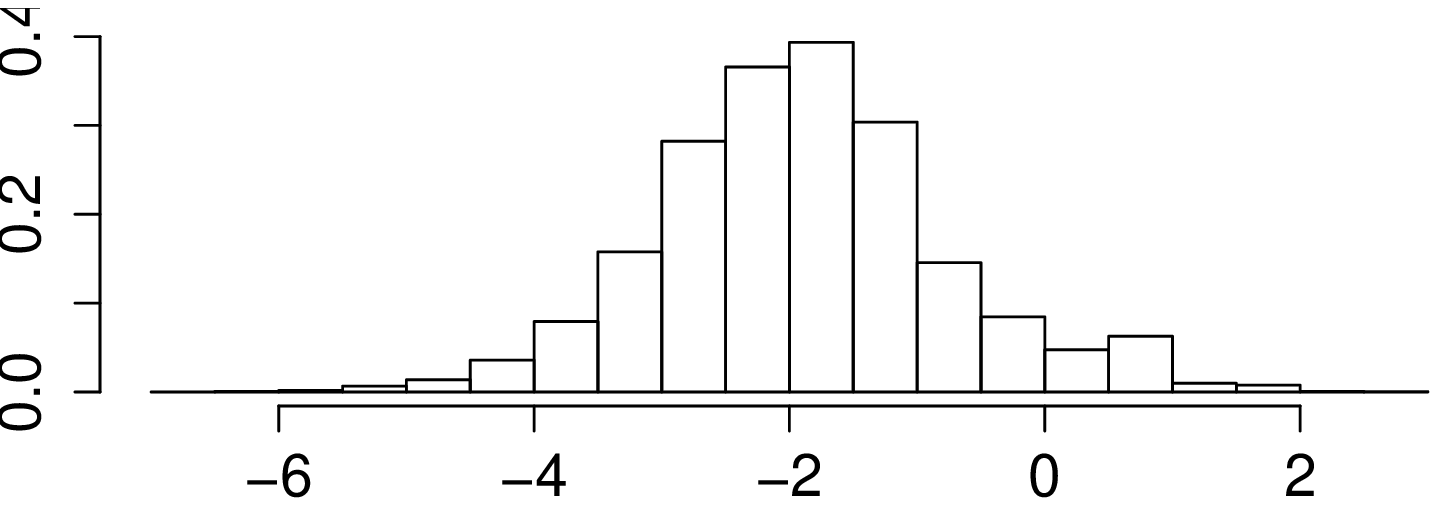}}
\caption[Caption For LOF]%
{\label{fig:interactionsDeer}Red deer example: 
  Estimated marginal posterior distribution for the
  interaction parameters. Estimated 95\% credibility 
  interval is given for each parameter.}
\end{figure}

\begin{figure}
  \centering
\subfigure[][$\theta^C_1 \ (-0.50,-0.21)$]{\label{fig:c1}\includegraphics[scale=0.5]{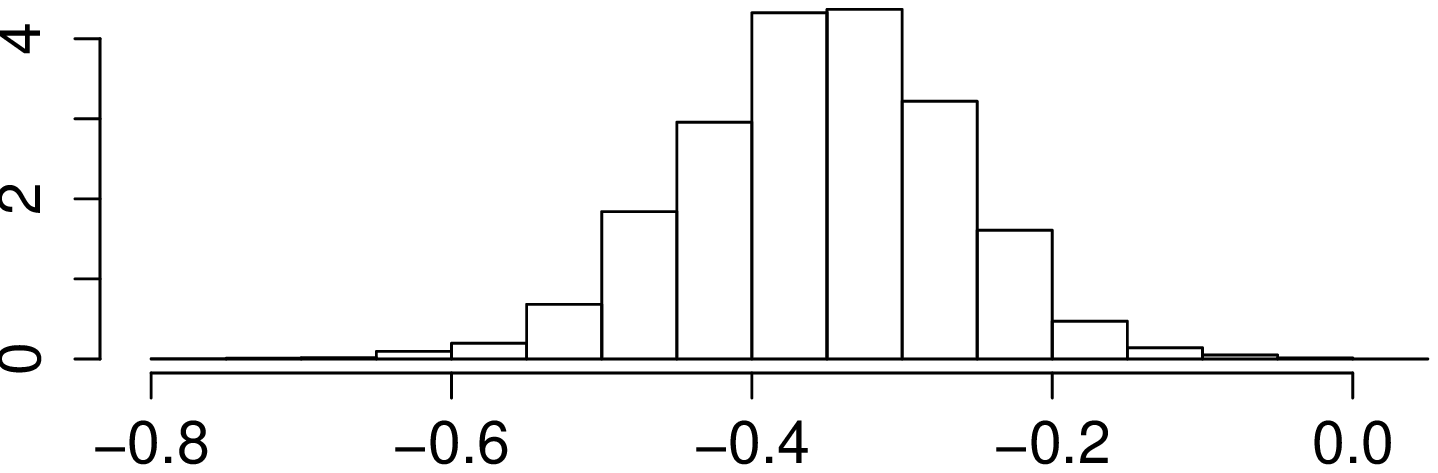}}
\subfigure[][$\theta^C_2 \ (-0.36,-0.07)$]{\label{fig:c2}\includegraphics[scale=0.5]{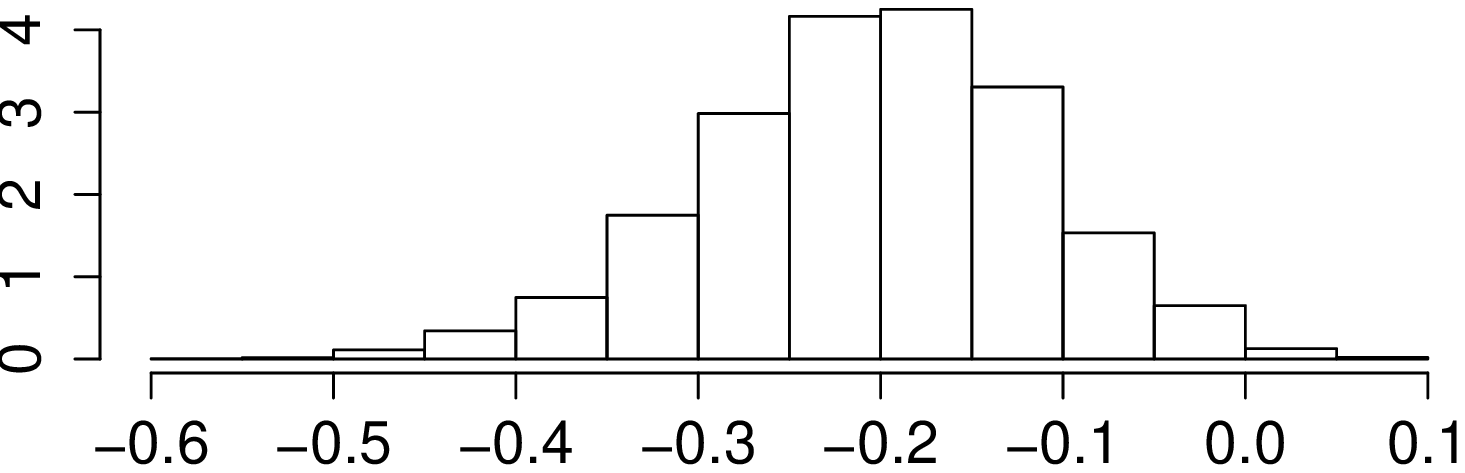}}\\
\subfigure[][$\theta^C_3 \ (-0.55,-0.27)$]{\label{fig:c3}\includegraphics[scale=0.5]{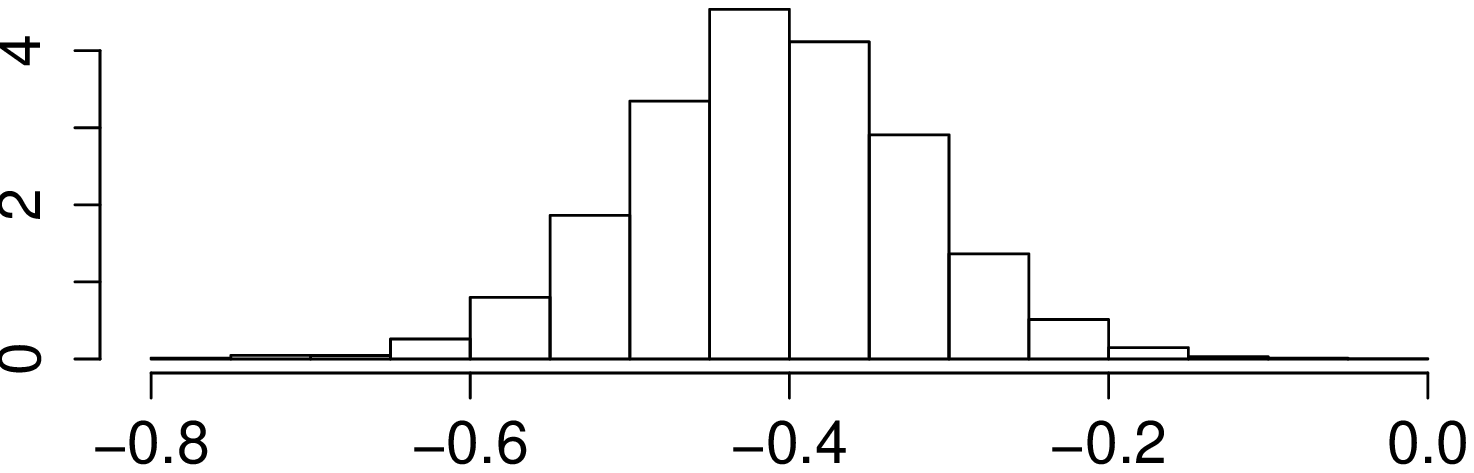}}
\subfigure[][$\theta^C_4 \ (-0.26,-0.05)$]{\label{fig:c4}\includegraphics[scale=0.5]{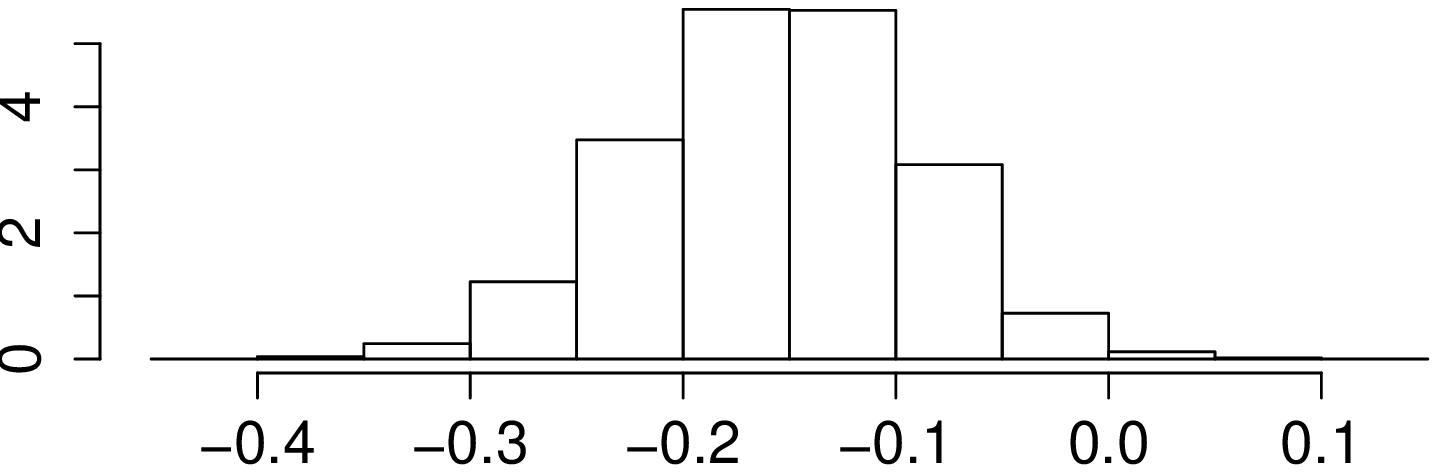}}
\caption{\label{fig:covariatesDeer}Red deer example: 
  Estimated marginal posterior distributions 
  for the parameters of the
  covariates. Estimated 95 \%
  credibility interval is given for each parameter.}
\end{figure}

\begin{figure}
  \centering
  \includegraphics[scale=0.5]{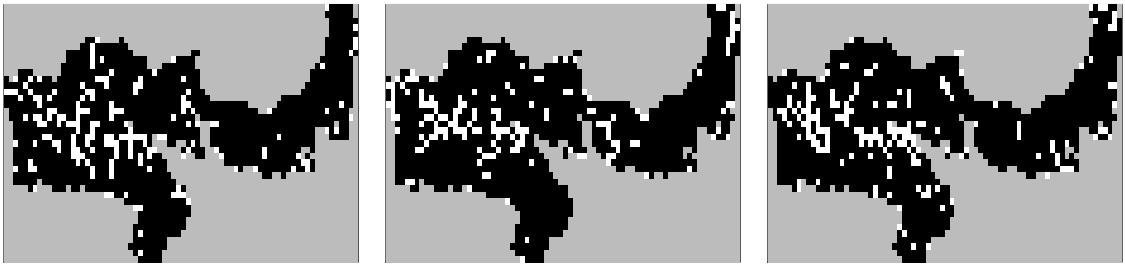} 
  \caption{\label{fig:simulationsDeer}Red deer example: 
    Three realisations from the likelihood for three random
    samples of $z$ from the posterior distribution.}
\end{figure}

We ran our algorithm for 50000 iterations, and the acceptance rates
for the parameter random walk proposal is 42\%, the group changing
proposal is 33\%, the trans-dimensional proposal is 5\%, and the
covariate proposal is 48\%. The posterior most probable grouping 
becomes $\{c^0\}$,
$\{c^1,c^{11},c^{\confcABAB},c^{\confcABBA},
c^{\confcBAAB},c^{\confcAAAB},c^{\confcAABA},
c^{\confcABAA},c^{\confcBAAA}\}$ and $\{c^{\confcAAAA}\}$ 
with probability 33.2\%. In total more
than 2500 different groupings
are visited, and except for the posterior most probable grouping the posterior probabilities of all other groupings are less than 5\%. The estimated
posterior probability distribution for the number of groups becomes
43\% for 3 groups, 48\% for 4 groups, 8\% for 5 groups and 1\% for
6 groups. In particular, the realisations with four or more groups are
mostly groupings where the set $\{c^1,c^{11},c^{\confcABAB},c^{\confcABBA},
c^{\confcBAAB},c^{\confcAAAB},c^{\confcAABA},
c^{\confcABAA},c^{\confcBAAA}\}$  is split in various
ways. This can also be seen in Figure
\ref{fig:configurationMatrixDeer}, which shows
the estimated posterior probability of two configuration sets
being assigned to the same group.
\begin{figure*}[h]\center
Figure \ref{fig:configurationMatrixDeer} approximately here.
\end{figure*}
The grey blocks in this figure show the estimated posterior most probable grouping described
above. Next
we estimate the posterior density for the interaction parameters, see
Figure \ref{fig:interactionsDeer}.
\begin{figure*}[h]\center
Figure \ref{fig:interactionsDeer} approximately here.
\end{figure*}
As we can see, most of the higher
order interaction parameters becomes significantly different from zero,
suggesting that a $2\times 2$ clique system is needed for this
data set. 
Figure \ref{fig:covariatesDeer} shows the estimated
posterior density for the covariate parameters.
\begin{figure*}[h]\center
Figure \ref{fig:covariatesDeer} approximately here.
\end{figure*}
As we can see from the
credibility intervals, all these parameters are significantly different
from zero, which justifies the need to include them. 
Simulations of $p(x|z,\theta^C,y)$ for
three randomly chosen posterior samples of $z$ and $\theta^C$ are
shown in Figure \ref{fig:simulationsDeer}. 
\begin{figure*}[h]\center
Figure \ref{fig:simulationsDeer} approximately here.
\end{figure*}
As we can see the spatial
dependency in these realisations looks similar to the data which
indicates that the features of this data set are captured with this
model.

\begin{figure}
  \centering
\subfigure[][$g(x)=\sum_ix_i$]{\label{fig:de1}\includegraphics[scale=0.5]{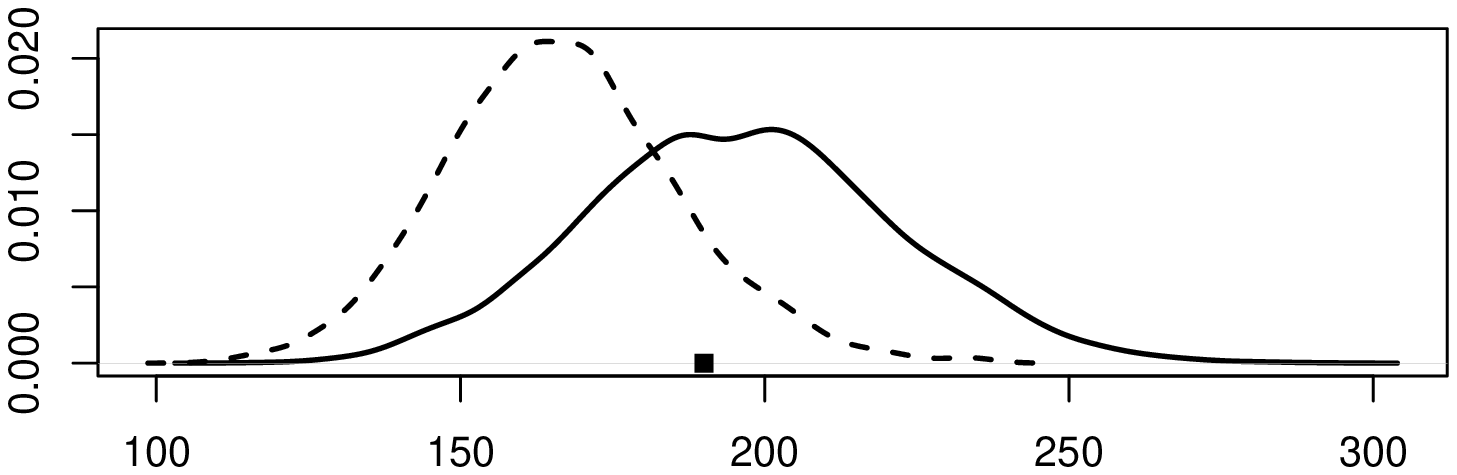}}
\subfigure[][$g(x)=\sum_{i,j:\text{vertical adjacent sites}}I(x_i=x_j)$]{\label{fig:de2}\includegraphics[scale=0.5]{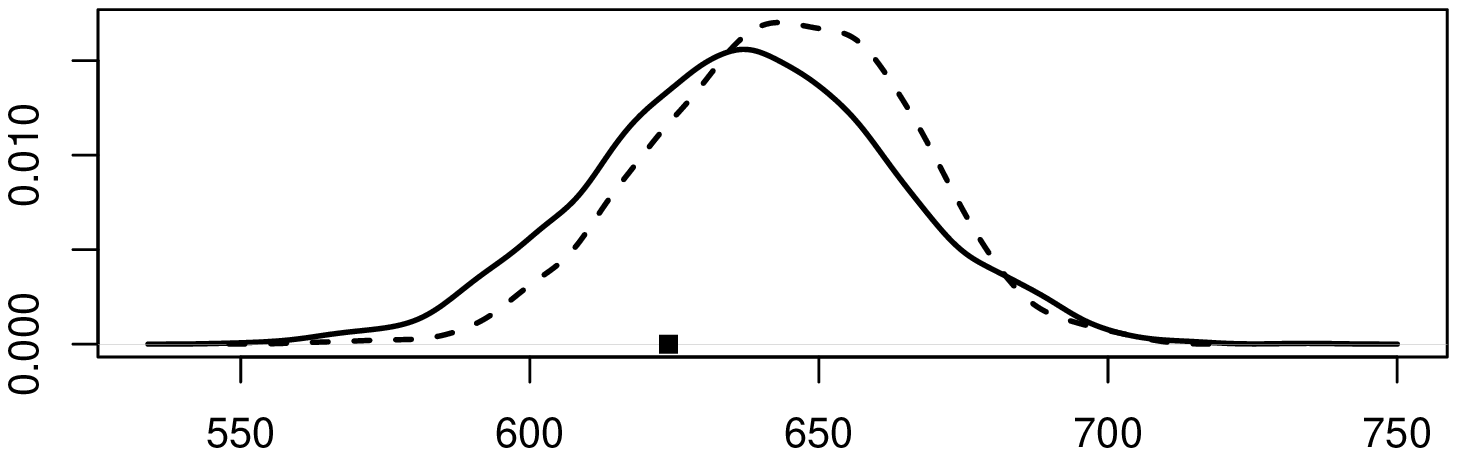}}\\
\subfigure[][$g(x)=\sum_{i,j:\text{horizontal adjacent sites}}I(x_i=x_j)$]{\label{fig:de3}\includegraphics[scale=0.5]{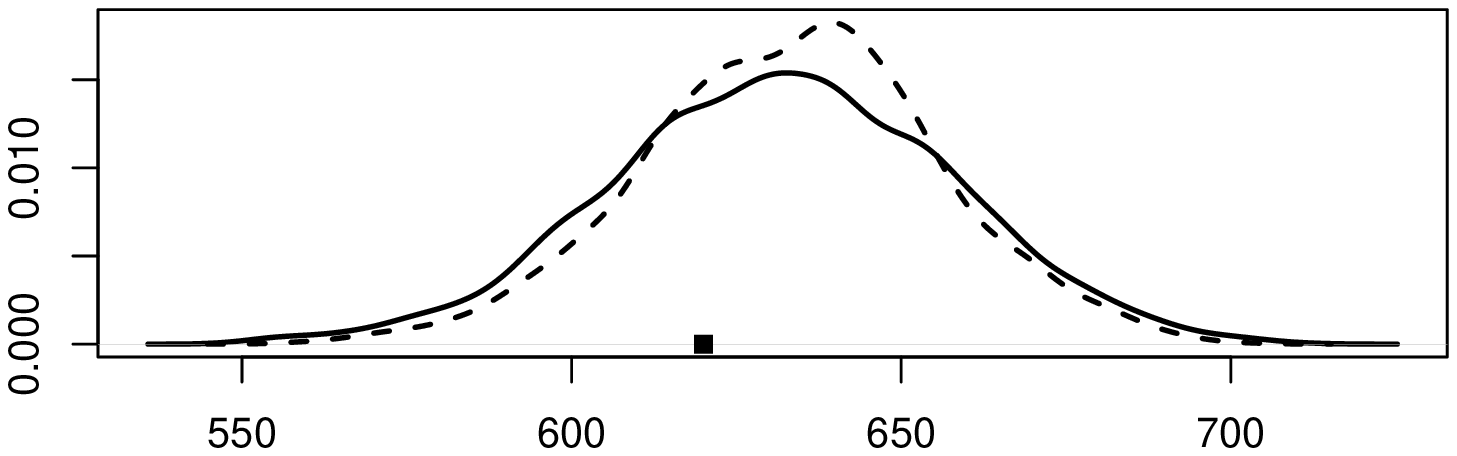}}
\subfigure[][$g(x)=\sum_{\Lambda\in\mathcal{L}_m}I\left ( x_{\Lambda}=
  { \left[\confBBBB \right]}\right)$]{\label{fig:de4}\includegraphics[scale=0.5]{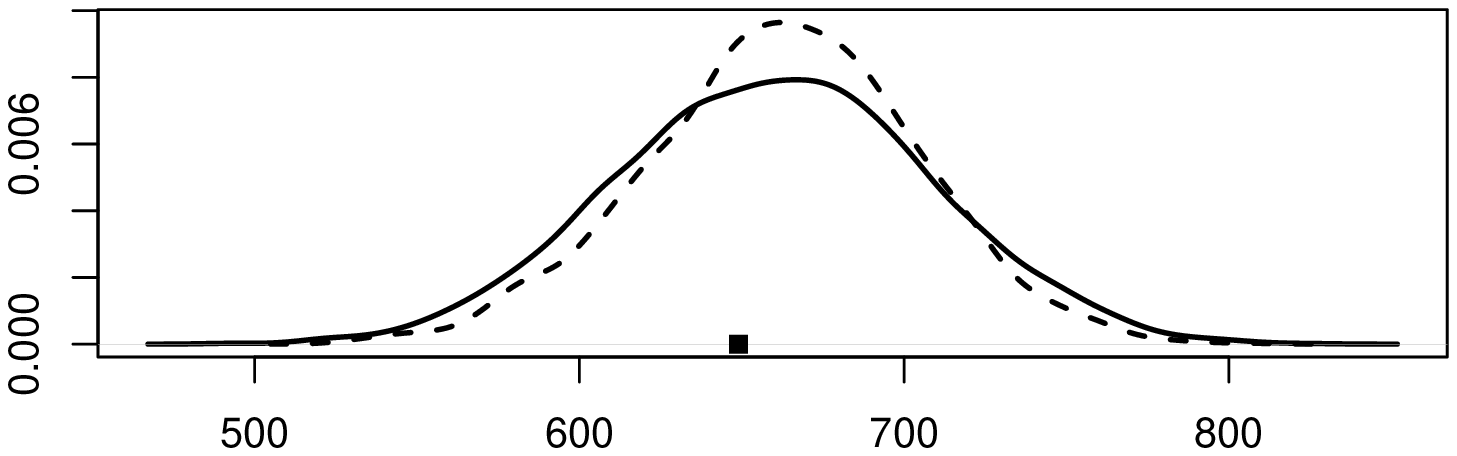}}\\
\subfigure[][$g(x)=\sum_{\Lambda\in\mathcal{L}_m}I\left ( x_{\Lambda}=
  {\left[\confBAAB\right]}\right)$]{\label{fig:de5}\includegraphics[scale=0.5]{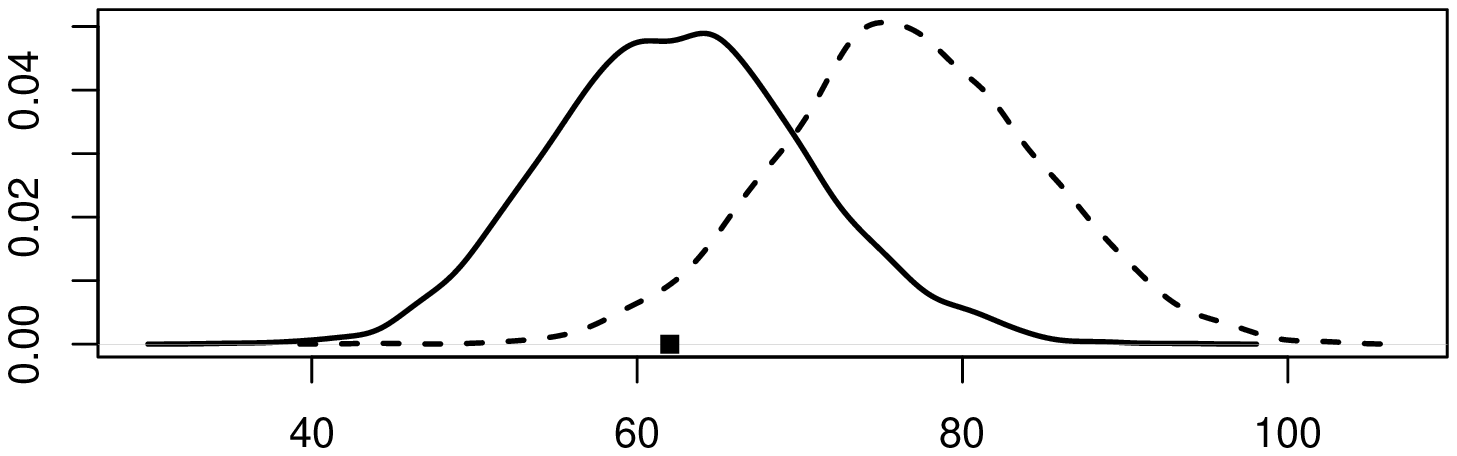}}
\subfigure[][$g(x)=\sum_{\Lambda\in\mathcal{L}_m}I\left ( x_{\Lambda}=
  {\left[\confAAAB\right]}\right)$]{\label{fig:de6}\includegraphics[scale=0.5]{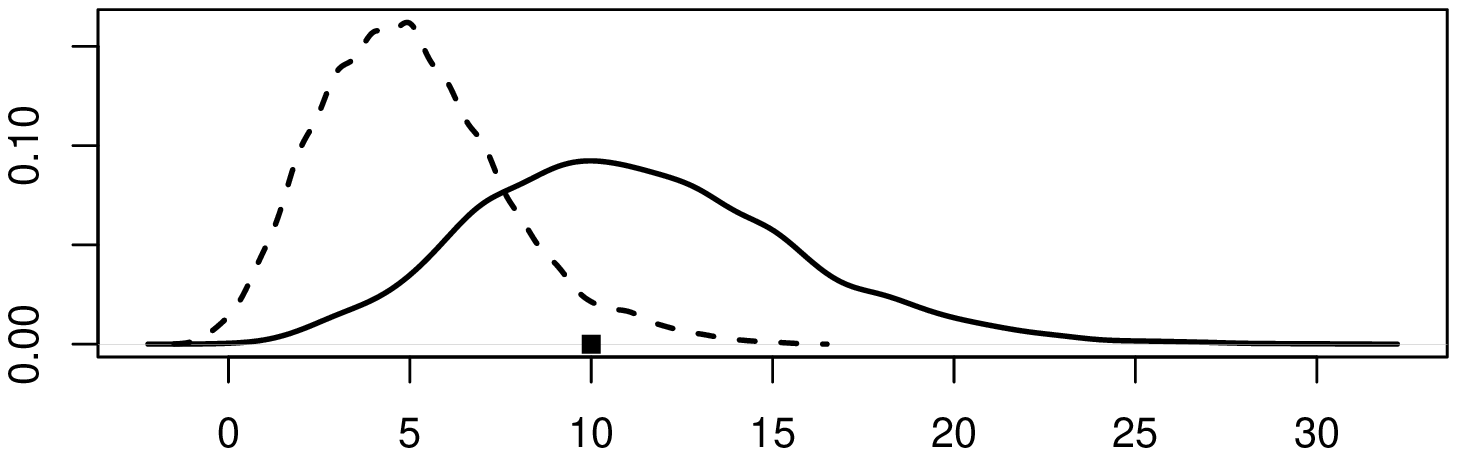}}
\caption{\label{fig:compareSimDeer}Red deer example: Distribution
  of six statistics of
  realisations from our $2\times 2$ model with posterior samples of
  $z$ (solid), and the nearest neighbour pairwise interaction model (dashed). The data evaluated with
each statistic is shown with a black dot.}
\end{figure}

As discussed above, the estimated marginal posterior densities 
for the interaction parameters in Figure \ref{fig:interactionsDeer}
indicate that higher order interaction parameters are needed for this
data set. To investigate this further we also run a corresponding MCMC simulation
with a prior where the spatial interaction is as in the nearest neighbour
autologistic model defined in \citet{Besag1972}, whereas the covariates
are included as in (\ref{eq:likelihoodDeer}). This pairwise interaction
prior has three interaction parameters, for first-order interactions
and for horisontal and vertical second-order interactions, respectively, 
and apriori
we assume these three parameters to be independent and 
Gaussian distributed with zero-mean
and standard deviations equal to ten. To simulate these three
parameters we randomly choose one and propose a zero mean Gaussian
change with standard deviation equal to 0.3 to the chosen parameter. For the $\theta_j^C$ parameters
we adopt the same prior and proposals as before. For the pairwise interaction prior and our original prior
in (\ref{eq:likelihoodDeer}), Figure \ref{fig:compareSimDeer} shows 
\begin{figure*}[h]\center
Figure \ref{fig:compareSimDeer} approximately here.
\end{figure*}
estimates of the resulting
marginal posterior distributions for the same six statistics studied in 
our Ising simulation example. For several of the statistics we see that 
there is a clear difference between the results for the two priors. 
The differences for the higher-order interaction statistics are perhaps
less surprising, but one should note that the distribution of the 
first-order statistic in 
Figure \ref{fig:compareSimDeer}(a) also changes quite much when
allowing higher order interactions. One should also note that 
our $2\times 2$ model fits better to the statistics of the data, 
shown as black dots in the figures.

Returning to the $2\times 2$ prior, using
$\gamma=0$ in the prior for this data set gives the estimated posterior probability
distribution 24\%, 63\%, 11\% and 2\% for 3, 4, 5 and 6 groups
respectively, whereas for $\gamma=1$ we obtain 60\%, 35\% and 5\% for 3, 4
and 5 groups respectively. Again we see that higher values of $\gamma$
results in more realisations with fewer number of
groups. However, for all the three values of $\gamma$ the estimated most probable
grouping is the same.

We end our discussion of this data set by mentioning that some results when assuming a clique size of $3\times 3$ is
included in the supplementary material of this paper. These results
indicate that no more significant structure is introduced in the $3\times
3$ case for this data set.

\section{Closing remarks}
\label{sec:discussion}
Our main focus in this paper is to design a generic prior distribution
for the parameters of an MRF. This is done by assuming a set of maximal 
cliques defined from a template maximal clique $\Lambda_0$, 
but as the number of free parameters grows quickly as a function of
the number of elements in $\Lambda_0$ we construct our prior
distribution such that it gives a positive 
probability for groups of parameters to have
exactly the same value. In that way we reduce the effective number of
parameters, still keeping the flexibility large cliques 
provides. Proposal distributions that enable
us to simulate from the resulting posterior distributed is also
presented. However, to evaluate the likelihood we use a
previously defined approximation to MRFs \citep{phd6}, and the trade
off between accuracy and computational complexity limits in practice the size of the cliques that can be assumed. An
alternative to approximations is perfect sampling \citep{propp1996}, but this was in all
our examples too computationally intensive. A third alternative would be
to use an MCMC sample of $x$ instead of a perfect sample, as described
in for instance \citet{Everitt2012}. An issue with this approach is the need to set a burn in period for the sampler of $x$, where a
too long burn in period would make the parameter sampler too
intensive. Lastly, we illustrate the effect of our prior distribution and
sampling algorithm on two examples. 

Our focus in this paper is on binary
MRFs. It is however possible to generalise our framework to discrete MRFs, i.e. where $x_i\in \{0,1,...,K\}$ for
$K \geq 2$. An identifiable parametrisation of a
discrete MRF using clique potentials can with a small effort be
defined in a similar way to what is done in the binary case, and ones this parametrisation is established, the prior distribution
presented in this paper can be used unchanged. The same apply
to our sampling strategy.    

With our prior distribution the size of the maximal cliques, and
thereby the number of configuration sets, act as a hyper parameter and must
be set prior to any sampling algorithm. One could imagine putting a
prior also on $\Lambda_0$, introducing the need to
construct algorithms for trans-dimensional sampling also for $\Lambda_0$. 
Another way to avoid the need to set the number of
configuration sets would be to construct a
prior distribution for the $\beta$ parameters. A natural choice would
be to construct a positive prior probability for these parameters to
be exactly zero, and in this way the significant interactions of an MRF can
be inferred from data. However, as discussed above, 
it is not clear to us how to design generic
prior distributions for the values of these interaction parameters, as
higher order interactions intuitively would be different from lower
order interaction. Also, grouping $\beta$ parameters together in order
to reduce the number of parameters would, for the same
reason as above, make little sense. An ideal solution would be somehow
to draw strength from both of the two
parametrisations in order to assign a prior distribution to both the 
appearance of different
cliques and the number of free parameters. This idea is currently work
in progress. 

\section*{Supporting Information}

Additional Supporting Information may be found in the online version
of this article:\\

{\noindent \bf Section S.1:} Proof of one-to-one relation
between $\phi$ and $\beta$.

{\noindent \bf Section S.2:} Proof of translational
invariance for $\beta$.

{\noindent \bf Section S.3:} Proof of translational invariance for $\phi$.

{\noindent \bf Section S.4:} Details for the MCMC sampling algorithm.

{\noindent \bf Section S.5:} The independence model with
check of convergence.

{\noindent \bf Section S.6:} Reed deer data with $3 \times 3$ maximal cliques.

{\noindent \bf Section S.7:} Parallelisation of the
  sampling algorithm.


\bibliographystyle{jasa}
\bibliography{bibl}	

\vspace{1cm} \noindent Petter Arnesen, Department of Mathematical Sciences,
Norwegian University of Science and Technology, Trondheim 7491, Norway.
\\
E-mail: petterar@math.ntnu.no

\clearpage

\def\arraystretch{0.5}
\begin{table}
\centering
\begin{tabular}{ c c c }
  $k\times l$ &$2^{kl}$ &$N_{\Lambda_0}$  \\
\hline
  $1\times 2$ & 4 &2  \\
  $2\times 2$ & 16 & 10  \\
$2\times 3$ & 64 & 44 \\
  $3\times 3$ & 512 & 400  \\
$3\times 4$ & 4096 & 3392 \\
  $4\times 4$ & 65536 & 57856 
\end{tabular}
 \caption{\label{tab:ktimesl}The number of configurations and the
   corresponding number of free parameters $N_{\Lambda_0}$ when
   $\Lambda_0$ is a $k\times l$ block of nodes.}
\end{table}

\begin{figure}
\centering
\begin{equation*}
  \arraycolsep=4.5pt\def\arraystretch{0.6}\begin{array}{cccccccccccc}
c^0 &\vv1.00&\vv 1.00&&&&&&&&&\\
c^{\confcAAAA} &\vv 1.00 &\vv1.00&&&&&&&&&\\
c^1 &&&\vv1.00&\vv 0.96&\vv 0.97&\vv 0.97&\vv0.97&\vv 0.97&\vv0.96&&\\
c^{11}&&&\vv0.96 &\vv 1.00&\vv 0.96&\vv 0.96&\vv
0.96&\vv 0.96&\vv0.96&&\\
c^{\confcABAB}&&&\vv 0.97&\vv 0.96&\vv 1.00&\vv 0.97&\vv 0.96&
\vv 0.96&\vv 0.95&&\\
c^{\confcAAAB}&&&\vv0.97&\vv0.96&\vv0.97&\vv1.00&\vv 0.96&\vv 0.96&\vv0.96&&\\
c^{\confcAABA}&&&\vv0.97&\vv0.96&\vv0.96&\vv0.96&\vv 1.00&\vv 0.97&\vv
0.97&&\\
c^{\confcABAA}&&&\vv0.97&\vv0.96&\vv 0.96&\vv 0.96&\vv
0.97&\vv 1.00&\vv 0.96&&\\
c^{\confcBAAA}&&&\vv0.96&\vv0.96&\vv0.95 &\vv 0.96&\vv0.97&\vv
0.96&\vv 1.00&&\\
c^{\confcABBA}&&&&&&&&&&\vv 1.00&\vv 1.00\\
c^{\confcBAAB}&&&&&&&&&&\vv 1.00&\vv 1.00\\
&c^0 & c^{\confcAAAA} & c^1 &c^{11} & c^{\confcABAB} & c^{\confcAAAB} & 
c^{\confcAABA} & c^{\confcABAA} & c^{\confcBAAA} & c^{\confcABBA} & c^{\confcBAAB}
\end{array}
\end{equation*}
\caption{\label{fig:groupingMatrixIsing}Ising model example: Estimated posterior probabilities for two configuration
    sets to be grouped together. The true grouping is shown
    in grey, and only probabilities larger than $5 \%$ are given. Note
  the permutation done to the ordering of the configuration sets $c_i$.}
\end{figure}


\begin{figure}
\centering
\begin{equation*}
  \arraycolsep=4.5pt\def\arraystretch{0.6}\begin{array}{cccccccccccc}
c^0 & \vv1.00&&&&&0.12&0.06&&&&\\
c^1 &&\vv1.00&\vv 0.69&\vv 0.68&\vv0.58& \vv 0.67&\vv
0.75& \vv 0.70 &\vv 0.68&\vv 0.60 &\\
c^{11} &&\vv 0.69&\vv1.00&\vv 0.80&\vv 0.69&\vv
0.62&\vv0.64&\vv 0.70&\vv 0.69&\vv 0.65 &\\
c^{\confcABAB} &&\vv 0.68 &\vv 0.80&\vv 1.00&\vv 0.69&\vv
0.61&\vv 0.64&\vv 0.70&\vv 0.69&\vv 0.65&\\
c^{\confcABBA} &&\vv 0.58&\vv0.69&\vv0.69&\vv 1.00&\vv0.55&\vv
0.58&\vv 0.64&\vv 0.64 &\vv 0.69 &0.05\\
c^{\confcBAAB} &0.12&\vv 0.67&\vv 0.62&\vv 0.61&\vv 0.55&\vv
1.00&\vv 0.62&\vv0.61&\vv 0.61&\vv 0.57&\\
c^{\confcAAAB} &0.06&\vv 0.75&\vv 0.64 &\vv0.64&\vv 0.58&\vv0.62&\vv1.00&\vv 0.65&\vv 0.64&\vv0.58&\\
c^{\confcAABA}&&\vv 0.70&\vv 0.70&\vv0.70&\vv 0.64&\vv0.61&\vv 0.65&\vv 1.00&\vv
0.67&\vv0.63 &\\
c^{\confcABAA}&&\vv0.68&\vv0.69&\vv 0.69&\vv0.64&\vv61&\vv0.64&\vv0.67&\vv 1.00&\vv 0.63&\\
c^{\confcBAAA}&&\vv0.60&\vv 0.65&\vv 0.65&\vv 0.69&\vv 0.57&\vv0.58&\vv0.63&\vv0.63 &\vv 1.00&0.06\\
c^{\confcAAAA}&&&&&0.05&&&&&0.06&\vv 1.00\\
&c^0 & c^1 & c^{11} & c^{\confcABAB} & c^{\confcABBA} & 
c^{\confcBAAB} & c^{\confcAAAB} & c^{\confcAABA} &
c^{\confcABAA} & c^{\confcBAAA} & c^{\confcAAAA}
\end{array}
\end{equation*}
\caption{\label{fig:configurationMatrixDeer}Red deer example: 
  Estimated posterior probabilities for two configuration
    sets to be grouped together. The
    estimated most probable grouping is shown in grey, and only
    probabilities larger than 5 \% are given.}
\end{figure}

\end{document}


	

{\noindent{\LARGE \textbf{Supplementary materials to the paper: \\Fully Bayesian binary Markov random field\\
      models: 
Prior specification and posterior\\ simulation} \vspace{1cm} \\} 
{\Large \textsc{Petter Arnesen}}\\{\it Department of
    Mathematical Sciences, Norwegian University of Science and
    Technology}\vspace{1cm}\\{\Large \textsc{H\aa kon Tjelmeland}}\\{\it Department of
    Mathematical Sciences, Norwegian University of Science and
    Technology}\vspace{1cm}
}

\renewcommand{\baselinestretch}{1.5} \small\normalsize

\renewcommand{\thesection}{S.\arabic{section}}
\renewcommand{\thesubsection}{\thesection.\arabic{subsection}}
                
\section{Proof of one-to-one relation between $\phi$ and $\beta$}
\label{sec:S1}

\begin{theorem}
Consider an MRF and constrain the $\phi$ parametrisation of the potential functions as described in
Definition 1. Then there is a one-to-one relation between 
$\{\beta^\lambda;\lambda\in\mathcal{L}\}$
and $\{\phi^\lambda;\lambda\in\mathcal{L}\}$. 
\end{theorem}

\begin{proof}We prove the theorem by
establishing recursive equations showing how to compute the $\beta^\lambda$'s from
the $\phi^\lambda$'s and vice versa.

Setting $x=\bold 1^\lambda$ for some $\lambda\in\mathcal{L}$ into the two 
representations of $U(x)$ in (2) and (4), we get
\begin{equation*}
\sum_{\Lambda\in \mathcal{L}_m}V_{\Lambda}(\bold
1^{\lambda}_{\Lambda})=\sum_{\lambda^\star\in \mathcal{L}}\beta^{\lambda^\star}
\prod_{(i,j)\in \lambda^\star} \bold 1^{\lambda}_{\{(i,j)\}}.
\end{equation*}
Using (3) and Definition 1 this gives
\begin{equation}
\sum_{\Lambda\in \mathcal{L}_m}\phi^{\lambda\cap
  \Lambda}=\sum_{\lambda^\star\subseteq \lambda}\beta^{\lambda^\star}.
\label{eq:equal}
\end{equation}
Splitting the sum on the left hand side into one sum over 
$\Lambda\in\mathcal{L}_m^\lambda$ and one sum over
$\Lambda\in\mathcal{L}_m\setminus\mathcal{L}_m^\lambda$,
and using that $\lambda\cap \Lambda=\lambda$ for 
$\Lambda\in \mathcal{L}_m^\lambda$ we get
\begin{equation*}
|\mathcal{L}_m^{\lambda}|\phi^{\lambda}+\sum_{\Lambda\in \mathcal{L}_m\setminus \mathcal{L}_m^\lambda}\phi^{\lambda\cap
  \Lambda}=\sum_{\lambda^\star\subseteq \lambda}\beta^{\lambda^\star}.
\end{equation*}
Solving for $\phi^\lambda$ gives
\begin{equation}
\phi^{\lambda}=\frac{1}{|\mathcal{L}_m^{\lambda}|}\left
  [\sum_{\lambda^\star\subseteq
    \lambda}\beta^{\lambda^\star}-\sum_{\Lambda\in \mathcal{L}_m\setminus
\mathcal{L}_m^\lambda}\phi^{\lambda\cap \Lambda}\right].
\label{eq:claim}
\end{equation}
Clearly $|\lambda\cap\Lambda| < | \lambda|$ when
$\Lambda\in \mathcal{L}_m\setminus \mathcal{L}_m^\lambda$,
so \eqref{eq:claim} implies that we can compute all 
$\{ \phi^\lambda;\lambda\in\mathcal{L}\}$ recursively from 
$\{ \beta^\lambda;\lambda\in\mathcal{L}\}$. First 
we can compute $\beta^\lambda$ for $|\lambda|=0$, i.e. 
$\phi^\lambda=\phi^\emptyset = \beta^\emptyset/|\mathcal{L}_m|$, then
all $\beta^\lambda$ for which $|\Lambda|=1$, thereafter
all $\beta^\lambda$ for which $|\Lambda|=2$ and so on until
$\beta^\lambda$ has been computed for all $\lambda\in\mathcal{L}$.
Thus, $\{ \phi^\lambda;\lambda\in\mathcal{L}\}$ is uniquely
specified by $\{ \beta^\lambda;\lambda\in\mathcal{L}\}$.

Solving \eqref{eq:equal} with respect to $\beta^\lambda$ we get
\begin{equation}
\beta^{\lambda}=\sum_{\Lambda\in \mathcal{L}_m}\phi^{\lambda
  \cap \Lambda}-\sum_{\lambda^\star \subset \lambda}\beta^{\lambda^\star},
\label{eq:claim2}
\end{equation}
and noting that clearly $|\lambda^\star|<|\lambda|$ when 
$\lambda^\star \subset \lambda$ we correspondingly  get 
that $\{ \beta^\lambda;\lambda\in\mathcal{L}\}$ can be 
recursively computed from $\{ \phi^\lambda;\lambda\in\mathcal{L}\}$.
One must first compute $\beta^\lambda$ for $|\lambda|=0$, i.e.
$\beta^\emptyset$, then all $\beta^\lambda$ for which 
$|\lambda|=1$, thereafter all $\beta^\lambda$ for which 
$|\lambda|=2$ and so on. Thereby $\{ \beta^\lambda;\lambda\in\mathcal{L}\}$
is also uniquely specified by $\{ \phi^\lambda;\lambda\in\mathcal{L}\}$
and the proof is complete.
\end{proof}

\section{Proof of translational invariance for $\beta$}
\label{sec:S2}
\begin{theorem}
An MRF $x$ defined on a rectangular lattice
$S=\{ (i,j);i=0,\ldots,n-1,j=0,\ldots,m-1\}$ with 
torus boundary conditions and $\mathcal{L}_m$ given
in (5) is stationary if and only 
if $\beta^{\lambda}=\beta^{\lambda\oplus (t,u)}$ for all 
$\lambda \in \mathcal{L}$, $(t,u)\in S$. We then say
that $\beta^\lambda$ is translational invariant.
\end{theorem}

\begin{proof}
We first prove the only if part of 
the theorem by induction on $|\lambda|$. Since $\emptyset
\oplus (t,u)=\emptyset$ we clearly have
$\beta^{\emptyset}=\beta^{\emptyset \oplus (t,u)}$ and thereby
$\beta^{\lambda}=\beta^{\lambda \oplus (t,u)}$ for $|\lambda|=0$. 
Now assume all interaction parameters up to order $o$
to be translational invariant, i.e. $\beta^{\lambda}=\beta^{\lambda \oplus
(t,u)}$ for $|\lambda|\leq o$. Now focusing on any $\lambda^\star\in \mathcal{L}$
with $|\lambda^\star|=o+1$, the assumed stationarity in particular gives 
that we must have $p(\bold 1^{\lambda^\star})=p(\bold 1^{\lambda^\star\oplus (t,u)})$.
Using (2) and (4) it follows that
\begin{equation*}
\beta^{\lambda^\star}+\sum_{\lambda \subset \lambda^\star}
\beta^{\lambda}=\beta^{\lambda^\star\oplus(t,u)}+\sum_{\lambda \subset \lambda^\star\oplus(t,u)}\beta^{\lambda}.
\end{equation*}
Rewriting the sum on the right hand side we get
\begin{equation*}
\beta^{\lambda^\star}+\sum_{\lambda \subset \lambda^\star}\beta^{\lambda}=\beta^{\lambda^\star\oplus(t,u)}+
\sum_{\lambda \subset \lambda^\star}\beta^{\lambda\oplus(t,u)},
\end{equation*}
where the induction assumption gives that the two sums must be equal, and 
thereby $\beta^{\lambda^\star}=\beta^{\lambda^\star\oplus(t,u)}$, which 
completes the only if part of the proof.

To prove the if part of the theorem we need to show that if the interaction parameters
are translational invariant then $U(\bold 1^A) = U( \bold 1^{A\oplus (t,u)})$ for any
$A\subseteq S$ and $(t,u)\in S$. For $U(\bold 1^A)$ we have
\begin{eqnarray*}
U(\bold 1^A) &=& \sum_{\lambda\in\mathcal{L}} \beta^\lambda \prod_{(i,j)\in\lambda} \bold 1^A_{i,j}\\
&=& \sum_{\lambda\in \mathcal{L}} \beta^{\lambda\oplus (t,u)} \prod_{(i,j)\in\lambda\oplus (t,u)}
\bold 1_{i,j}^A\\
&=& \sum_{\lambda\in\mathcal{L}} \beta^{\lambda\oplus (t,u)} \prod_{(i,j)\in \lambda}
\bold 1_{i,j}^{A\oplus (t,u)} \\
&=& \sum_{\lambda\in\mathcal{L}} \beta^{\lambda} \prod_{(i,j)\in \lambda}
\bold 1_{i,j}^{A\oplus (t,u)} = U(\bold 1^{A\oplus (t,u)}),
\end{eqnarray*}
where the first equality follows from (4), the second equality
is true because $\{ \lambda\oplus (t,u);\lambda\in \mathcal{L}\} =
\mathcal{L}$ for any $(t,u)\in S$, the third equality follows 
from the identity $\bold 1^A_{(i,j)\oplus (t,u)} = 
\bold 1_{i,j}^{A\oplus (t,u)}$, and the fourth equality is using the assumed 
translational invariance of the interaction parameters. Thereby the proof
is complete.
\end{proof}

\section{Proof of translational invariance for $\phi$}
\label{sec:S3}

\begin{theorem}
An MRF $x$ defined on a rectangular lattice 
$S=\{ (i,j);i=0,\ldots,n-1,j=0,\ldots,m-1\}$ with torus boundary
conditions and $\mathcal{L}_m$ given in 
(5) is stationary if and only if
$\phi^{\lambda}=\phi^{\lambda\oplus (t,u)}$ for all 
$\lambda \in \mathcal{L}$ and $(t,u)\in S$. We then say
that $\phi^\lambda$ is translational invariant. 
\end{theorem}

\begin{proof}
Given the result in Theorem 2 it is 
sufficient to show that $\phi^\lambda$ is translational invariant
if and only if $\beta^\lambda$ is translational invariant. We first
assume $\beta^\lambda$ to be translational invariant for all $\lambda\in\mathcal{L}$ 
and need to show that then also $\phi^\lambda$ must be translational invariant for 
all $\lambda\in\mathcal{L}$. Starting with 
\eqref{eq:claim}, repeatedly using the specific form we are using for $\mathcal{L}_m$ and the assumed
translational invariance for $\beta^\lambda$, we get for any $\lambda\in\mathcal{L}$, $(t,u)\in S$ 
\begin{eqnarray*}
\phi^{\lambda\oplus (t,u)} &=& \frac{1}{\left|\mathcal{L}_m^{\lambda\oplus (t,u)}\right|}
\left[ \sum_{\lambda^\star\subseteq \lambda\oplus (t,u)}\beta^{\lambda^\star} -
\sum_{\Lambda\in\mathcal{L}_m} \phi^{(\lambda\oplus (t,u))\cap\Lambda}
+ \sum_{\Lambda\in\mathcal{L}_m^{\lambda\oplus (t,u)}} \phi^{(\lambda\oplus (t,u))\cap \Lambda}\right] \\
&=& \frac{1}{|\mathcal{L}_m^\lambda|}\left[ \sum_{\lambda^\star\subseteq \lambda}
\beta^{\lambda^\star\oplus (t,u)} - 
\sum_{\Lambda\in\mathcal{L}_m} \phi^{(\lambda\oplus (t,u))\cap (\Lambda\oplus (t,u))}
+ \sum_{\Lambda\in\mathcal{L}_m^\lambda} \phi^{(\lambda\oplus (t,u))\cap (\Lambda\oplus (t,u))}\right]\\
&=& \frac{1}{|\mathcal{L}_m^\lambda|} \left[ \sum_{\lambda^\star\subseteq \lambda} \beta^{\lambda^\star}
- \sum_{\Lambda\in\mathcal{L}_m} \phi^{(\lambda\cap \Lambda)\oplus (t,u)} +
\sum_{\Lambda\in\mathcal{L}_m^\lambda}\phi^{(\lambda\cap\Lambda) \oplus (t,u)}\right] \\
&=& \frac{1}{|\mathcal{L}_m^\lambda|}\left[ \sum_{\lambda^\star\subseteq\lambda} \beta^{\lambda^\star} - 
\sum_{\Lambda\in\mathcal{L}_m\setminus\mathcal{L}_m^\lambda} \phi^{(\lambda\cap\Lambda)\oplus (t,u)}\right].
\end{eqnarray*}
From this we can use induction on $|\lambda|$ to show that $\phi^\lambda$ is 
translational invariant for all $\lambda\in\mathcal{L}$. Setting $\lambda=\emptyset$
we get $\phi^{\lambda\oplus (t,u)} = (1/|\mathcal{L}_m^\emptyset|)\beta^\emptyset$ which 
is clearly not a function of $(t,u)$. Then assuming $\phi^{\lambda\oplus (t,u)}=\phi^\lambda$
for all $\lambda\in \mathcal{L}$ with $|\lambda| \leq o$, considering the above 
relation for a $\lambda$ with $|\lambda|=o+1$, and observing that 
$|\lambda\cap \Lambda|\leq o$ when $|\lambda|=o+1$ and $\Lambda\in \mathcal{L}_m\setminus
\mathcal{L}_m^\lambda$, we get that also $\phi^{\lambda\oplus (t,u)}=\phi^\lambda$ for 
$|\lambda|=o+1$, and the induction proof is complete.

Next we assume $\phi^\lambda$ to be translational invariant for all $\lambda\in\mathcal{L}$
and need to show that then also $\beta^\lambda$ is translational invariant for all 
$\lambda\in\mathcal{L}$. Starting with \eqref{eq:claim2}, using the assumed translational 
invariance of $\beta^\lambda$, 
and again repeatedly using the specific form of $\mathcal{L}_m$, we get
\begin{eqnarray*}
\beta^{\lambda\oplus (t,u)} &=& \sum_{\Lambda\in\mathcal{L}_m}\phi^{(\lambda\oplus (t,u))\cap \Lambda}
-\sum_{\lambda^\star\subset \lambda\oplus (t,u)}\beta^{\lambda^\star} \\
&=& \sum_{\Lambda\in\mathcal{L}_m} \phi^{(\lambda\oplus (t,u))\cap (\Lambda\oplus (t,u))}
- \sum_{\lambda^\star \subset \lambda} \beta^{\lambda^\star \oplus (t,u)}\\
&=&\sum_{\Lambda\in\mathcal{L}_m} \phi^{(\lambda\cap \Lambda)\oplus (t,u))}
- \sum_{\lambda^\star \subset \lambda} \beta^{\lambda^\star \oplus (t,u)}.
\end{eqnarray*}
Using this we can easily use induction on $|\lambda|$ to show that we must have 
$\beta^{\lambda\oplus (t,u)}=\beta^\lambda$. The proof is thereby complete.
\end{proof}   

\section{Details for the MCMC sampling algorithm}
\label{sec:S4}

In this section we provide details of the proposal distributions that we use when sampling from the posterior distribution
\begin{equation*}
p(z|x) \propto p(x|z) p(z),
\end{equation*}
where $p(x|z)$ and $p(z)$ are the MRF
and the prior given in the paper, respectively.

To simulate from this posterior distribution we adopt a reversible 
jump Markov chain Monte Carlo (RJMCMC) algorithm with 
three types of updates. The first update type uses a random walk proposal
for one of the $\varphi$ parameters, the second proposes to move one 
configuration set to a new group, and the third proposes to 
change the number of groups, $r$, in the partition of the 
configuration sets. In the following we describe the 
proposal mechanisms for each of the three update types. 
The corresponding acceptance probabilities are given by standard formulas. It should be noted that only the 
last type of proposal produces a change in the dimension of 
the parameter space.
   
\subsection{\it Random walk proposal for parameter values}

The first proposal in our algorithm is simply to propose
a new value for an already existing parameter using a random walk proposal, but
correcting for the fact that the parameters should sum to zero. More precisely, we first draw a change $\varepsilon \sim \mbox{N}(0,\sigma^2)$,
where $\sigma^2$ is an algorithmic tuning parameter. Second, we
uniformly draw one element from the current state
$z=\{ (C_i,\varphi_i),i=1,\ldots,r\}$, $(C_i,\varphi_i)$ say, and
define the potential new state as
\begin{equation*}
z^* = \left \{ \left (C_j,\varphi_j - \frac{1}{r}\varepsilon \right ),
j=1,\ldots,i-1,i+1,\ldots,r \right \} \cup \left \{
\left (C_i,\varphi_i+\varepsilon - \frac{1}{r}\varepsilon \right )
\right \}.
\end{equation*}

\subsection{\it Proposing to change the group for one configuration set}
\label{sec:swappingProposal}
Letting the current state be $z=\{ (C_i,\varphi_i),i=1,\ldots,r\}$,
we start this proposal by drawing a pair of groups, 
$C_i$ and $C_j$ say, where the first set $C_i$ is restricted to 
include at least two configuration sets. We draw $C_i$ and $C_j$ so 
that the difference between the 
corresponding parameter values, $\varphi_i - \varphi_j$, tend to be small.
More precisely, we draw $(i,j)$ from the joint distribution
\begin{eqnarray*}
q(i,j)\propto
\begin{cases}
\exp\left(-(\varphi_i-\varphi_j)^2\right ) & \mbox{if $i\neq
  j$ and group $C_i$ contains at least two configuration sets,}\\
0  & \mbox{otherwise.}
\end{cases}
\end{eqnarray*}
Thereafter
we draw uniformly at random one of the configuration sets in $C_i$, 
$c$ say. Our potential new state is then obtained by moving $c$ from 
$C_i$ to $C_j$. Thus, our potential 
new state becomes 
\begin{equation*}
z^* = \left ( z \setminus 
\{ (C_i,\varphi_i),(C_j,\varphi_j)\}\right ) 
\cup \left \{ (C_i\setminus c,\varphi_i ), (C_j\cup
c,\varphi_j )\right \}. 
\end{equation*}

\subsection{\it Trans-dimensional proposals}
\label{sec:transDimProposal}
Let again the current state be $z=\{ (C_i,\varphi_i),i=1,\ldots,r\}$. In
the following we describe how we propose a new state by either
increasing or reducing the number of groups, $r$, with one. There
will be a one-to-one transition in the proposal, meaning that the
opposite proposal, going from the new state to the old state has a non-zero
probability. We make no attempt to jump between states where the
difference between the dimensions are larger than one. 

First we draw whether to increase or to decrease the number of
groups. If the number of groups are equal to the number of
configurations sets, no proposal to increase the number of groups can
be made due to the fact that empty groups have zero prior
probability. In that case we propose to decrease the number
of dimensions with probability 1. In our proposals we also make the
restriction that only groupings containing at least one group with
only one configuration set can be subject to a dimension reducing
proposal. In a case where no such group exists, a proposal of
increasing the number of dimensions are made with probability 1. In a
case where both proposals are allowed we draw at random which to do with
probability 1/2 for each. Note that at least one of the two proposals
is always valid. 

We now explain how to propose to increase the number of groups by one.
We start by drawing uniformly at random one of the groups with more 
than one configuration set, $C_i$ say, which we want to split into two 
new groups. Thereafter we draw uniformly at random one of the 
configuration sets in $C_i$, $c$ say, and form a new partition of 
the configuration sets by extracting $c$ from $C_i$ and adding a 
new group containing only $c$. Next we need to draw a parameter value for 
the new group $\{ c\}$, and the parameter values for the other groups
also need to be modified for the proposal to conform with the requirement
that the sum of the (proposed) parameters should equal zero. 
We do this by 
first drawing a change $\varepsilon\sim \mbox{N}(0,\sigma^2)$ in 
the parameter value for $c$, where $\sigma^2$ is the same tuning
parameter as in the random walk proposal. We then define the potential new state as
\begin{eqnarray*}
z^* &=& \left \{ \left (C_j,\varphi_j - \frac{1}{r+1}(\varphi_i +
    \varepsilon)\right ),
j=1,\ldots,i-1,i+1,\ldots,r\right \} \cup \\
&&\left \{
\left (C_i\setminus c,\varphi_i - \frac{1}{r+1}(\varphi_i + \varepsilon)\right),
\left(\{ c\},\varphi_i + \varepsilon - \frac{1}{r+1}(\varphi_i +
  \varepsilon)\right)\right \}.
\end{eqnarray*}

Next we explain the proposal we make when the dimension is to be decreased
by one. Since we need a one-to-one transition in our proposals, we
get certain restrictions for these proposals. In particular, the fact that
only groupings containing at least one group with only one
configuration set are possible outcomes from a dimension increasing
proposal dictates that dimension decreasing proposals only can be made
from such groupings. Assume again our current model to be $z=\{
(C_i,\varphi_i),i=1,\ldots,r\}$, where at least one group contains
only one configuration set. The strategy is to propose to merge one group consisting of only one 
configuration set into another group. As in Section \ref{sec:swappingProposal}, we draw the two configuration
sets to be merged so that the difference between the corresponding
parameter values tend to be small. More precisely, we let the two 
groups be $C_i$ and $C_j$ where $(i,j)$ is sampled according to the 
joint distribution
\begin{eqnarray*}
q(i,j)\propto
\begin{cases}
\exp\left(-(\varphi_i-\varphi_j)^2\right ) & \mbox{if $i\neq j$ and $C_i$ consists of 
only one configuration set,}\\
0  & \mbox{otherwise.}
\end{cases}
\end{eqnarray*}
Next we need to specify potential new parameter values. As
these must conform with how we generated potential new values in the split
proposal, we have no freedom left in how to do this. The potential 
new state must be
\begin{equation*}
z^* = \left \{ \left (C_k,\varphi_k + \frac{1}{r-1}\varphi_i\right
 ), k\in \{ 1,\ldots,r\} \setminus \{ i,j\}\right \}
\cup \left \{ \left (C_j\cup C_i,\varphi_j +
    \frac{1}{r-1}\varphi_i\right )\right \}.
\end{equation*}
The split and merge steps produce a change in the dimension of 
the parameter space, so to calculate the acceptance probabilities 
for such proposals we need corresponding Jacobi determinants.
It is straightforward to show that the Jacobi determinants for 
the merge and split proposals become $\frac{r}{r-1}$ and
$\frac{r}{r+1}$, respectively.

\section{The independence model with check of convergence}
\label{sec:S5}

Consider a model where the variables are all independent of each other
and $p(x_{i,j})=p^{x_{i,j}}(1-p)^{1-x_{i,j}}$ for each $(i,j)\in S$ and where $p$ is the
probability of $x_{i,j}$ being equal to $1$. We get 
\begin{equation}\label{eq:independence}
p(x)=\prod_{(i,j)\in S}p^{x_{i,j}}(1-p)^{1-x_{i,j}}\propto\exp\left(\alpha \sum_{(i,j)\in S}x_{i,j}\right),
\end{equation}
where
\begin{equation*}
\alpha=\ln \left ( \frac{p}{1-p}\right ).
\end{equation*}
In this section we use the independence model as an example, and in particular we fit an MRF with $2 \times 2$ maximal cliques to
data simulated from this model. Therefore it is helpful to know
how one
can represent the independence model using $2 \times 2$ maximal
cliques, and this can be done using the same strategy that was used
for the Ising model in Section 2.2 in the paper. We get $\phi^0=\eta$, $\phi^1=\frac{\alpha}{4}+\eta$, $\phi^{11}=\phi^{\confcABAB}=\phi^{\confcABBA}=
\phi^{\confcBAAB}=\frac{\alpha}{2}+\eta$, $\phi^{\confcAAAB}=\phi^{\confcAABA}=\phi^{\confcABAA}=
\phi^{\confcBAAA}=\frac{3\alpha}{4}+\eta$ and $\phi^{\confcAAAA}=\alpha+\eta$, where 
$\eta$ is an arbitrary value coming from the arbitrary value for $\beta^\emptyset$.
If $p=0.5$ we see that $\alpha=0$ and all the configuration set
parameters are equal.  

We generate a realisation from the independence model with $p=0.3$ on a $100\times 100$
lattice, consider this as our observed data $x$, and simulate by the
MCMC algorithm defined in Section 4 from the
resulting posterior distribution. Using the notation for the configuration
sets in a $2\times 2$ maximal clique and the results above, we ideally want our algorithm to produce realisations
with the groups $\{c^0\}$, $\{c^1\}$, $\{c^{11},c^{\confcABAB},
c^{\confcABBA},c^{\confcBAAB}\}$,
$\{c^{\confcAAAB},c^{\confcAABA},c^{\confcABAA},c^{\confcBAAA}\}$ and
$\{c^{\confcAAAA}\}$. Note that due to our
identifiability restriction in
(11) the configuration set parameters
should be close to the solution above with
$\eta=-\alpha/2$. To check convergence we investigated trace plots of
various statistics, see Figure \ref{fig:tracePlotsMarginal}, and the
conclusion was that the algorithm converges very quickly. 
\begin{figure*}[h]\center
Figure \ref{fig:tracePlotsMarginal} approximately here.
\end{figure*}
The acceptance rate for the parameter value proposals
is $24\%$, whereas the acceptance rates for the other two types of 
proposals are both around $2\%$. We run
our sampling algorithm for 20000 iterations, and estimate the
posterior probability of the number of groups. The configuration sets
are organised into 4 (77\%), 5
(21\%) or 6 (2\%) groups, so for these data the grouping tends to be a little bit
too strong compared to the correct number of groups. This can also be seen
from the estimated posterior probability of two configuration sets being
assigned to the same group, shown in Figure \ref{fig:pairMatrixMarginal}.
\begin{figure*}[h]\center
Figure \ref{fig:pairMatrixMarginal} approximately here.
\end{figure*}
This figure suggests the four groups
$\{c^0\},\{c^1\},\{c^{11},c^{\confcABAB},c^{\confcABBA},c^{\confcBAAB},
c^{\confcAABA}\}$, $\{c^{\confcAAAB},c^{\confcABAA},c^{\confcBAAA},
c^{\confcAAAA}\}$
which is also calculated to be the most probable
grouping estimated by counting the number of occurrences. In fact the
posterior probability for this grouping is as high as $55 \%$. In Figure \ref{fig:pairMatrixMarginal} we also see how the most probable
grouping differ from the correct model grouping, shown in grey. The group $\{ c^{\confcAAAB},c^{\confcAABA},c^{\confcABAA},
c^{\confcBAAA}\}$ in the correct model is split in two
in the most probable grouping, and the subsets 
$\{ c^{\confcAABA}\}$ and 
$\{ c^{\confcAAAB},c^{\confcABAA},c^{\confcBAAA}\}$ are inserted into the correct model groups 
$\{ c^{11},c^{\confcABAB},c^{\confcABBA},c^{\confcBAAB}\}$  
and $\{ c^{\confcAAAA}\}$, respectively.

As in the Ising model example in Section 5.1 we estimate the posterior distribution for the
interaction parameters, see Figure \ref{fig:marginalInteractionParameters}.
\begin{figure*}[h]\center
Figure \ref{fig:marginalInteractionParameters} approximately here.
\end{figure*}
As we can
see, the true value of the interaction parameters are mostly within the
credibility intervals, but the tendency to group the configurations too
much is in this case forcing some of the true values into a tail of
the marginal posterior distributions.  

As in the Ising example we compare the distribution of the same six
statistics from simulations from our $2\times 2$ model with
posterior samples of $z$, the independence model with correct parameter
value, and the independence model with parameter value obtained by posterior
sampling, see Figure \ref{fig:compareSimIndependence}.
\begin{figure*}[h]\center
Figure \ref{fig:compareSimIndependence} approximately here.
\end{figure*}
As we can see, our model captures approximately the
correct distribution of the chosen statistics. It is interesting to
note that for some statistics the realisations from the independence
model with simulated $\alpha$ values follows our model tightly
whereas for the other statistics it is close to the correct model.

Also for this data set we investigated the case where $\gamma=0$ and
$\gamma=1$. For $\gamma=0$ the configuration
sets are organised into 4 (75\%), 5
(23\%) or 6 (2\%) groups, and for $\gamma=1$ we get 4 (93\%), 5 (6\%),
6 (1\%) groups.  As expected we again see the tendency
towards stronger grouping when $\gamma$ is increased. 

We also did experiments were the value of $p$ was changed. If the
value of $p$ is close to 0.5 the tendency to group the
configurations too much becomes stronger. This makes perfectly sense,
since the correct grouping for $p=0.5$ is to put all
configuration sets into only one group. In the other end, choosing $p$
closer to 0 or 1 gives a stronger tendency to group the configurations
according to the correct solution. This illustrates the fact that
the algorithm tries to find a good model for the data using as few
groups as possible, but as the difference between the true parameter values
of the groups becomes larger the price to pay for choosing a model with fewer
parameters increases.

\section{Red deer data with $3\times 3$ maximal cliques}
\label{sec:S6}

In this section we present some results when assuming maximal clique size of
$3\times 3$ for the red deer data set presented in Section 5.2 in the
paper. The main drawback
with our approach is computational time, which is very dependent on
the approximation parameter $\nu$. One also needs to keep in mind that even data from simple models
will need many groups in the $3 \times 3$ case to be modelled correctly. For instance, for the
independence model the 401 configuration sets would need to be separated
into 10 groups, while for the Ising model one would need 11 groups to
get the correct model grouping. Similarly, the posterior most
probable grouping found for the $2 \times 2$ case for the reed deer
example would need 38 groups to be modelled in the $3\times 3$
case. Thus it is important not to assume larger maximal cliques than
needed. However for this data set it is possible to run the sampling algorithm
with $3\times 3$ clique, even though this is computationally
expensive. 

To get convergence we need a small generalisation to the
proposal distribution for the trans-dimensional sampling step
presented in Section \ref{sec:transDimProposal}. In particular we
allow for several configuration sets to be split out into a new group
at a single proposal,  and correspondingly allow for the
possibility of several configuration sets to be merge into another
group in one single proposal. The estimated marginal
distribution of the number of groups is 1\%, 65\%, 33\%, and 1\%
for 29, 30, 31 and 32 groups respectively. Three realisations from
the likelihood for
three randomly chosen realisations of $z$ are shown in Figure
\ref{fig:3t3}(a), and comparing with the realisations for
the $2\times 2$ case, see Figure 8 in the paper, it is hard
to see any differences in the spatial structure of the realisations.
\begin{figure*}[h]\center
Figure \ref{fig:3t3} approximately here.
\end{figure*}

We also investigated the distribution of three statistics for 5000
realisation from the likelihood of each of the two clique sizes, see Figure \ref{fig:3t3}(b), and it appears to be
little difference also here. These results indicate that $2\times
2$ maximal cliques might have sufficient complexity to explain this data set.          

\section{Parallelisation of the sampling algorithm}
\label{sec:S7}

Most of the computing time for running our sampling algorithm is used
to evaluate the likelihood in (2). In order to reduce
the running time we adopt a scheme that do multiple updates of the Markov chain by evaluating likelihoods in parallel.

Assume we are in a state $z$ and propose to split/merge into a new
state $z_1$. Now there are two possible outcomes for this
proposal. Either we reject the proposal, which results in state $z$, or
we accept the proposal, which results in state $z_1$. Either way we
always propose a parameter update in the next step, and proposing this
step from both the two states $z$ and $z_1$ before evaluating the
acceptance probability for the split/merge step is possible. The
possible outcomes for these three proposals are $z$, $z_1$, $z_2$
and $z_{12}$, where $z$ is the outcome where neither the split/merge
proposal nor the following parameter proposal is accepted, $z_1$ is the
outcome where the split/merge proposal is accepted but not the following
parameter proposal, $z_2$ is the outcome where the split/merge
proposal is not accepted but the parameter proposal is, and $z_{12}$ is the
outcome where both the split/merge proposal and the following parameter proposal are accepted. If we continue the argument we can do the same to
propose updates where configurations are moved from one group to
another group, and in the red deer example we even include a
level where updates of covariates are proposed. After making all
proposals we evaluate the likelihoods for each possible state in
parallel. The result is that we do need to evaluate too many
likelihoods, but if the number of CPUs that are available is larger
than or equal to the number of likelihoods we need to evaluate, a
computational gain close to the number of levels is obtained. The
updating scheme is illustrated in Figure \ref{fig:parallellScheme}.
\begin{figure*}[h]\center
Figure \ref{fig:parallellScheme} approximately here.
\end{figure*}

\counterwithin{figure}{section}
\setcounter{section}{5}
\setcounter{figure}{0}

\begin{figure}
  \centering
\subfigure[][Trace plot for the number of groups.]{\includegraphics[scale=0.65]{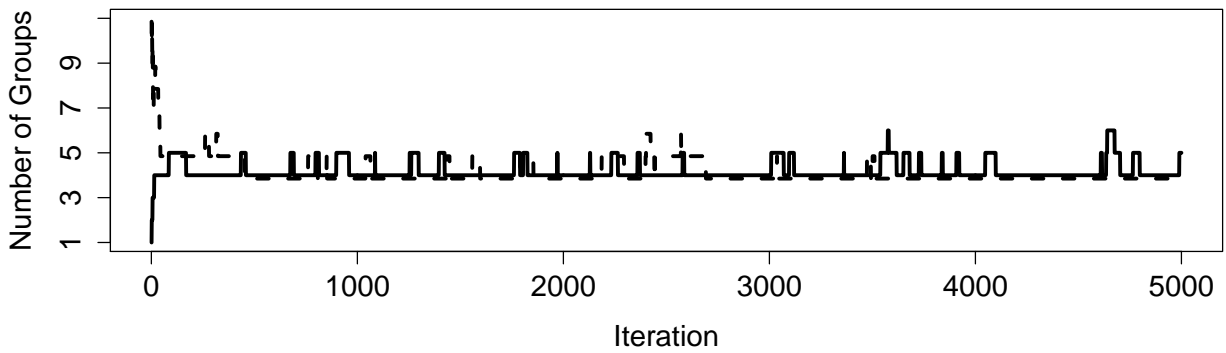}\label{fig:g1}}\\
\subfigure[][Trace plot for $\phi^0$(black), $\phi^1$(dark grey) and
$\phi^{\confcAAAA}$(light grey).]{\includegraphics[scale=0.65]{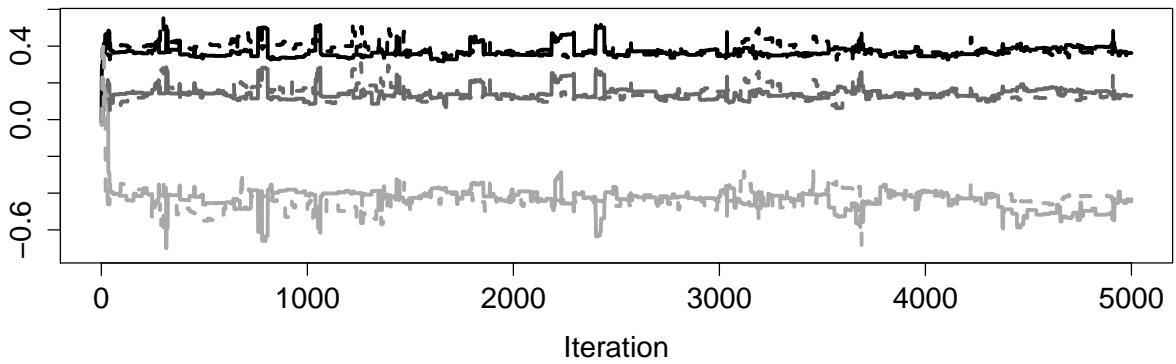}\label{fig:g2}}
\caption{Independence model example: Trace plots for the first quarter
  of the posterior simulation run. Solid curves are the result from a
  simulation where the initial number of groups is 1, and dashed
  curves are from a run with an initial value of 11 (maximal) number of groups.}
\label{fig:tracePlotsMarginal}
\end{figure} 

\begin{figure}
\centering
\begin{equation*}
  \arraycolsep=4.5pt\def\arraystretch{0.6}\begin{array}{cccccccccccc}
c^0 &\vv1.00&&&&&&&&&&\\
c^1 &&\vv1.00&&&&&&&&&\\
c^{11}&&&\vv1.00&\vv 0.94&\vv 0.88&\vv 0.90&0.13&0.77&&&\\
c^{\confcABAB} &&&\vv 0.94&\vv 1.00&\vv 0.85&\vv 0.93&0.12&0.75&&&\\
c^{\confcABBA} &&&\vv0.88&\vv0.85&\vv 1.00&\vv0.85&0.19&0.83&&&\\
c^{\confcBAAB} &&&\vv 0.9&\vv 0.93&\vv 0.85&\vv 1.00&0.12&0.75&&&\\
c^{\confcAAAB}&&&0.13&0.12&0.19&0.12&\vv 1.00&\vv 0.28&\vv0.78&\vv0.78&0.74\\
c^{\confcAABA}&&&0.77&0.75&0.83&0.75&\vv 0.28&\vv 1.00&\vv
0.12&\vv0.12 &0.09\\
c^{\confcABAA}&&&&&&&\vv0.78&\vv0.12&\vv 1.00&\vv 0.93&0.93\\
c^{\confcBAAA}&&&&&&&\vv0.78&\vv0.12&\vv0.93 &\vv 1.00&0.93\\
c^{\confcAAAA}&&&&&&&0.74&0.09&0.93&0.93&\vv 1.00\\
&c^0 & c^1 & c^{11} & c^{\confcABAB} &
c^{\confcABBA} & c^{\confcBAAB} & c^{\confcAAAB} & c^{\confcAABA} & 
c^{\confcABAA} & c^{\confcBAAA} & c^{\confcAAAA}
\end{array}
\end{equation*}
\caption{\label{fig:pairMatrixMarginal}Independence model example: Estimated posterior probabilities for two configuration
    sets to be grouped together. The correct grouping is shown
    in grey, and only probabilities larger than $5 \%$ are given.}
\end{figure}

\begin{figure}
  \centering
\subfigure[][$\betaABBB\ (-1.08,-0.80)$.]{\label{fig:b1}\includegraphics[scale=0.5]{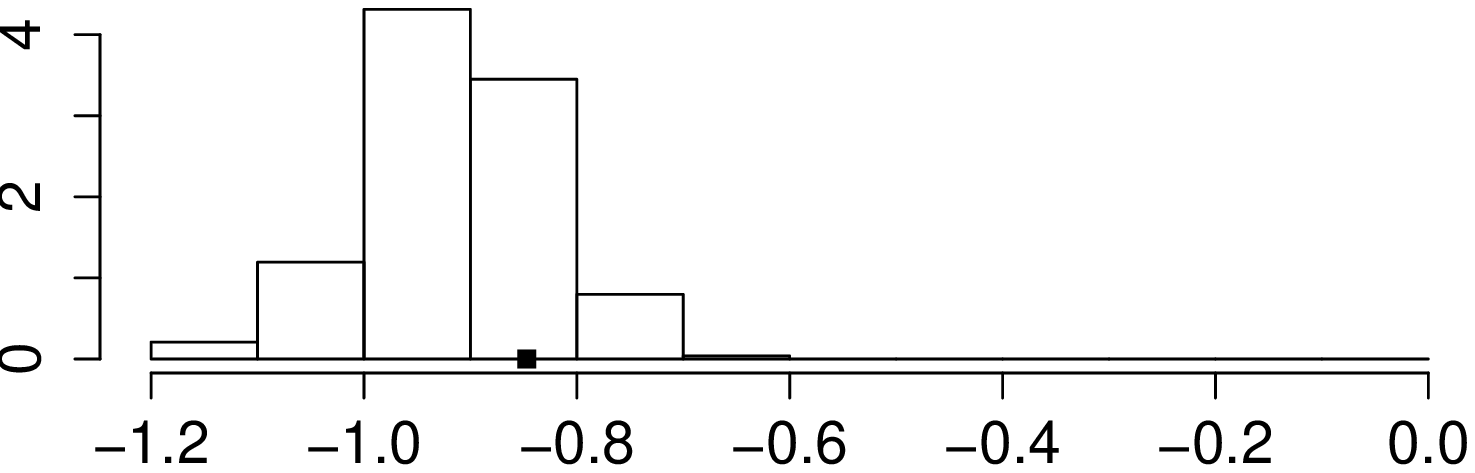}}
\subfigure[][$\betaAABB\ (-0.05,0.22)$.]{\label{fig:b2}\includegraphics[scale=0.5]{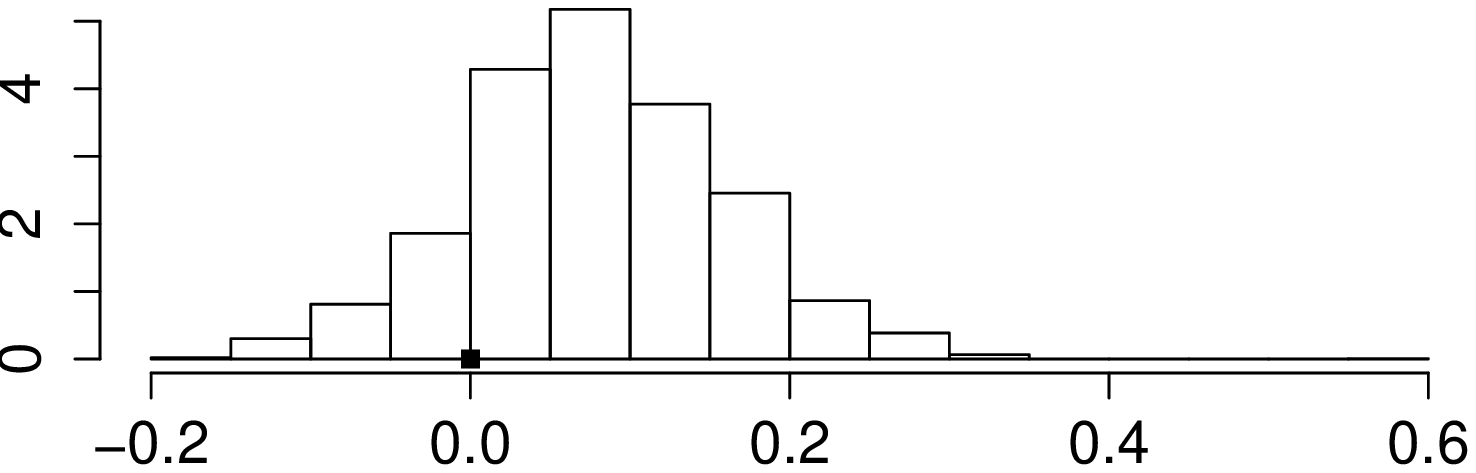}}\\
\subfigure[][$\betaABAB\ (-0.05,0.23)$.]{\label{fig:b3}\includegraphics[scale=0.5]{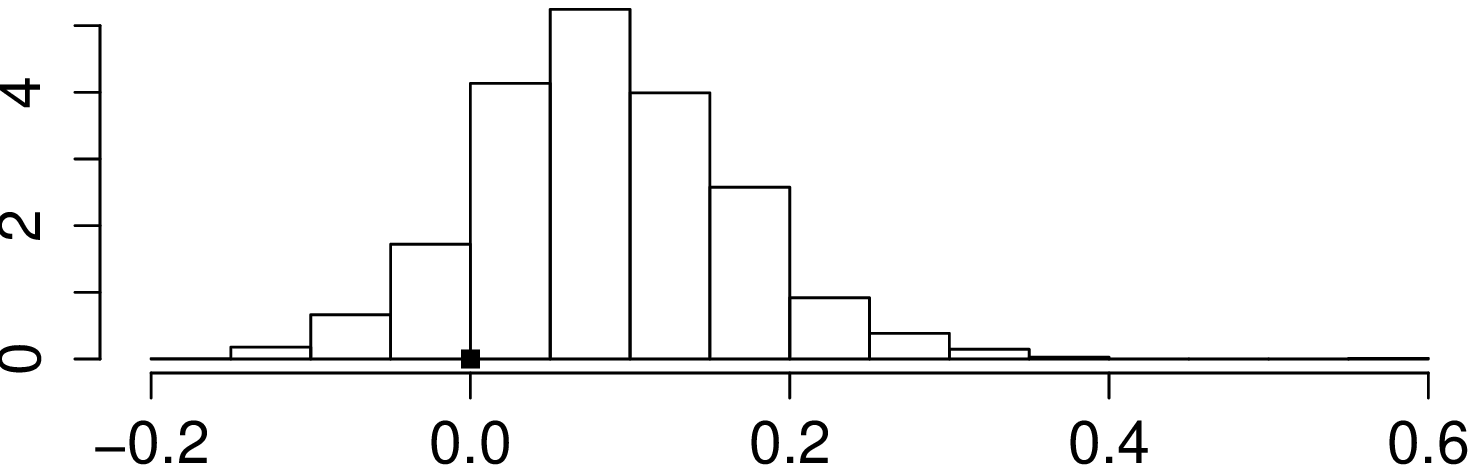}}
\subfigure[][$\betaABBA\ (-0.11,0.12)$.]{\label{fig:b4}\includegraphics[scale=0.5]{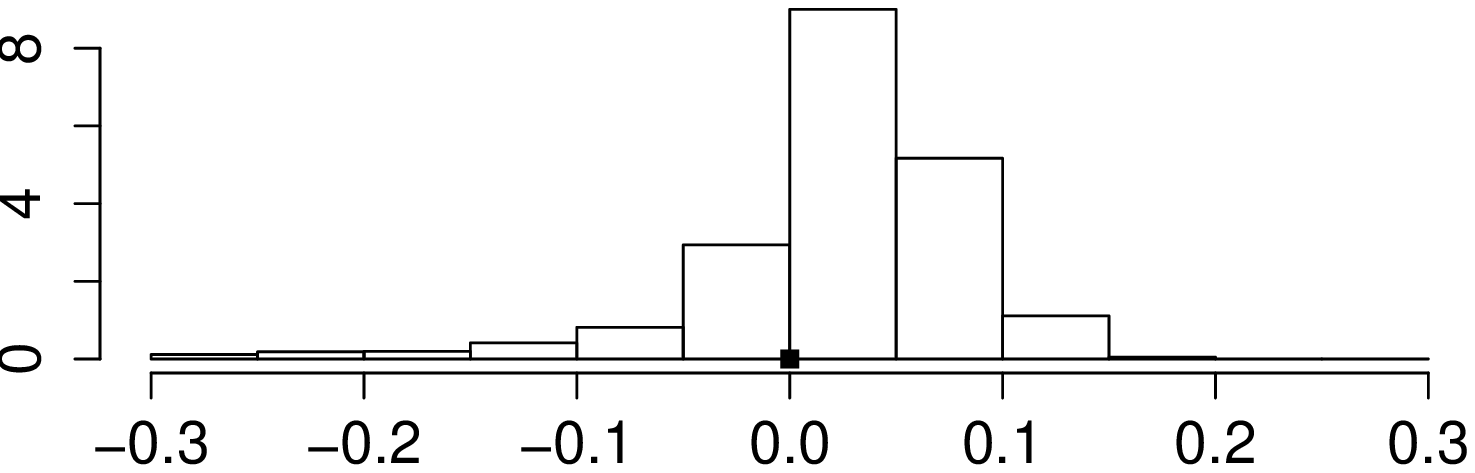}}\\
\subfigure[][$\betaBAAB\ (-0.03,0.17)$.]{\label{fig:b5}\includegraphics[scale=0.5]{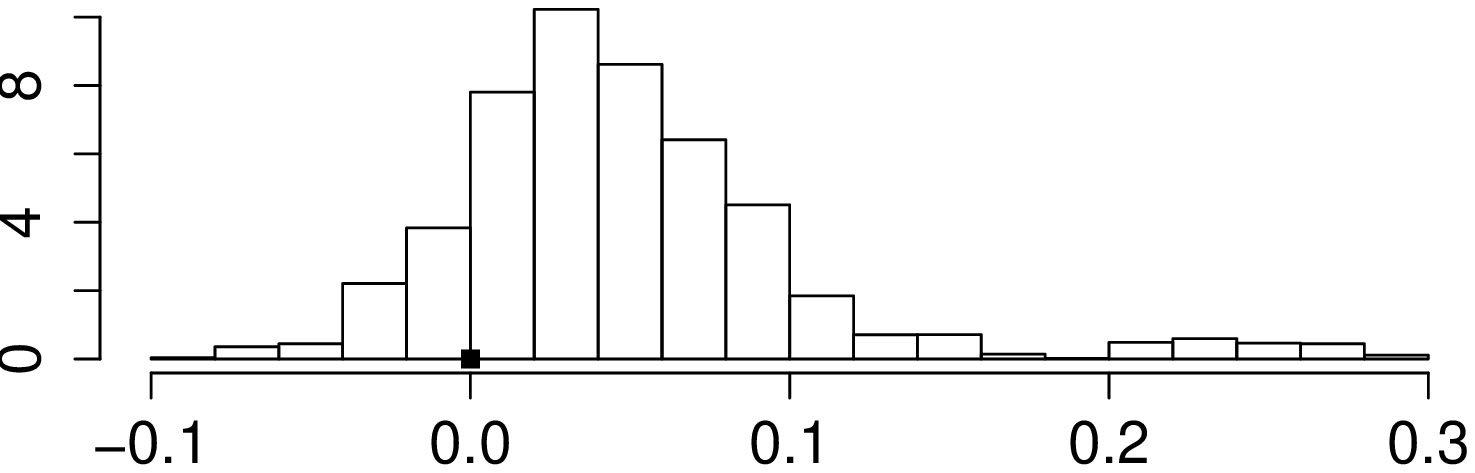}}
\subfigure[][$\betaAAAB\ (-0.37,0.21)$.]{\label{fig:b9}\includegraphics[scale=0.5]{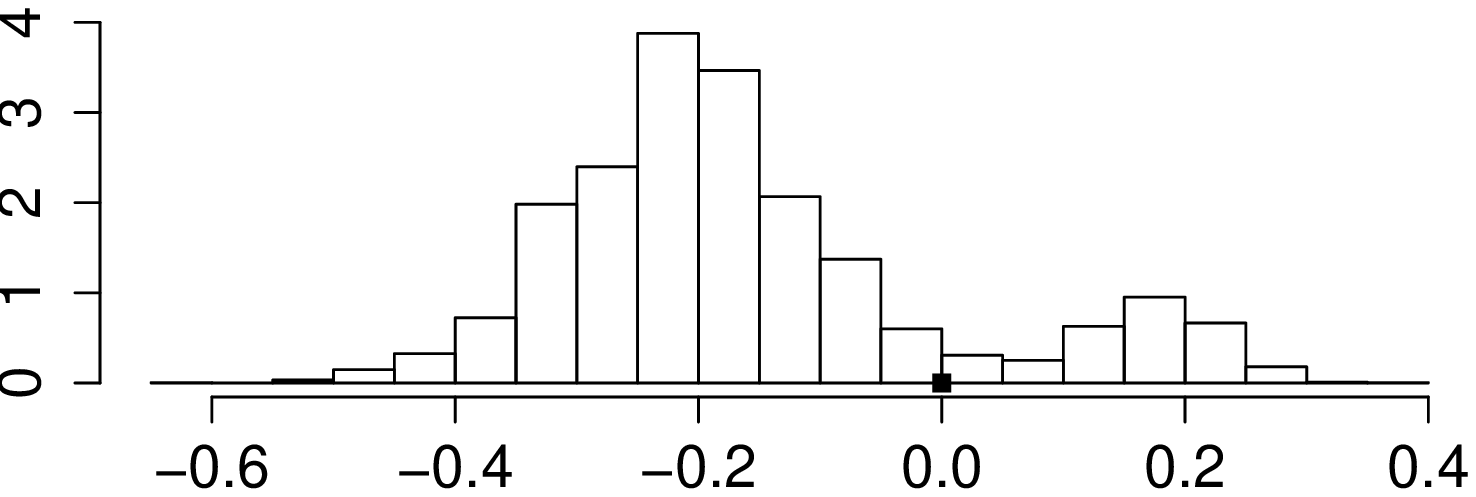}}\\
\subfigure[][$\betaAABA\ (-0.16,0.26)$.]{\label{fig:b8}\includegraphics[scale=0.5]{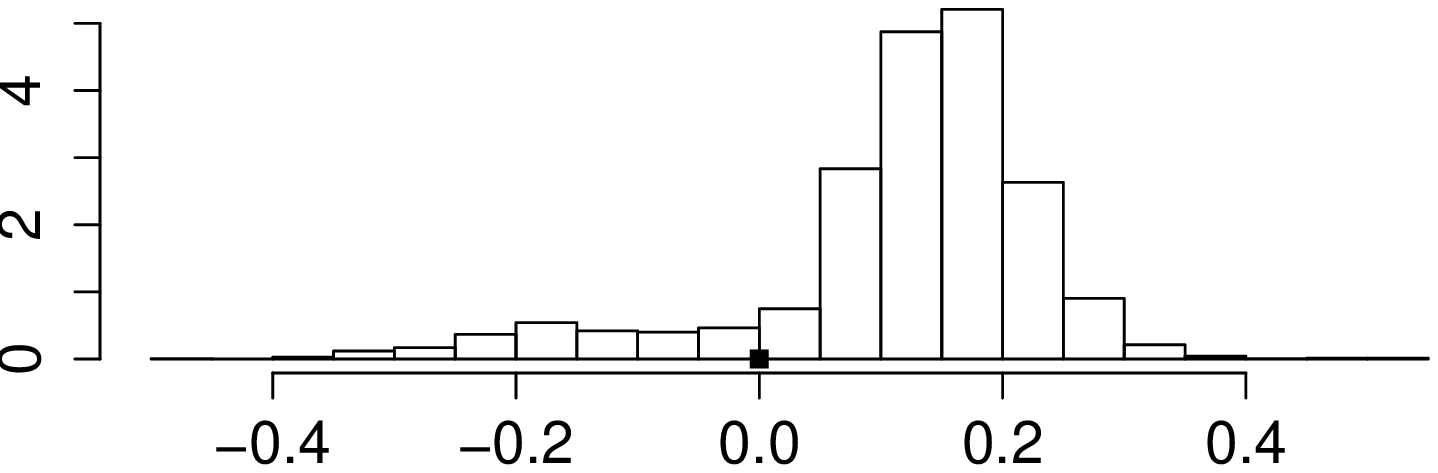}}
\subfigure[][$\betaABAA\ (-0.37,-0.05)$.]{\label{fig:b7}\includegraphics[scale=0.5]{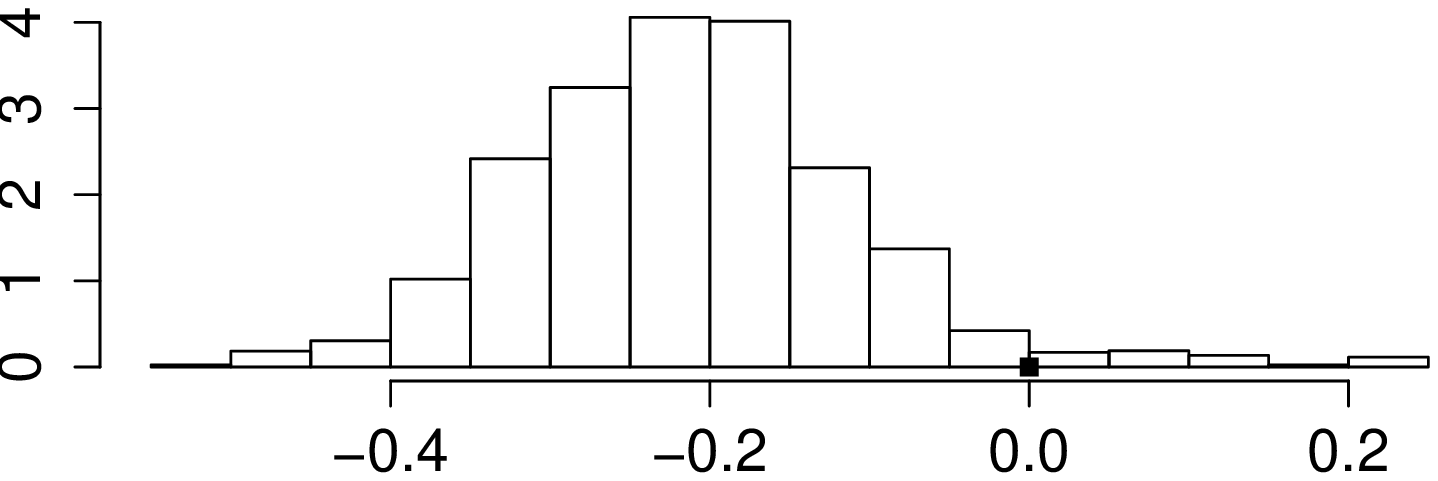}}\\
\subfigure[][$\betaBAAA\ (-0.42,-0.06)$.]{\label{fig:b6}\includegraphics[scale=0.5]{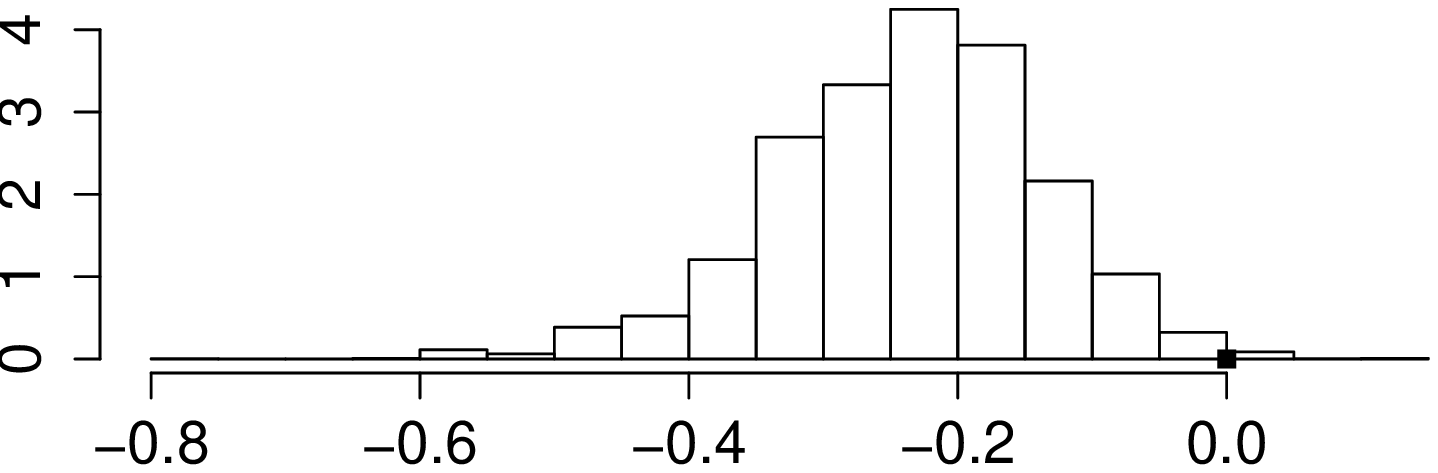}}
\subfigure[][$ \betaAAAA\ (-0.05,0.69)$.]{\label{fig:b10}\includegraphics[scale=0.5]{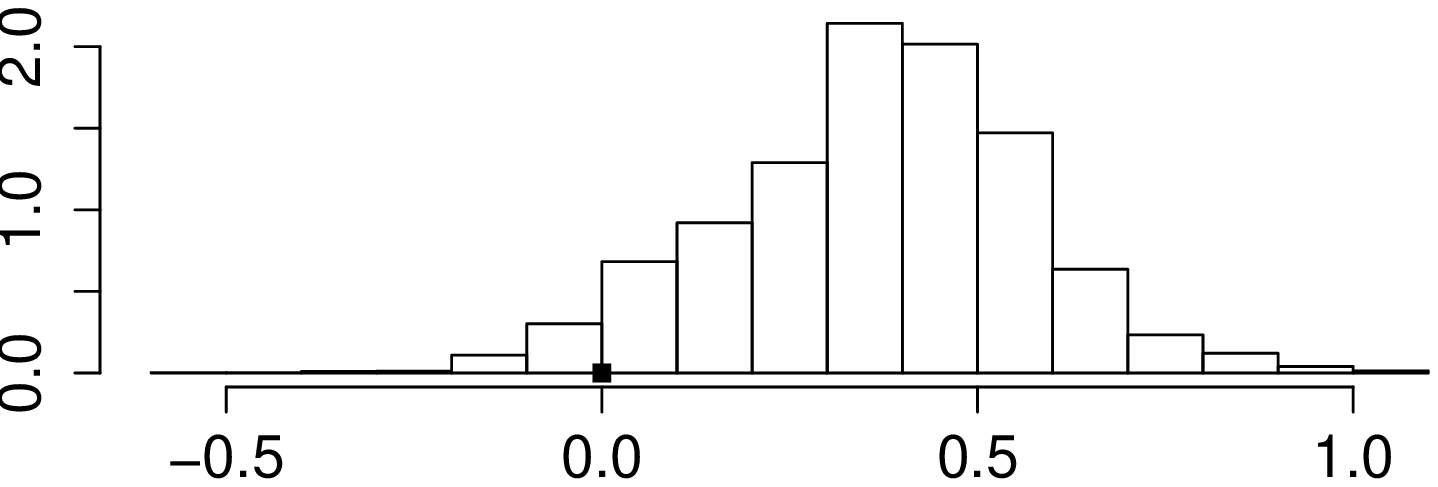}}
\caption[Caption For LOF]%
{\label{fig:marginalInteractionParameters}Independence model example: Estimated marginal posterior distribution of the interaction parameters. True
  values are shown with a black dot and estimated 95\% credibility interval is given for
  each parameter.}
\end{figure}

\begin{figure}
  \centering
\subfigure[][$g(x)=\sum_ix_i$]{\label{fig:i1}\includegraphics[scale=0.5]{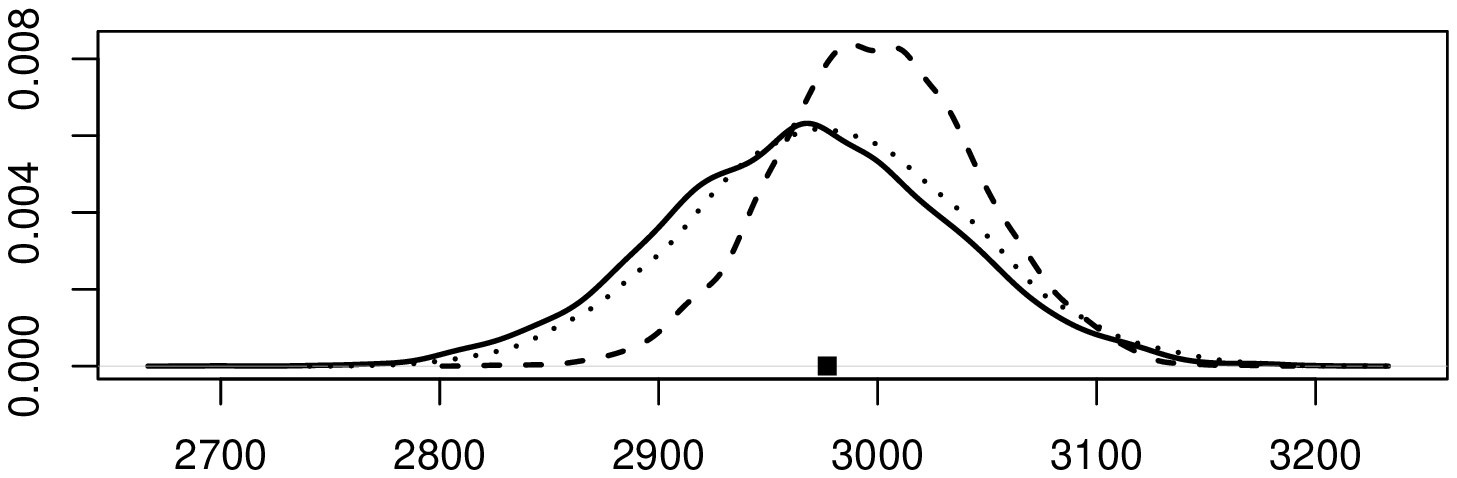}}
\subfigure[][$g(x)=\sum_{i,j:\text{vertical adjacent sites}}I(x_i=x_j)$]{\label{fig:i2}\includegraphics[scale=0.5]{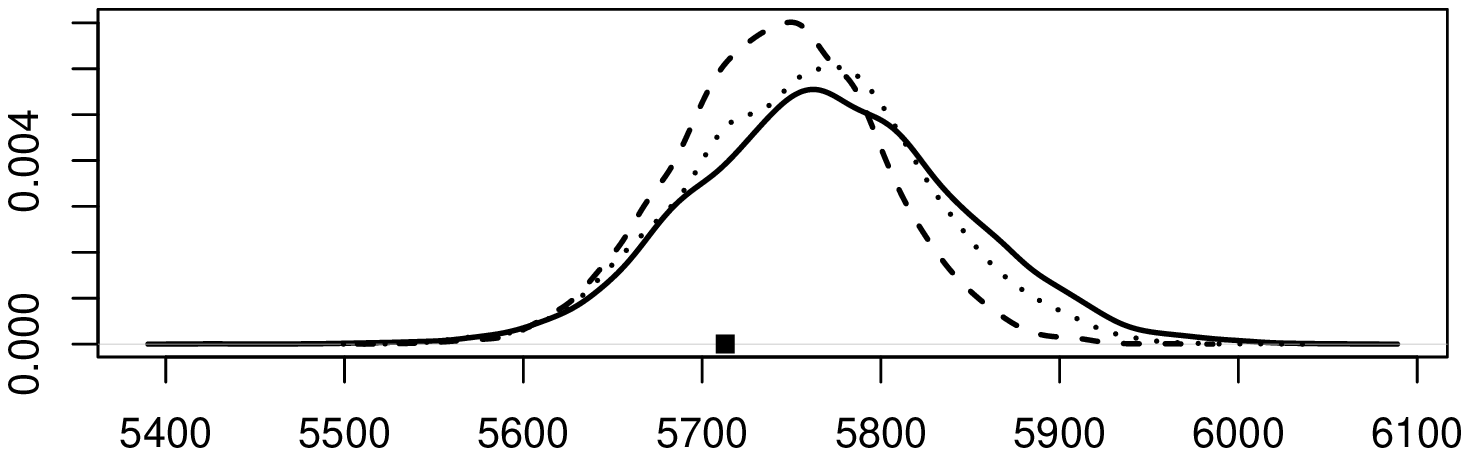}}\\
\subfigure[][$g(x)=\sum_{i,j:\text{horizontal adjacent sites}}I(x_i=x_j)$]{\label{fig:i3}\includegraphics[scale=0.5]{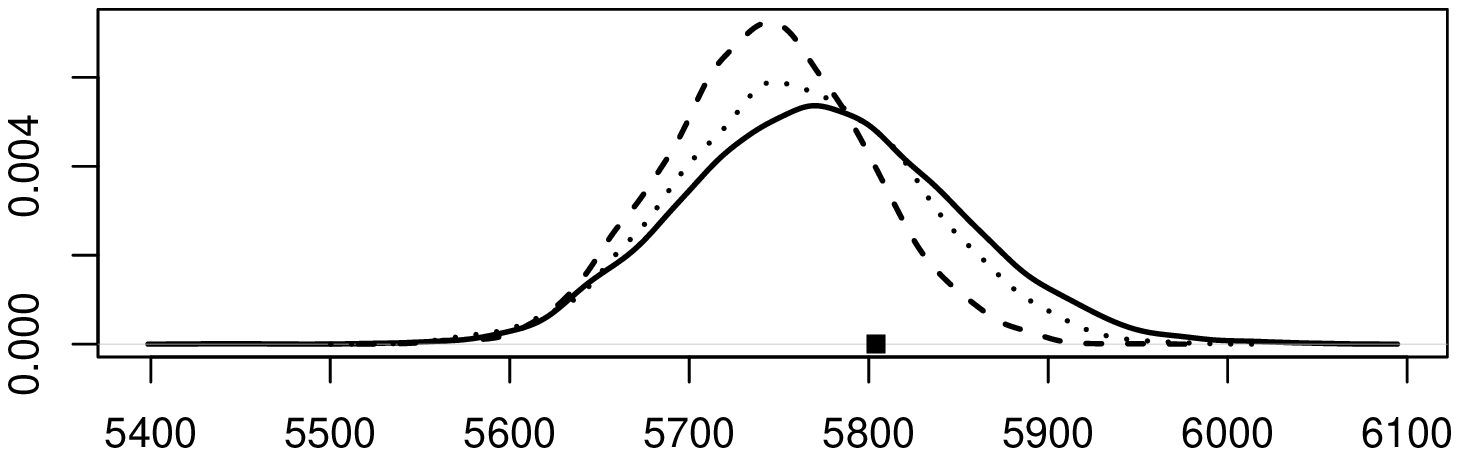}}
\subfigure[][$g(x)=\sum_{\Lambda\in\mathcal{L}_m}I\left( x_{\Lambda}=
  {\left[\confBBBB\right]}\right)$]{\label{fig:i4}\includegraphics[scale=0.5]{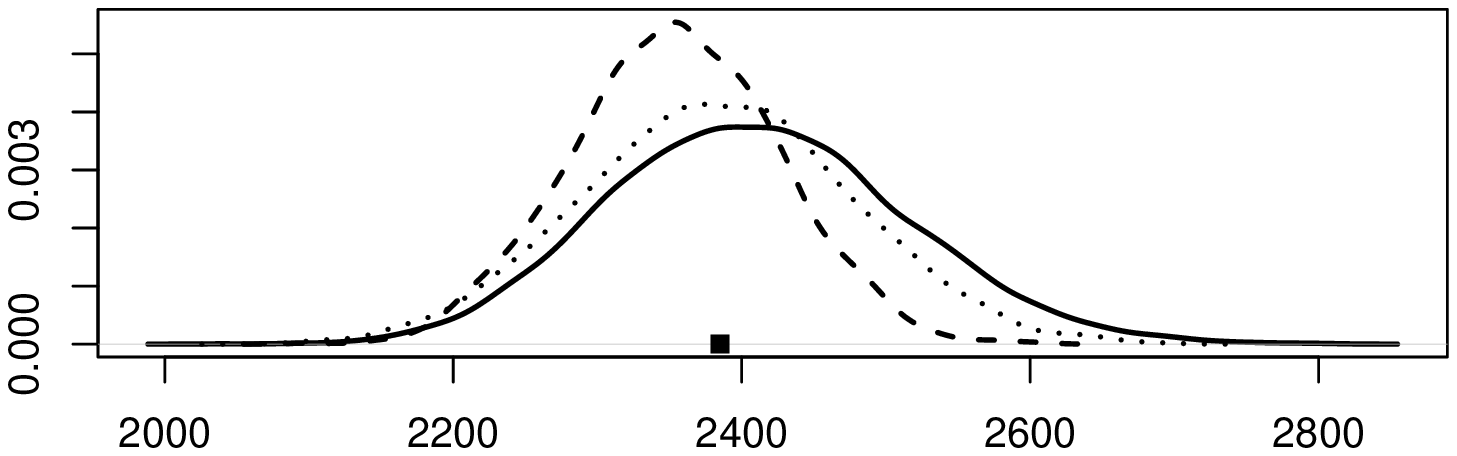}}\\
\subfigure[][$g(x)=\sum_{\Lambda\in\mathcal{L}_m}I\left ( x_{\Lambda}=
  {\left[\confBAAB\right]}\right)$]{\label{fig:i5}\includegraphics[scale=0.5]{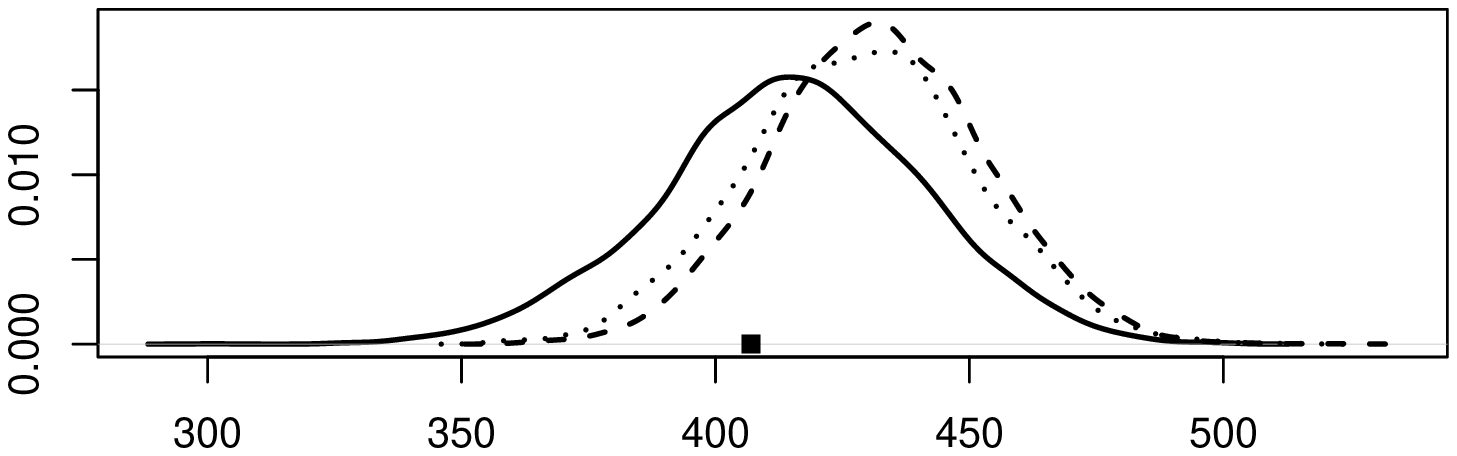}}
\subfigure[][$g(x)=\sum_{\Lambda\in\mathcal{L}_m}I\left ( x_{\Lambda}=
  {\left[\confAAAB\right]}\right)$]{\label{fig:i6}\includegraphics[scale=0.5]{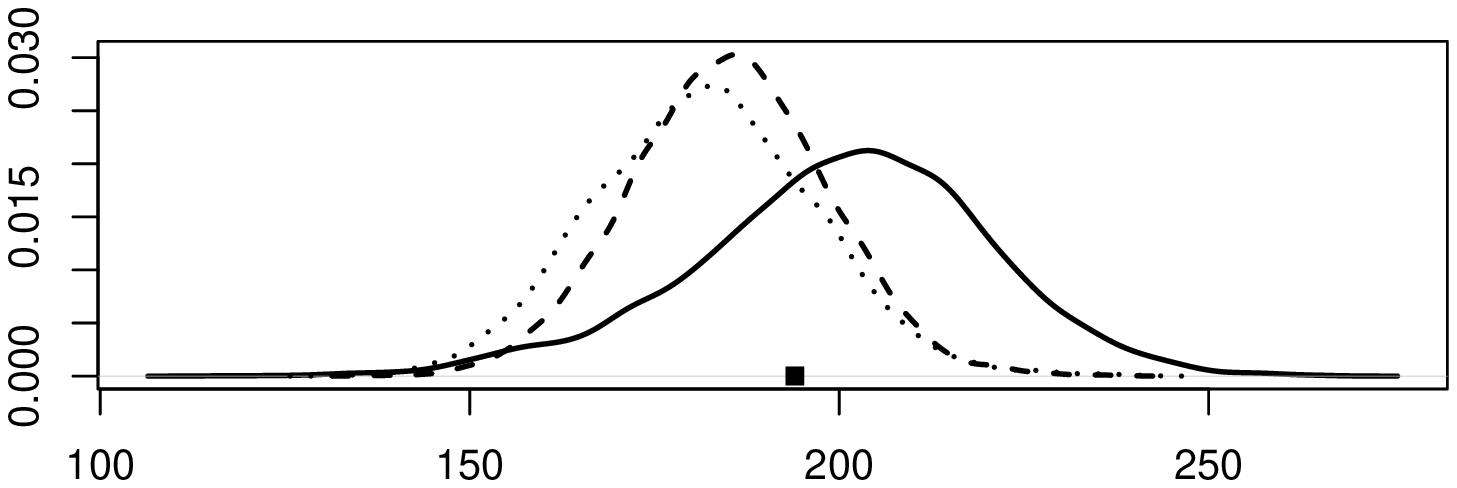}}
\caption{\label{fig:compareSimIndependence}Independence model example: Distribution of six statistics of realisations from our $2\times 2$ model with
posterior samples of $z$ (solid), the independence model with correct
parameter value (dashed), and the independence model with posterior samples of the
parameter value (dotted). The data evaluated with
each statistic is shown with a black dot.}
\end{figure}

\counterwithin{figure}{section}
\setcounter{section}{6}
\setcounter{figure}{0}

\begin{figure}
  \centering
\subfigure[Three realisations from the likelihood for three random
samples of $z$ from the posterior distribution.]{\label{fig:simulate_3t3}\includegraphics[scale=0.5]{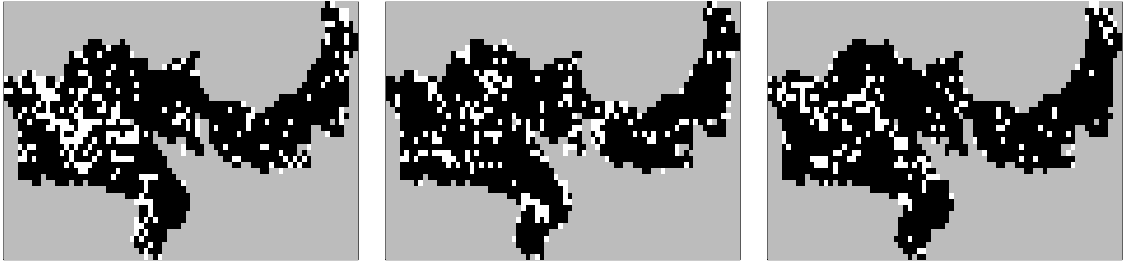}}\\
\subfigure[][Distribution of three functions of realisations from the
likelihood with $3\times 3$ cliques (solid) and $2\times 2$ cliques (dashed). The three functions are $g(x)=\sum_{\Lambda\in\mathcal{L}_m}I\left ( x_{\Lambda}={\left[\confBBBBBBBBB\right ]}\right )$ (left), $g(x)=\sum_{\Lambda\in\mathcal{L}_m}I\left ( x_{\Lambda}=
  {\left[\confBBBBBBBAA\right ]}\right )$ (middle), and $g(x)=\sum_{\Lambda\in\mathcal{L}_m}I\left ( x_{\Lambda}=
  {\left[\confAAAAABABB\right ]}\right )$ (right). ]{\label{fig:gfunc_Deer_3t3}\includegraphics[scale=0.5]{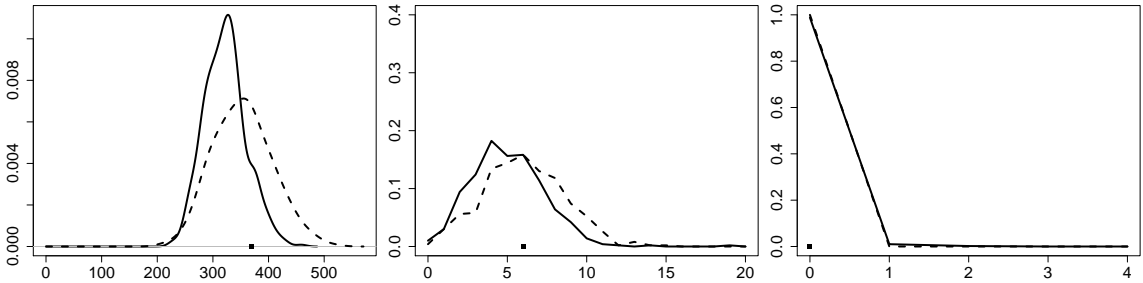}}
\caption{Red deer $3\times 3$ example: Posterior results with $\gamma=0.5$.}
                \label{fig:3t3}
\end{figure}

\counterwithin{figure}{section}
\setcounter{section}{7}
\setcounter{figure}{0}
 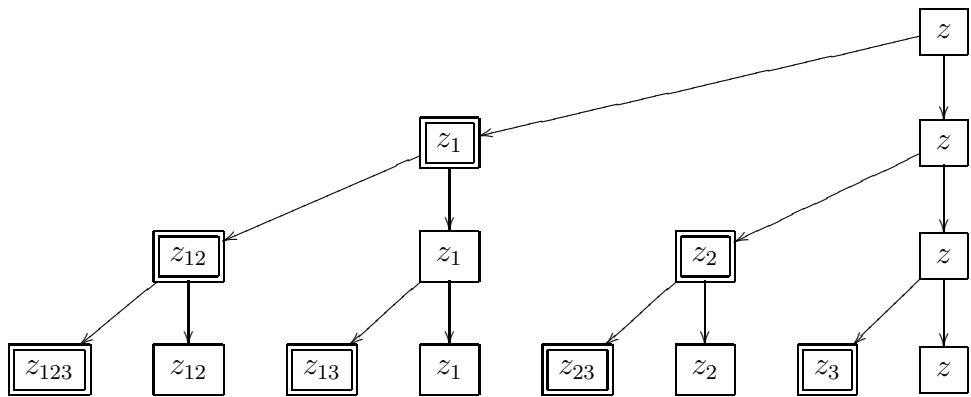
\begin{figure}
  \centering
\centerline{
\xymatrix{
  \\
&&&&&&&*++[F]{z}\ar[d]\ar[dllll]\\
&&&*++[F=]{z_1}\ar[d]\ar[dll]&&&&*++[F]{z}\ar[d]\ar[dll]\\
&*++[F=]{z_{12}}\ar[d]\ar[dl]&&*++[F]{z_1}\ar[d]\ar[dl]&&*++[F=]{z_2}\ar[d]\ar[dl]&&*++[F]{z}\ar[d]\ar[dl]\\
*++[F=]{z_{123}}&*++[F]{z_{12}}&*++[F=]{z_{13}}&*++[F]{z_1}&*++[F=]{z_{23}}&*++[F]{z_2}&*++[F=]{z_3}&*++[F]{z}
  }}
\vspace{0.25cm}
  \caption{Proposal scheme for parallel likelihood evaluations. Starting in model $z$, proposals are made down the graph. Arrows pointing straight down represents rejection of proposal while arrow pointing down and left represent acceptance. Double squares are used to represents states where a new likelihood evaluation is needed.}
  \label{fig:parallellScheme}
\end{figure}